\documentclass[]{emulateapj}

\def\mathbi#1{\textbf{\em{\hspace{-2pt}#1}}}

\shorttitle{DISCUSSION OF THE ELECTROMOTIVE FORCE TERMS IN THE MODEL OF GALACTIC DISKS}
\shortauthors{OTMIANOWSKA-MAZUR, KOWAL \& HANASZ}

\begin{document}

\title{Discussion of the electromotive force terms in the model of Parker
       unstable galactic disks with cosmic rays and shear}
\author{Katarzyna Otmianowska-Mazur\altaffilmark{1},
        Grzegorz Kowal\altaffilmark{2} and Micha{\l} Hanasz\altaffilmark{3}}
\altaffiltext{1}{Astronomical Observatory, Jagiellonian University,\\ ul. Orla 171, 30-244 Krak\'{o}w, Poland}
\altaffiltext{2}{Department of Astronomy, University of Wisconsin, 475 North Charter Street, Madison, WI 53706, USA}
\altaffiltext{3}{Toru\'{n} Centre for Astronomy, Nicholas Copernicus University,\\ 87-148 Piwnice/Toru\'{n}, Poland}

\begin{abstract}
We analyze the electromotive force (EMF) terms and basic assumptions of the linear and nonlinear dynamo theories in our three-dimensional (3D) numerical model of the Parker instability with cosmic rays and shear in a galactic disk. We also apply the well known prescriptions of the EMF obtained by the nonlinear dynamo theory (Blackman \& Field 2002 and Kleeorin et al. 2003) to check if the EMF reconstructed from their prescriptions corresponds to the EMF obtained directly from our numerical models. We show that our modeled EMF is fully nonlinear and it is not possible to apply any of the considered nonlinear dynamo approximations due to the fact that the conditions for the scale separation are not fulfilled.
\end{abstract}


\section{Introduction}
\label{sec:intro}

It seems that the issue of the magnetic field amplification in galaxies may be well explained  by the two main physical mechanisms: the Parker instability (PI), which takes into account the cosmic rays (CR) and the shear \cite[e.g.][and references therein]{hanasz03,hanasz04}, and the magneto-rotational instability \cite[MRI, e.g.][]{dziourkevitch04,kitchatinov04}. The possible scenario of the magnetic field evolution could be presented as follows: when the protogalaxy starts to rotate differentially, the MRI mechanism occurs, and this results in a very efficient magnetic field amplification even to the level of $\mu$G with the global e-folding time of 100~Myr or even less \citep{dziourkevitch04}. Simultaneously, the quadrupole symmetry of the large-scale magnetic field is being created. MRI also causes the turbulent motions in the galactic disks. The process of supernovae (SN) explosions, that arises in young galactic objects \cite[e.g.][and references therein]{widrow02} suppresses the MRI mechanism. In the same time, the cosmic rays produced in SN remnants may induce the Parker instability process \cite[e.g.][]{parker92,hanasz04}. Hence, we may conclude that during the early stage of galaxy evolution the MRI process is being replaced by the PI mechanism.

The local simulations of the large-scale magnetic field took into account the turbulent dynamo theory and the  magnetic back-reaction onto the turbulent motions \citep{piddington70,piddington72a,piddington75a,kulsrud95}. The authors drew the conclusion that it was difficult to obtain the amplification of the total magnetic field \cite[e.g.][and references therein]{widrow02}. It might be explained by the equipartition of the random magnetic field component with the random turbulent motions. Such process suppresses the dynamo action at later times. Moreover, the magnetic field amplification might be easily stopped even by the presence of the weak large-scale magnetic field \citep{cattaneo91,vainshtein92,cattaneo94,cattaneo96,ziegler96}. \cite{blackman99} explained that papers analyzing the analytically strong suppression of the dynamo coefficient $\alpha$ should distinguish between different state orders of the turbulent quantities. However, their analysis did not completely solve the problem of the quenching of the dynamo coefficients. In their next paper \citep{blackman00}, they proved that the results from the \cite{cattaneo96} model were based on the assumption about the periodicity of the boundary conditions. Nevertheless, the quenching effects could also appear even when the open boundary conditions were applied \cite[e.g.][]{brandenburg01b}.

The classical dynamo theory does not conserve the total magnetic helicity \cite[see e.g.][BF02]{blackman02}. In media characterized by the high magnetic Reynolds number ($R_m\gg1$) the total helicity should remain constant in closed regions \cite[e.g.][]{berger84,brandenburg02b,brandenburg05b,subramanian02}. The permanent helicity is also an additional factor, which suppresses the dynamo activity \citep[e.g.][]{brandenburg02b}. The following papers \citep{blackman00,kleeorin00,blackman03,kleeorin99,kleeorin02,kleeorin00, kleeorin03,rogachevskii00,rogachevskii01} presented the two methods that allowed the modeling of the dynamo action evading the problem of the constant helicity. The first method is based on the ejection of the magnetic helicity through boundaries. The second one uses the creation of the negative and positive helicity at the large and small scales respectively \cite[see][]{brandenburg02b,kleeorin02,kleeorin03}. \cite{blackman02} in their next paper analyzed the nonlinear prescription of both dynamo coefficient $\alpha$ and $\beta$. Both factors were obtained without any linearization and took into account all terms in the equation of the evolution of the fluctuating part of the magnetic field (see Eq.~\ref{eqn:fluct_field_evol}). The results were similar to the dynamo coefficients obtained by \citep{pouquet76}, but the units were different (without the time integration). They also solved the small- and large-scale helicity dynamo equations numerically simultaneously with the equation for the EMF time evolution. That allowed them to obtain the growth of the large-scale magnetic field in the kinematic phase.

The new form of the dynamo coefficients for anisotropic turbulent motions with
the presence of the large-scale magnetic field was presented by
\cite{rogachevskii01}. They calculated dynamo coefficients according to the
\cite{raedler80} EMF prescription \cite[see also][]{kowal05}, which neglected all
quadratic  terms in the mean field in the EMF. \cite{kleeorin03} used those forms
of the dynamo coefficients to solve numerically the dynamo equation in the local
thin-disc approximation. The authors took into account the quenching of both
coefficients, $\alpha$ and $\beta$, and helicity flux through the boundary. They
found that it was  possible to obtain the growth of the large-scale magnetic
field when $\alpha$-quenching was only analyzed \citep{kleeorin02}. If the model
included also $\beta$-quenching, no growing solution of the dynamo
\citep{kleeorin03} could be obtained.  Both in Blackman-Field and in
Kleeorin-Rogachevskii approaches there is an $\alpha_{\rm m}$ term that
quantifies the small scale helicity current. This term depends on the current
helicity flux, and there are different theories for this flux. In Kleeorin et al.
a heuristically motivated expression for the flux was used, in
\cite{brandenburg05b}  the \cite{vishniac01} flux was used, and in
\cite{shukurov06} a simple advective flux was used. In all these cases the
current helicity flux allows the field to saturate at high levels.

The latest research on the turbulent enforcement in the solar convective zone calculated in the local cube with the shear has shown that the open boundaries help to obtain the amplification of the large-scale magnetic field even without the helicity of turbulent motions \cite[end references therein]{brandenburg05a,brandenburg05b}. We have to stress that their result, an increase of the total magnetic energy, was obtained without any assumption considering the additional EMF of dynamo. The authors applied isotropic and homogeneous turbulence with and without the helical forcing in their model. On the other hand, Brandenburg and his collaborators \citep{brandenburg05a} interpreted their results in terms of the mean field dynamo theory. The time evolution of $\alpha$ was obtained from the calculated electromotive force \citep{brandenburg04,kowal05,brandenburg02a,brandenburg01a}. \cite{brandenburg04} explained that thanks to the flux of the current helicity flowing out of the cube the process of $\alpha$-quenching tends to be not as disastrous as forseen. The only conditions are the intermediate level of $R_m$ and open boundaries. When the high value of $R_m$ is applied to the model, the comparatively lower value of the large-scale magnetic field strength are obtained \citep{brandenburg04}. However, in astrophysical objects $R_m$ is always high. That is why this result seems to be peculiar. On the other hand, the authors made it clear that the total magnetic energy grows mainly in the kinematic phase of the dynamo and their results do not depend strongly on $R_m$. Futhermore, they calculated the value of $\alpha$ coefficient based on the modeled EMF. The obtained factor was similar to the same coefficient calculated according to the \cite{kleeorin00,kleeorin02,kleeorin03} prescriptions. We believe that the fact that such similarity occurs results from the isotropic and homogeneous turbulence in both models \citep{brandenburg05a,brandenburg05b,brandenburg04}. It may also happen due to the fact that the authors applied standard dynamo approximations, which neglect the quadratic terms in the mean field in EMF.

The previously mentioned results indicate that the realistic physical simulations are of the great importance when the MRI \citep{dziourkevitch04} and the PI \citep{hanasz04,hanasz03,kowal05} processes are considered. Both models, which meet enumerated requirements, showed that it was possible to amplify galactic magnetic field efficiently. \cite{hanasz03} and \cite{hanasz04,hanasz05} presented that the following two processes: the Parker instability driven by cosmic rays from supernovae and the shear from the differential rotation enable the magnetic field amplification (with the e-folding time scale of 250~Myr or even 140~Myr). The model also applied realistic gravity according to the \cite{ferriere98} prescriptions.

The idea of obtaining the dynamo coefficients ($\alpha$ and $\beta$) from the electromotive force calculated in the local numerical simulations proved to be essential for many other authors too \cite[e.g.][etc]{ziegler96,brandenburg02a}. We included the calculations of the dynamo coefficients, which we obtained from the calculated EMF in our previous paper \citep{kowal05}. The application of the statistical methods provided us with the acceptable values of the dynamo $\alpha$-tensor. On the other hand, the values of $\beta$ coefficients were negative. Such values are inconsistent with the R\"adler prescription \citep{raedler80}. This may be caused either by the applied statistical method, which does not take into account physical differentiation, or by the linear EMF approximation \citep{kowal05}. For this reason we decided to analyze that matter in our present study. We search for the conditions, which should be fullfiled in order to make linearization of the electromotive force in the mean field dynamo theory possible. We would like to examine the following problems: the scale separation, the ratios of the terms in the equation for the small-scale magnetic
field evolution \cite[e.g.][]{raedler80}, the magnitude of the turbulent kinetic energy in comparison to the large-scale magnetic one. Next, we plan to apply the estimations of the dynamo coefficients derived by
\cite{blackman00,rogachevskii01} and \cite{brandenburg04} to our models. We would like to check if their
approximations fit into the calculated electromotive force in our models \cite{hanasz04}.   In this part of this
work we do not include the explicit analysis involving the conservation of the magnetic helicity. The investigation
of the magnetic helicity conservation in our models is already advanced, but it is complex enough to
be described in separate paper, which is under preparation.
We cannot apply the approximations of \cite{ruediger93} and \cite{kitchatinov94} derived from the EMF quenching by
the usage of the Second Order Correlation Approximations. They assumed that the Strouhal number $S$ is essentially
smaller than 1 ($S\ll1$, where $S = \tau_c \times \rm v/ \rm l_c$). This assumption is not fullfiled in our
numerical experiments, where $S$ is about 1. Finally, we discuss our results.

\begin{table*}
 \begin{center}
 \caption{Parameters of models examined in this paper. (*) The conversion rate,
          presented as 10\% in \cite{hanasz04}, was in fact equal to 100\%,
          due to a trivial calculation mistake. Therefore, the overall
          injection rate of cosmic ray energy was equivalent to a realistic
          one, corresponding to SN rate=~20~kpc$^{-2}$Myr$^{-1}$, with the
          energy conversion factor = 10\%.
\label{table-params}}
  \begin{tabular}{|l|c|c|c|c|}
   \hline
    Model   & A & B & C & D\\
   \hline
    Domain sizes [kpc] & 0.5 $\times$ 1 $\times$ 1.2 & \multicolumn{3}{c}{0.5 $\times$ 1 $\times$ 4} \vline \\
    Resolution         & 50  $\times$ 100 $\times$ 120& \multicolumn{3}{c}{50 $\times$ 100 $\times$ 400} \vline \\
    Vertical gravity at $R$ [kpc] = & 8.5 & \multicolumn{3}{c}{5} \vline\\
    Gas column density [cm$^{-2}$] & ... & \multicolumn{3}{c}{27$\times$10$^{20}$} \vline\\
    Angular velocity $\Omega$ [Myr$^{-1}$]  & 0.05 & \multicolumn{3}{c}{0.05}  \vline \\
    SN rate [kpc$^{-2}$Myr$^{-1}$]           & $2$    & \multicolumn{3}{c}{130} \vline \\
    Initial $\alpha=e_{\rm mag}/e_{\rm gas}$ & $10^{-8}$ & \multicolumn{3}{c}{$10^{-4}$}\vline \\
    Diffusion coefficients $K_\parallel$, $K_\perp$ [cm$^2$s$^{-1}$]  & $3 \times 10^{27}$, $3 \times 10^{26}$ & \multicolumn{3}{c}{$3 \times 10^{27}$, $3 \times 10^{26}$}\vline \\
    Conversion rate of SN kinetic to CR energy  &  10\%$^{(*)}$  & \multicolumn{3}{c}{10\%}\vline \\
\hline
    Resistivity $\eta$ [cm$^2$s$^{-1}$]    & $3\times 10^{24}$ & $0\times 10^{24}$ &
                        $3\times 10^{24}$ & $30\times 10^{24}$  \\
\hline
\end{tabular}
\end{center}
\end{table*}

%
\section{Numerical models of the cosmic-ray driven dynamo}
\label{sec:numerical_model}

The first complete 3D numerical model of the CR-driven dynamo has been demonstrated by \cite{hanasz04,hanasz05}. The presented numerical simulations were performed with the Zeus-3D MHD code \citep{stone92a,stone92b} extended with: the cosmic rays \citep{hanasz03}, supplied by the supernova remnants randomly exploding in the disk  volume, the resistivity of the ISM leading to the magnetic reconnection, the shearing-periodic boundary conditions, the rotational pseudo-forces and the realistic vertical disk gravity. The full set of equations describing the model includes the set of resistive MHD equations completed by the cosmic ray transport  equation \cite[see][]{hanasz04}, where anisotropic diffusion of cosmic rays is implemented following \cite{ryu03}.

The principle of the action of the CR-driven dynamo is based on the cosmic ray energy supplied continuously by SN remnants. Due to the anisotropic diffusion of cosmic rays along the horizontal magnetic field lines, cosmic rays tend to accumulate within the disc volume. However, the configuration stratified by the vertical gravity is unstable against the Parker instability. Buoyancy effects induce the vertical and horizontal motions of the fluid and the formation of the undulatory patterns -- magnetic loops in the frozen-in, predominantly horizontal magnetic fields. The presence of rotation in galactic disks implies a coherent twisting of the loops by means of the Coriolis force, which leads to the generation of the small-scale radial magnetic field components. The next phase is merging the small-scale loops by the magnetic reconnection process to form the large scale radial magnetic fields. Finally, the differential rotation stretches the radial magnetic field to amplify the large-scale azimuthal magnetic field. The coupling of amplification processes of the radial and azimuthal magnetic field components gives rise to an exponential growth of the large scale magnetic field. The timescale of the magnetic field amplification, resulting from the action of CR-driven dynamo, has been found to be about 250 Myr in typical galactic conditions \citep{hanasz04}, close to the galactic rotation timescale.

In our  model of the cosmic rays driven Parker instability with shear \citep{hanasz04,hanasz05} we have included the following physical elements: the cosmic ray component described by the diffusion-advection transport equation \cite[see][for the details of the numerical algorithm]{hanasz03}, cosmic rays diffusing anisotropically along magnetic field lines \citep{giacalone99,jokipii99}, supernova remnants exploding randomly in the disk volume \cite[see][]{hanasz00}, the finite (currently uniform) resistivity of the ISM \cite[see][]{hanasz02,hanasz03,tanuma03} and the realistic vertical disk gravity \citep{ferriere98}. The system of coordinates $x,y,z$ corresponds locally to the global galactic cylindrical system $r, \phi, z$. The disk rotation was defined by the values of the angular velocity $\Omega=$0.05~Myr$^{-1}$ and the value of the shearing parameter is $q=1$. The boundary conditions are periodic in the Y-direction, sheared in the X-direction \cite[following][]{hawley95} and open in the Z-direction.

The open boundary conditions in the Z-direction are constructed in such a way that fluid, along with the magnetic field frozen in the fluid, can move out through the upper and lower box boundaries. These boundary conditions rely on the copying of fluid variables and magnetic field components from the last cell-layer inside the computational domain to the ghost zones.

\begin{figure*}  
 \epsscale{1.1}
 \plottwo{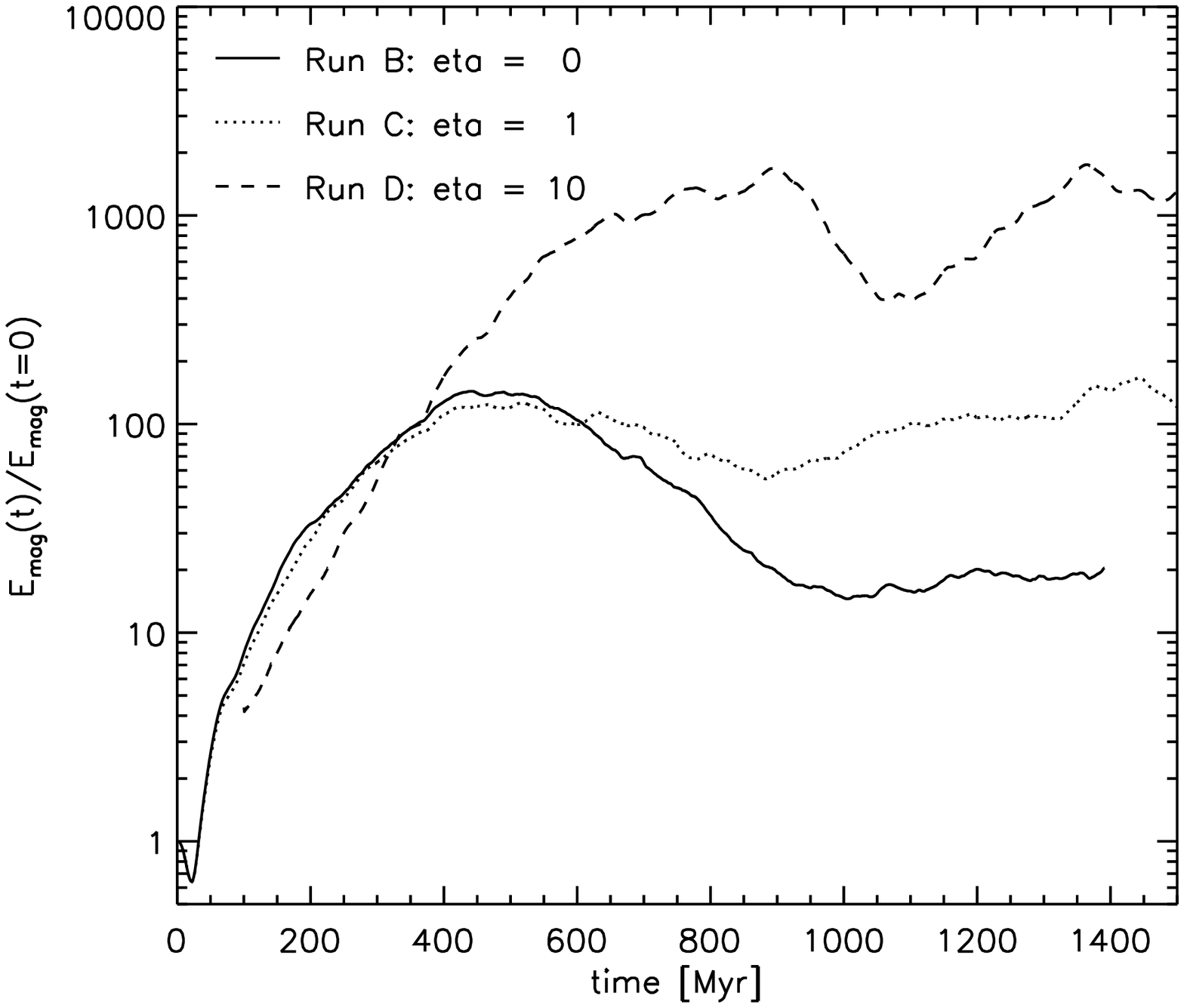}{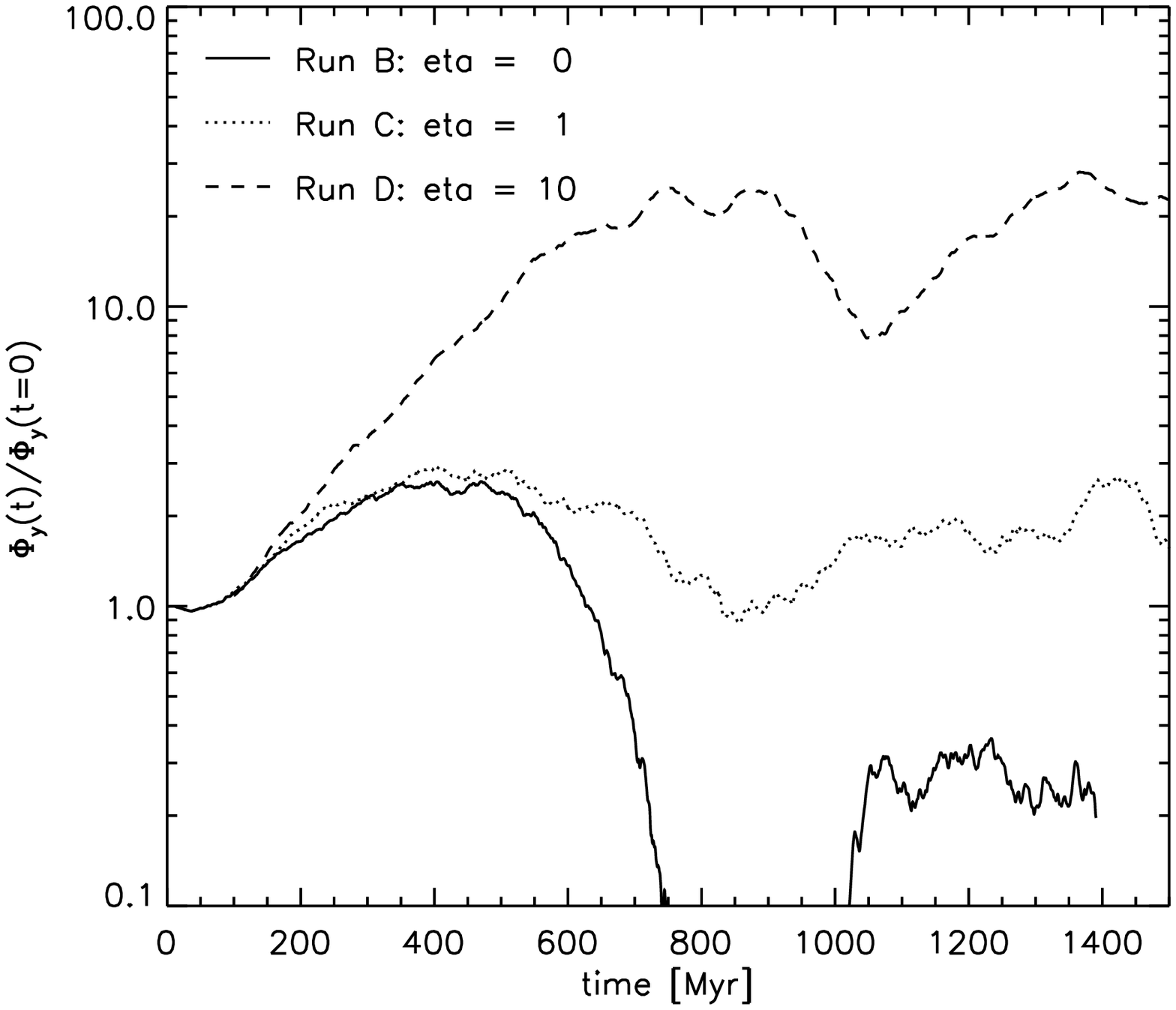}
 \caption{Time evolution of the magnetic flux (left plot) and magnetic energy (right plot) for the simulations B, C and D. We applied the normalization of both quantities to their values at $t=0$. \label{fig:mag-flux-ener}}
\end{figure*}

In this paper we analyze the numerical model of cosmic-ray driven dynamo by \cite{hanasz04} together with a series of the new three models that form a subset of the new extensive series, which are described in detail in two complementary papers \citep{hanasz06,hanasz07}. In the presented simulations we adopt the basic parameters (the vertical gravity model and the thermal gas column density) from the global model of ISM in Milky Way \citep{ferriere98} for the galactocentric radius R=8.5~kpc \cite[model A][]{hanasz04} and R=5~kpc (new models B, C and D). Moreover, models B, C and D rely on improved boundary conditions for cosmic rays (with respect to Model A), imposing $e_{\rm cr}=0$ on outer Z-boundaries. The initial magnetic field in the azimuthal direction corresponds to a small fraction $10^{-8}$ of thermal energy for model A and $10^{-4}$ for models B--D. The initial gas temperature is equal to 7000 K.

In all models we assume that supernovae explode randomly in the disk and supply cosmic rays with energy equal to  10\% of 10$^{51}$~erg SN kinetic energy output. We assume that the cosmic ray energy is injected instantaneously into the ISM, as a consequence of each SN event, with a Gaussian radial profile ($r_\mathrm{SN}=50$~pc) around the explosion center. Simulations of all four models examined in this paper have been performed in a Cartesian domain with parameters  summarized in Table~\ref{table-params}.

The amplification of the regular magnetic field is identified with the amplification of the azimuthal magnetic flux component averaged over all $XZ$ slices through the discretized computational domain. In Figure~\ref{fig:mag-flux-ener} we present the time evolution of the azimuthal component of magnetic flux and the total magnetic energy. Analogous plots for run A have been presented in \cite{hanasz04}. It is apparent in Figure~\ref{fig:mag-flux-ener} that the amplification of magnetic flux and magnetic energy depends on the resistivity as it has been reported by \cite{hanasz06}. Strictly speaking the efficiency of magnetic field amplification grows with increasing the resistivity, within the assumed range of variations of the resistivity parameter. Both magnetic flux and energy start to grow at $t\simeq 100$~Myr and continue until a maximum (which is resistivity dependent) is reached. After reaching the maximum the curves of magnetic flux and magnetic energy behave rather chaotically for all values of resistivity.

\begin{figure*}  
 \epsscale{0.20}
 \plotone{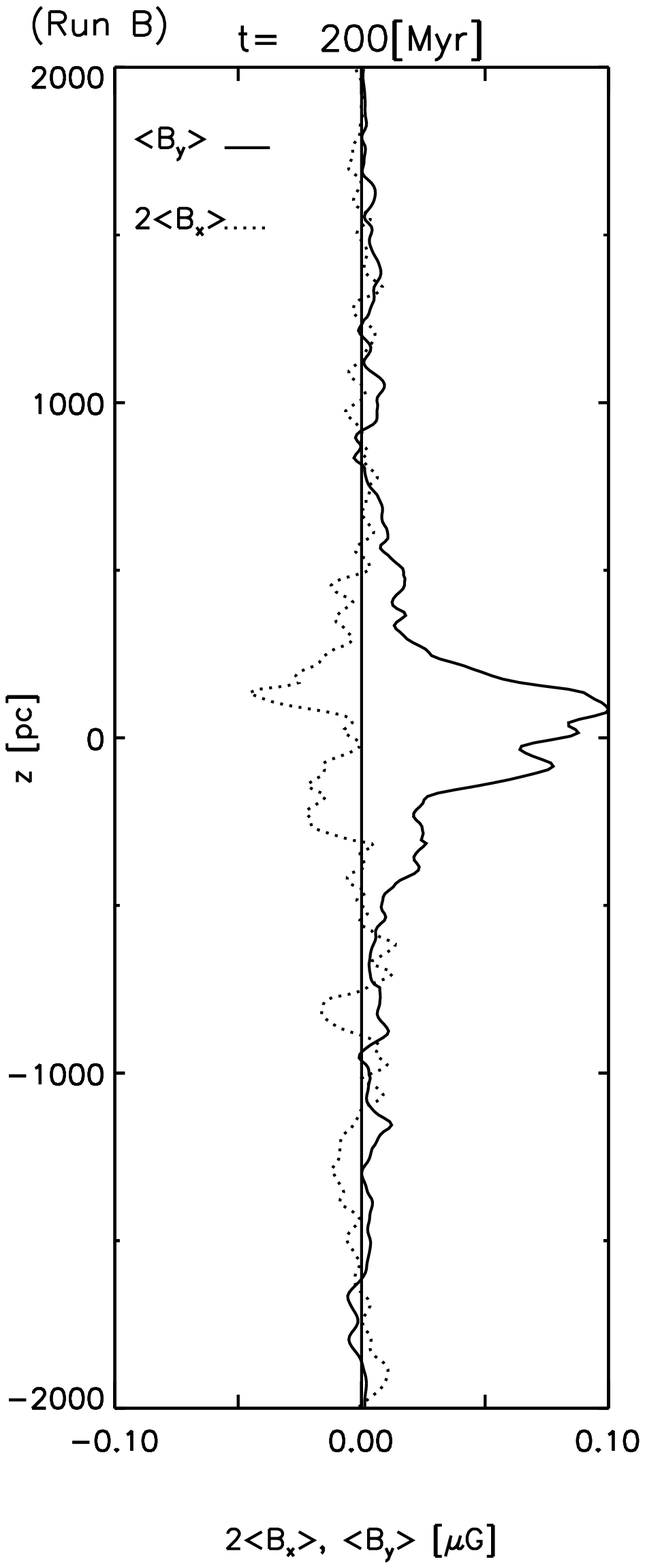}
 \plotone{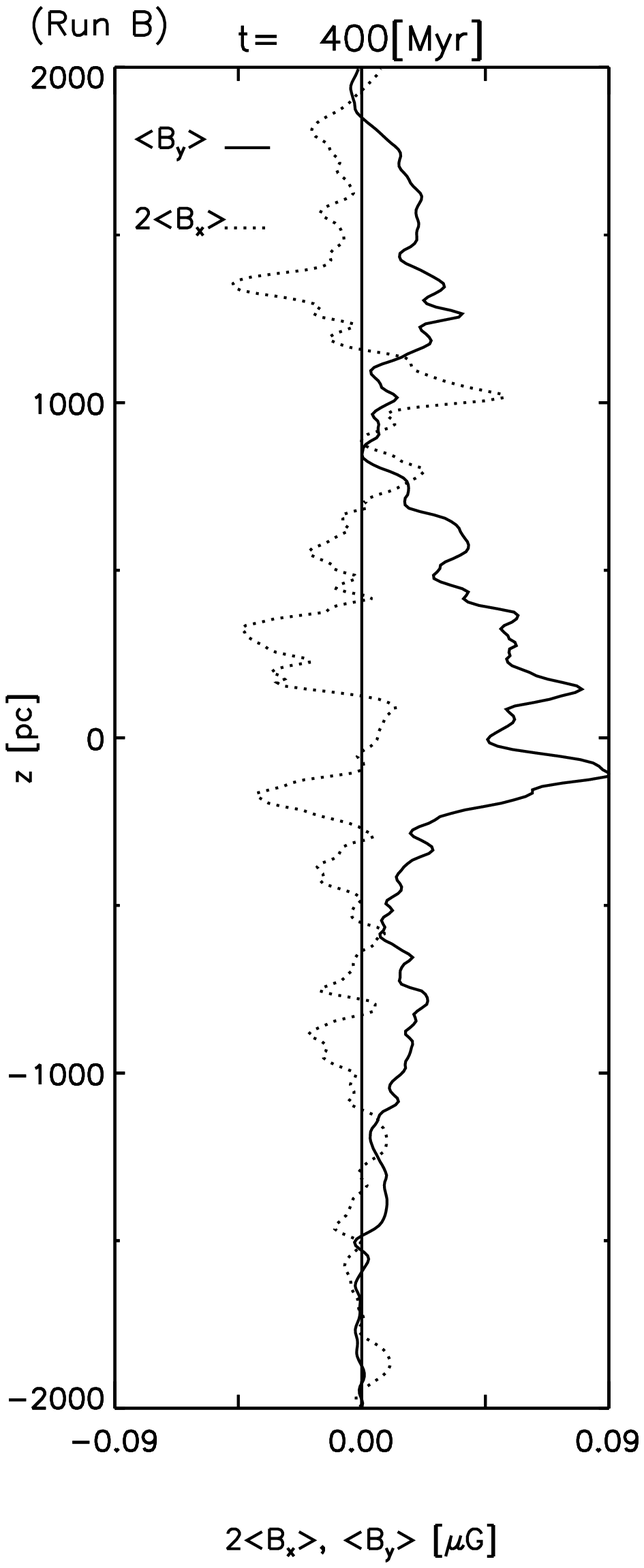}
 \plotone{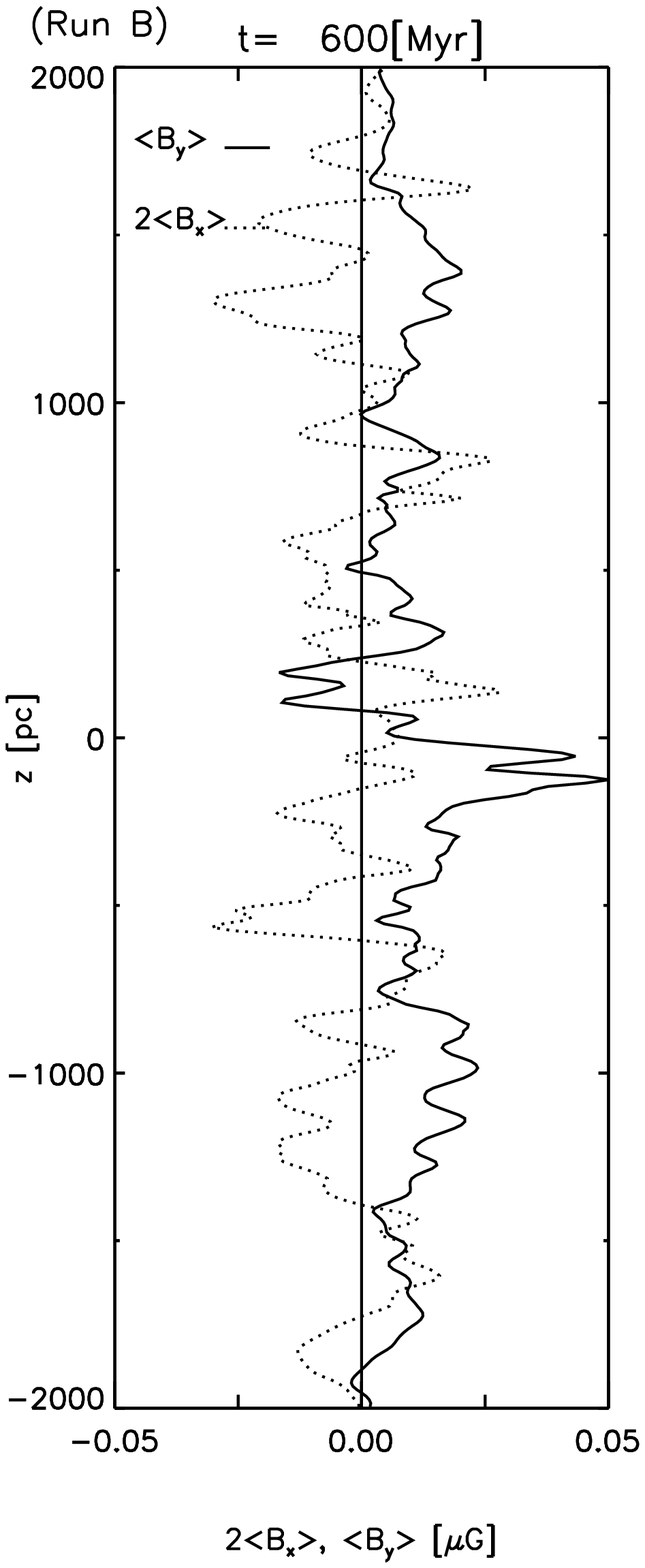}
 \plotone{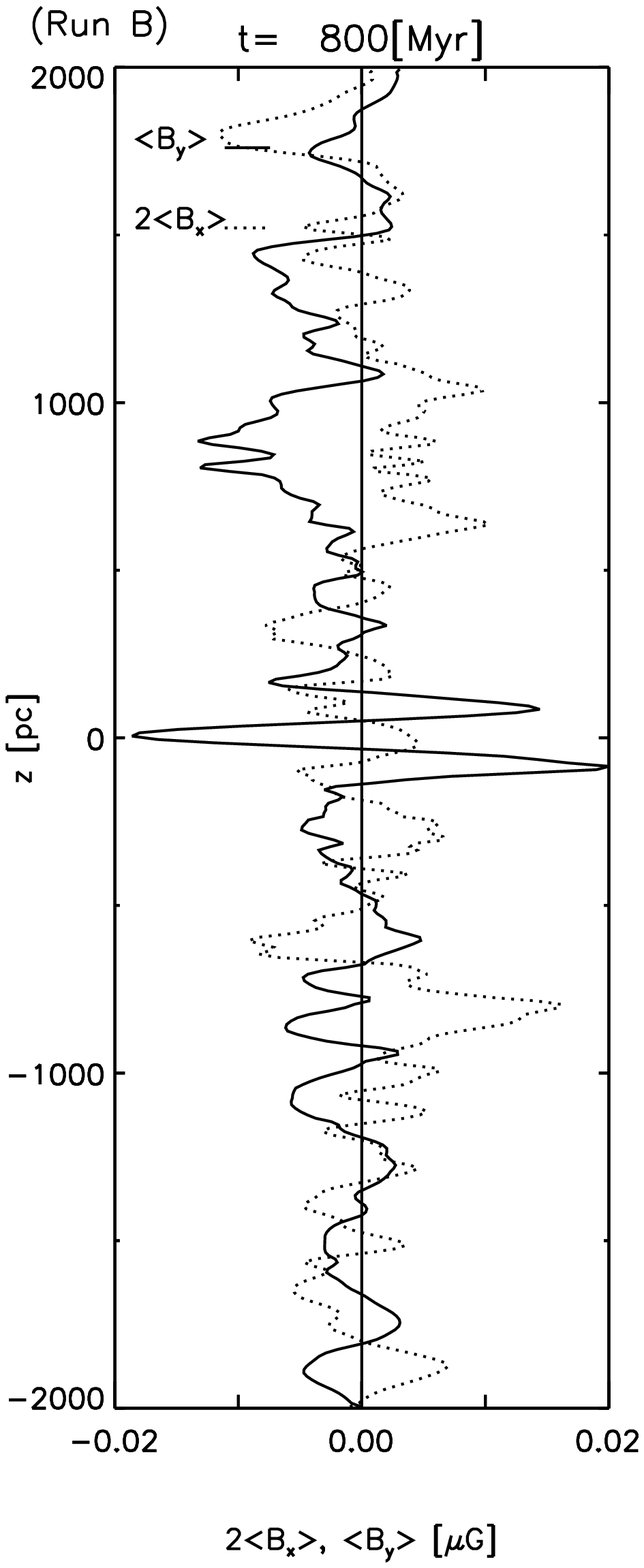} \\
 \plotone{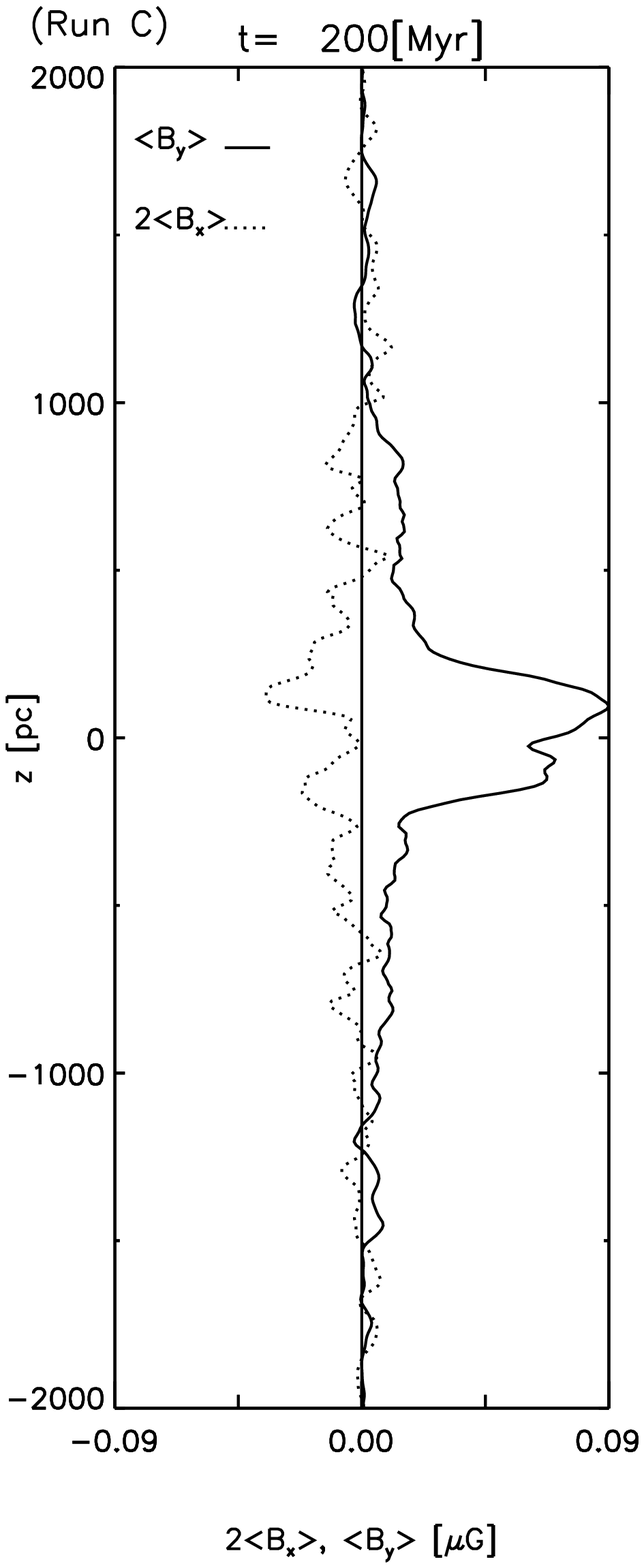}
 \plotone{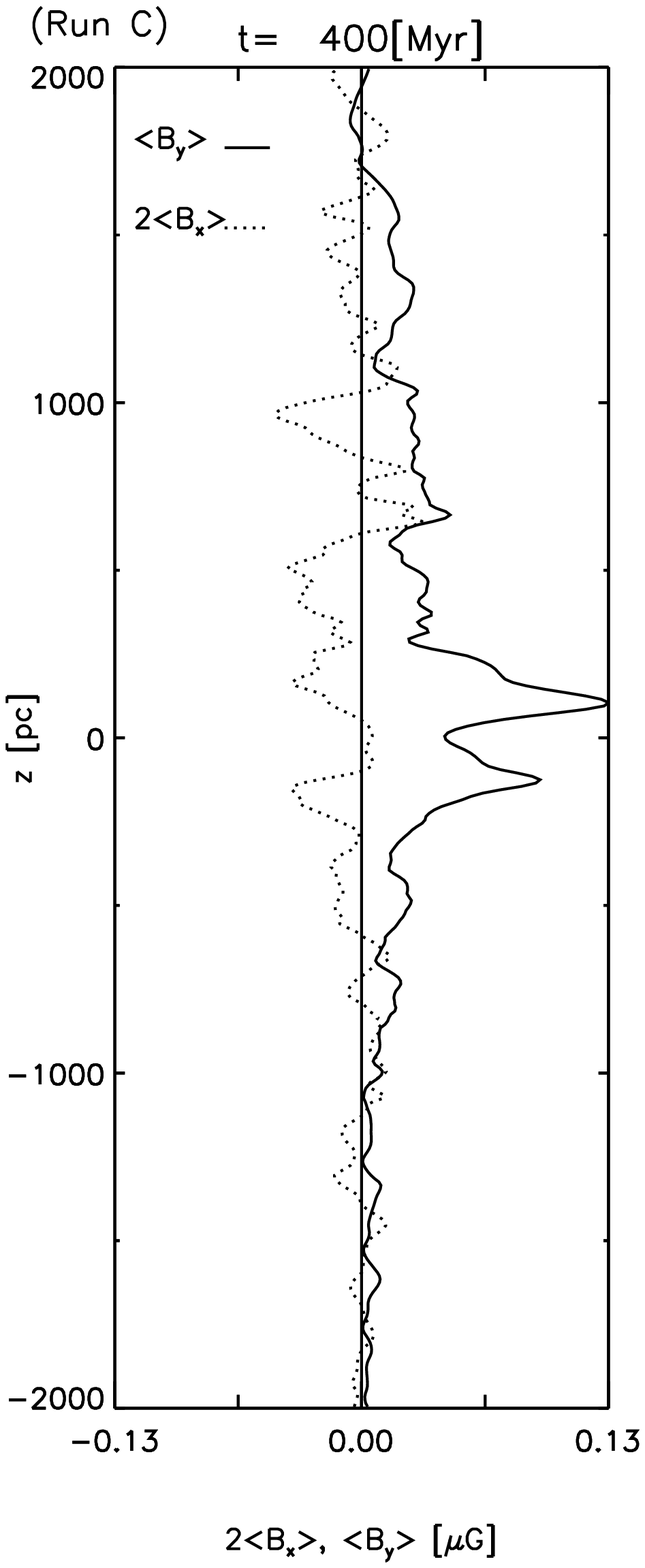}
 \plotone{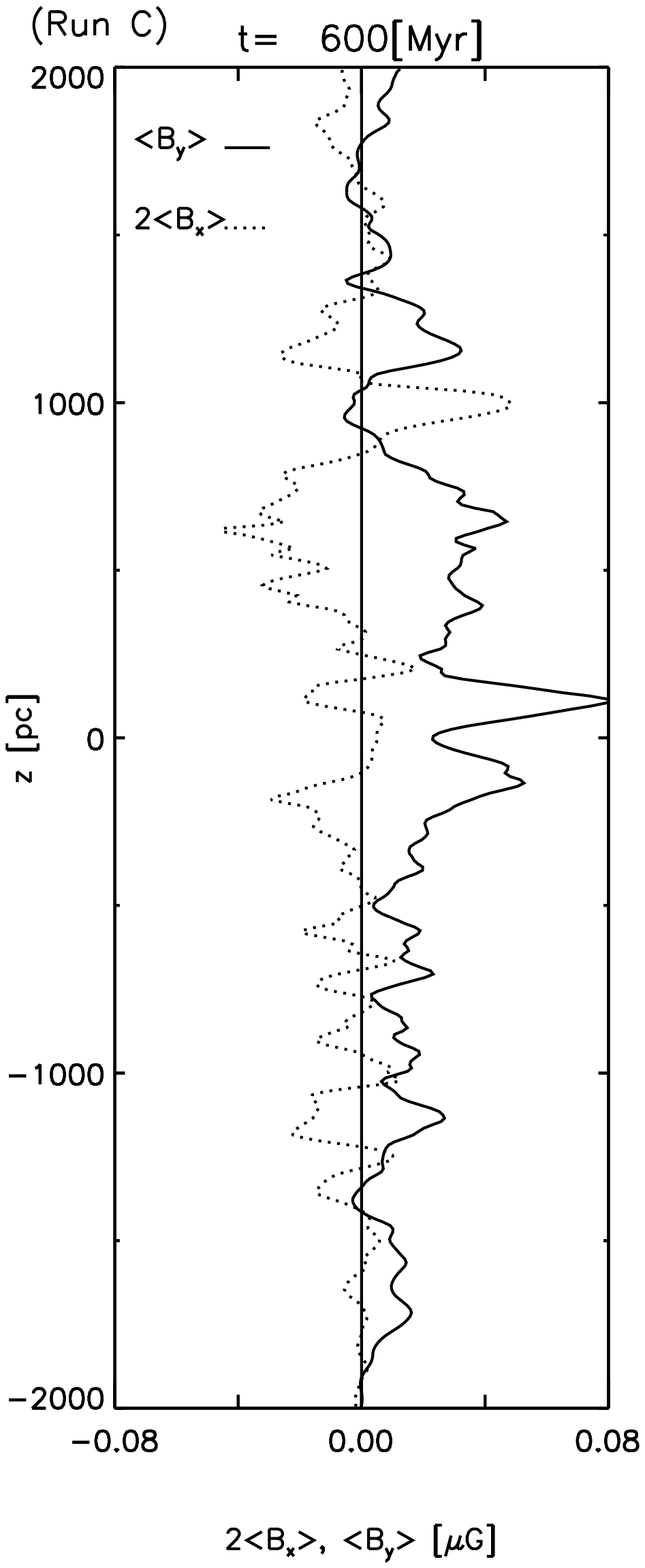}
 \plotone{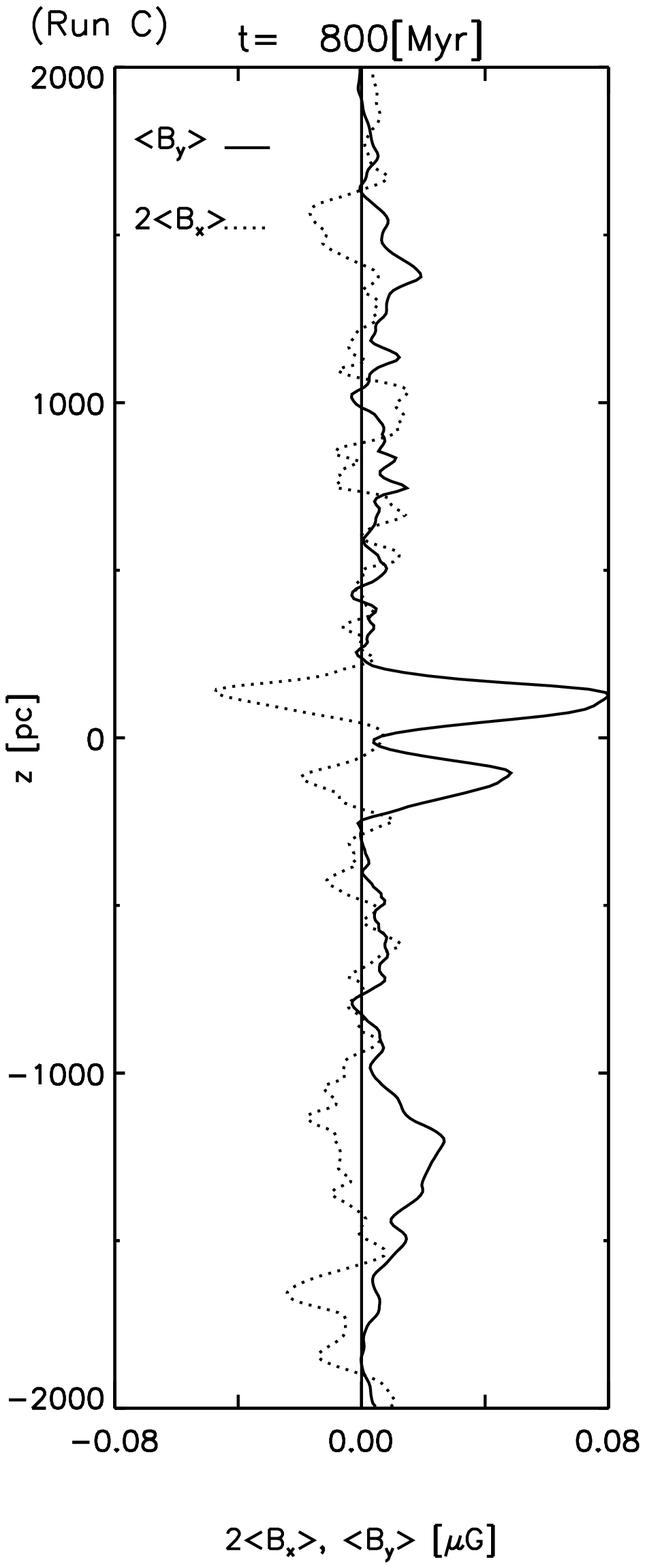} \\
 \plotone{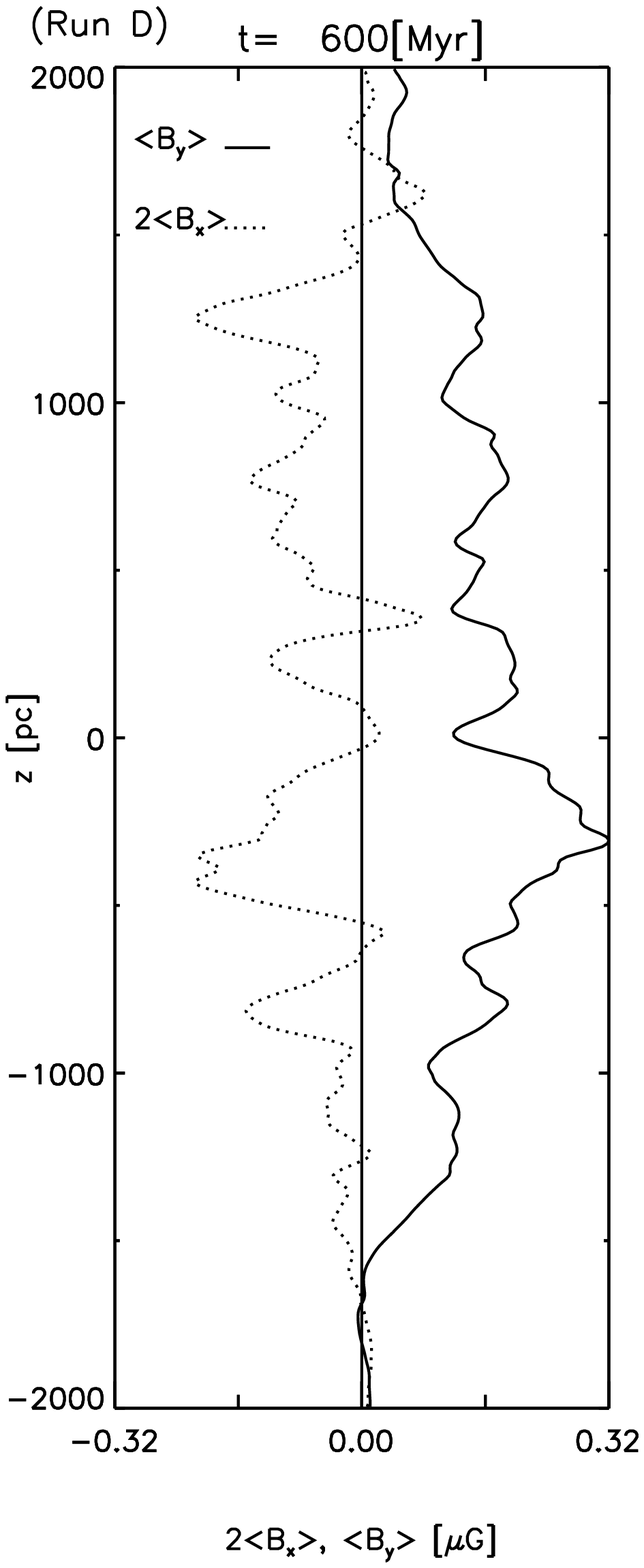}
 \plotone{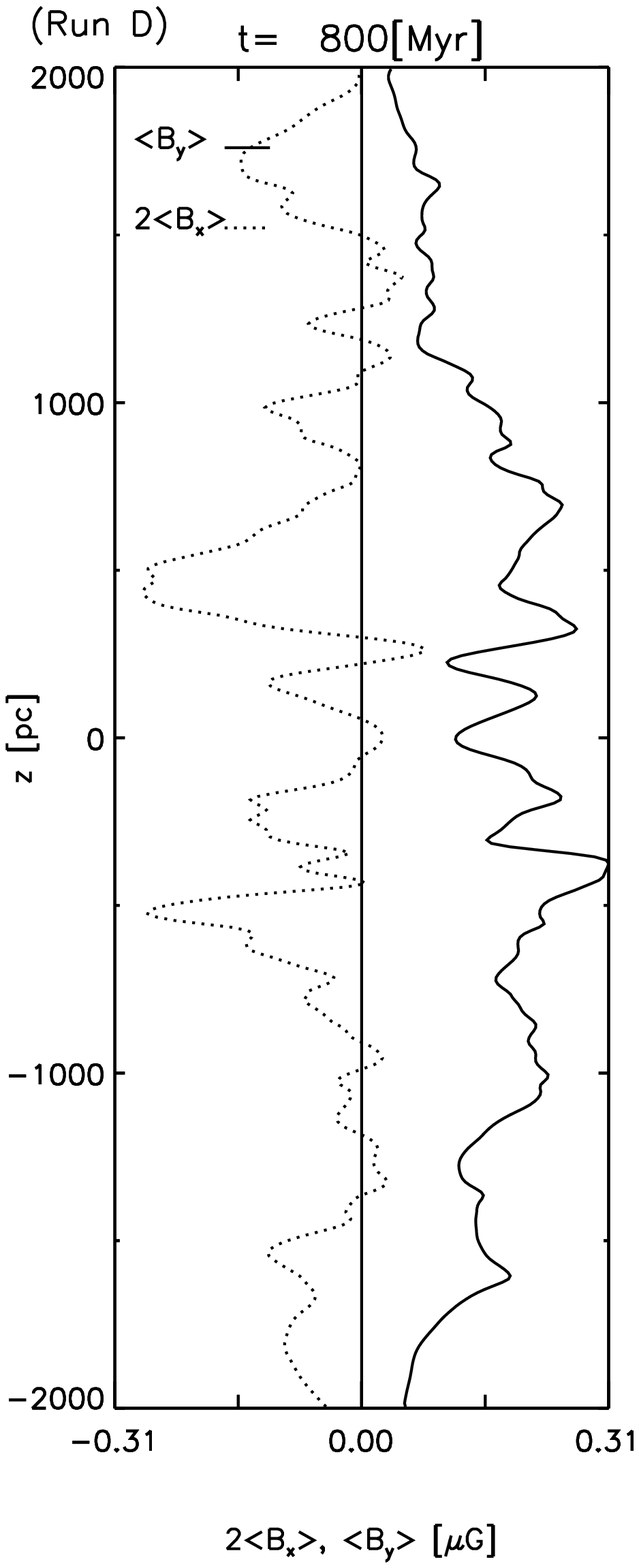}
 \plotone{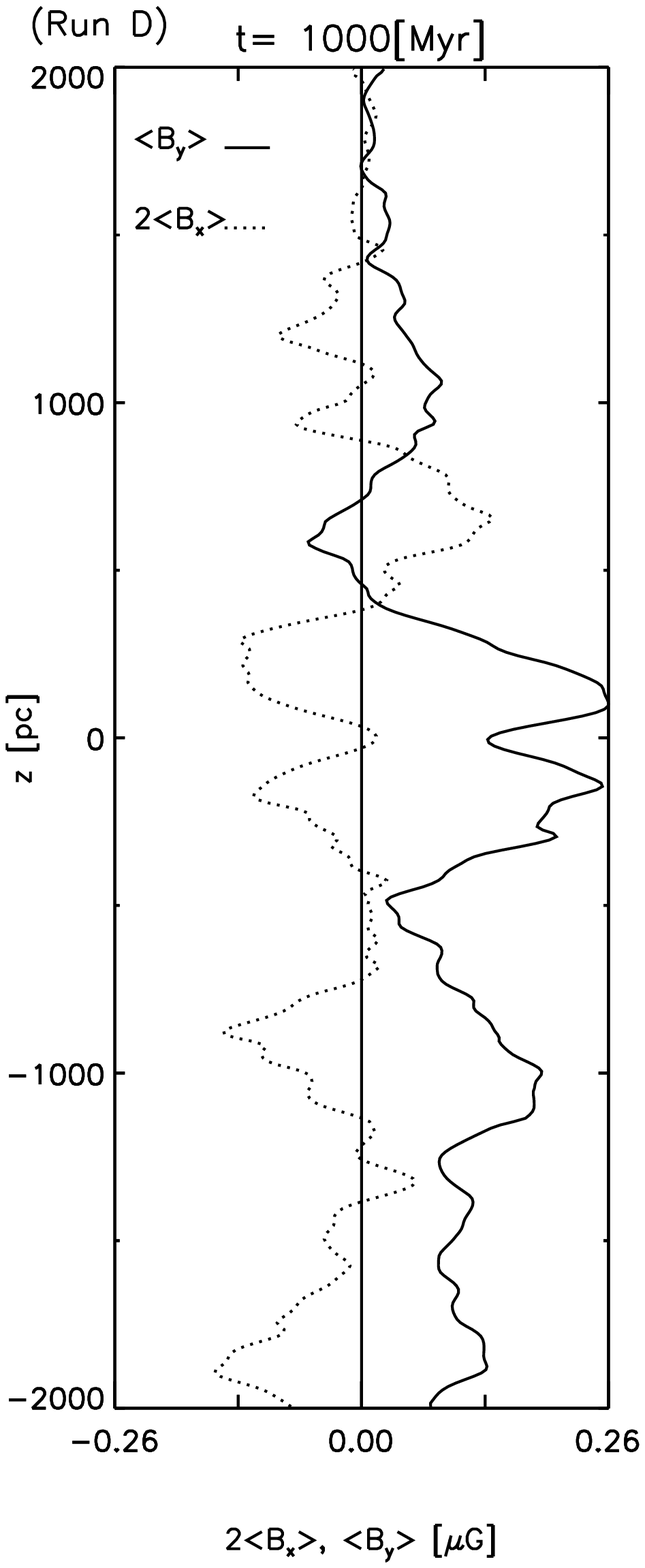}
 \plotone{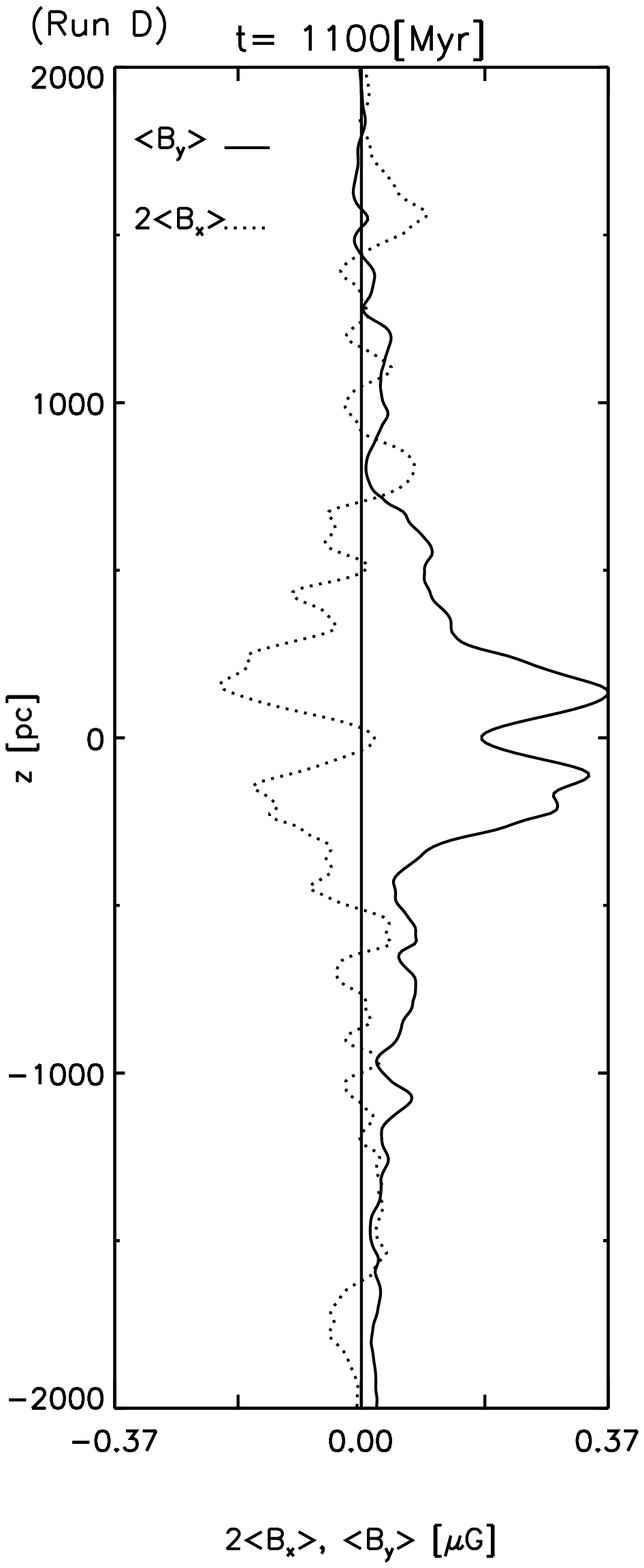}
 \caption{Evolution of the horizontally averaged magnetic field  (solid line -- azimuthal component, dotted line -- radial component) for runs B, C and D. It is noticeable that in the growth phases of the both magnetic flux and magnetic energy, shown in Fig.~\ref{fig:mag-flux-ener} (100~Myr$\leq t \leq$500~Myr for runs B and C, and 100~Myr$\leq t \leq$800~Myr and $ t \geq$1050~Myr for run D), the averaged magnetic field components $\bar{B}_x (z)$ and $\bar{B}_y (z)$ are apparently more regular, while at the decay phases vertical fluctuations of the mean field dominate. \label{fig:mean_b}}
\end{figure*}

In order to measure the amplification of the mean magnetic field we perform an averaging of $B_x$ and $B_y$ magnetic field components as it has been done by \cite{hanasz04} for run A. We plot the mean magnetic field $\bar{\bf B}(z)$ for runs B, C and D in Figure~\ref{fig:mean_b} at different times. We find that in the growth phases of both magnetic flux and magnetic energy, shown in Figure~\ref{fig:mag-flux-ener} (100~Myr$\leq t \leq $500~Myr for runs B and C, and 100~Myr$\leq t \leq$800~Myr and $ t \geq$1050~Myr for run D), the averaged magnetic field components $\bar{B}_x (z)$ and $\bar{B}_y (z)$ are apparently more regular, while at the decay phase the vertical fluctuations of the mean field dominate.

A close inspection of the mean, horizontally averaged azimuthal magnetic field component along the simulation time for run B, shows that the magnitude of the mean field does not grow in the period 0~Myr$\leq t \leq $500~Myr, but only the width of the central layer, of coherent (or uni-directional) azimuthal field, increases slightly. In the opposite case of run D, one can note the growth of the magnitude of the mean azimuthal field, till $t=800$ Myr, as well as the growth of the width of the central layer of coherent azimuthal field. The radial component of the averaged field displays much more temporal and spatial fluctuations than the azimuthal component in runs B--D.

Generally speaking fluctuations of the averaged field appear to be much more pronounced in runs B--D than in run A. This may be related to essentially more vigorous dynamics, related to a significantly larger SN rate (at the galactocentric radius 5 kpc, see Table 1)  of the system in simulations B--D as compared to simulation A corresponding to the SN rate at the galactocentric radius of Sun. This difference gives rise to a vertical wind, powered by CR pressure gradient,  reaching velocities of the order of 100 km/s at the altitude of 2 kpc.

The oscillations present in the averaged field migrate out of the disk midplane at large altitudes. On the other hand the behavior of the averaged field is almost stationary (coherent in space and time) near the disk midplane in the large resistivity run D. We shall associate the mentioned coherence with the presence of the mean field. In this sense the mean field is clearly generated in runs A and D, while its generation in runs B and C is questionable. The net azimuthal magnetic flux, apparent in the growth phases of magnetic field for run D, shown in Figure~\ref{fig:mag-flux-ener}, is a clear signature of the growth of the mean field.

The set of models B--D forms an interesting sample of experiments to be examined from the point of view of dynamo theories. In the forthcoming parts of the paper we are going to address the question to which extent the available dynamo theories are capable of reproducing the magnetic field evolution relying on dynamo coefficients. It is quite obvious that the truncation of the power series of the electromotive force, at linear or quadratic terms should lead to some discrepancy between magnetic fields measured in experiments and the ones reconstructed on the base of dynamo coefficients. This discrepancy can be directly measured and quantified in a statistic way. Another simple criterion for the applicability of dynamo theories can be proposed, however, in a more qualitative fashion.  We shall propose a criterion, which relies on a simple observation wheather the growth and decay of reconstructed magnetic field correlates with the growth and decay of the original field.

%
\section{Mean field dynamo theory}
\label{sec:dynamo_theory}

The evolution of magnetic fields in a dynamic system can be investigated in the mean field approach, where the separation of the total magnetic field into two parts, the mean large-scale and fluctuating small-scale components, is the fundamental assumption \citep{parker55,moffatt78,raedler80}). In the mean field dynamo theory the main interest is to determine the influence of the small-scale fluctuations upon the mean magnetic field. Following the basic steps of the mean field theory one can derive the dynamo equation.

We start from the equation of induction with the diffusion term:
\begin{equation}
 \frac{\partial \it \mathbi{B}}{\partial t} = \nabla \times \left( \mathbi{u} \times \mathbi{B} - \eta \nabla \times \mathbi{B} \right) .
\end{equation}
Next, we separate the velocity and magnetic fields into the mean and fluctuating
components:
\begin{equation}
\mathbi{B} = \bar{\mathbi{B}} + \mathbi{b}, \quad \bar\mathbi{b} = 0 \, ,
\end{equation}
\begin{equation}
\mathbi{u} = \bar{\mathbi{V}} + \mathbi{v}, \quad \bar{\mathbi{v}} = 0 \, ,
\end{equation}
where overline designates the mean field obtained from the averaging procedure. To obtain the mean field component properly, the averaging procedure should fulfill the required conditions called the Reynolds rules
\cite[see][e.g.]{brandenburg05b}:
\begin{equation}
 \overline{U_1 + U_2} = \bar{U}_1 + \bar{U}_2, \quad \overline{\bar{U}} = \bar{U}, \quad \overline{\bar{U} u} = 0, \quad \overline{\bar{U}_1 \bar{U}_2} = \bar{U}_1 \bar{U}_2,
 \label{eqn:reynolds1}
\end{equation}
\begin{equation}
 \overline{\partial U / \partial t} = \partial \bar{U} / \partial t, \quad \overline{\partial U / \partial x_i} = \partial \bar{U} / \partial x_i \, .
 \label{eqn:reynolds2}
\end{equation}
After the averaging operation we get two equations describing the evolution of the mean and the fluctuating parts of the magnetic field, $\bar\mathbi{B}$ and $\mathbi{b}$ respectively:
\begin{equation}
 \frac{\partial \bar\mathbi{B}}{\partial t} = \nabla \times \left[ \overline{\mathbi{v} \times \mathbi{b}} + \bar\mathbi{V} \times \bar\mathbi{B} - \eta \nabla \times \bar\mathbi{B} \right] \, ,
 \label{eqn:mean_field_evol}
\end{equation}
\begin{equation}
\frac{\partial \mathbi{b}}{\partial t} = \nabla \times \left[ \bar\mathbi{V} \times \mathbi{b} + \mathbi{v} \times \bar\mathbi{B} + (\mathbi{v} \times \mathbi{b} - \overline{\mathbi{v} \times \mathbi{b}}) - \eta \nabla \times \mathbi{b} \right] \, ,
 \label{eqn:fluct_field_evol}
\end{equation}
where the large-scale part of the vector product of fluctuating parts of the velocity and magnetic fields is the electromotive force:
\begin{equation}
 {\cal E} \equiv \overline{\mathbi{v} \times \mathbi{b}}.
 \label{eqn:emf}
\end{equation}

The investigation of $\cal E$ is the crucial point in the dynamo theory. It is considered a functional of $\bar\mathbi{v}$, $\mathbi{v}$ and $\bar\mathbi{B}$, which can form nonlinear higher order moments \cite[see][and the discussion in \S\ref{sec:conditions} and \S\ref{sec:nonlinear}]{raedler80,brandenburg05b}.

%
\subsection{Conditions for the linear approximation of the electromotive force}
\label{sec:conditions}

The simplest approximation of the dynamo equations in order to obtain the electromotive force prescribed in the form of Eq.~(\ref{eqn:emf}) contains the linearization of the equations for the fluctuating quantities. The first step of approximation is an assumption that the fluctuations are small, so we can neglect all nonlinear terms, which consist of fluctuating fields of orders higher than one. At first, this is done by neglecting the term $\mathbi{v} \times \mathbi{b}$ in Eq.~(\ref{eqn:fluct_field_evol}). This can be true for example if $R_m$ is small, what cannot be fullfiled in the case of ISM. The term $\bar\mathbi{V} \times \mathbi{b}$ is usually neglected because it describes an advection of the small scale magnetic field. However, this assumption cannot be justified in the case of the presence of a shear in the system. Strong shears could lead to a new dynamo effect, the shear-current effect \cite[see][]{rogachevskii03}. In the case of a small $R_{m}$ one can neglect both the nonlinear terms $\nabla \times (\mathbi{v} \times \mathbi{b} - \overline{\mathbi{v} \times \mathbi{b}})$. In most astrophysical applications, these terms can still be neglected if the correlation time $\tau_{corr}$ of the turbulence is small, $\tau_{corr} v_{rms} k_f \ll 1$, where $v_{rms}$ and $k_f$ are typical velocities and wave numbers associated with the random velocity field ${\bf v}$ \cite[see][]{brandenburg05b}. Finally, we result in the simplified equation describing the evolution of the fluctuating fields:
\begin{equation}
 \frac{\partial \mathbi{b}}{\partial t} = \nabla \times \mathbi{v} \times \bar\mathbi{B} \, .
 \label{eqn:fluct_field_evol_simpl}
\end{equation}
The absence of the diffusive term in Eqs.~(\ref{eqn:emf})-(\ref{eqn:fluct_field_evol_simpl}) follows from the assumption, that in the correlation time the advection strongly dominates over diffusion effects, so $R_{m}\gg0$. This is valid for most of astrophysical applications where $R_m\sim10^{10} - 10^{20}$.  To calculate ${\cal E}$ we integrate Eq.~(\ref{eqn:fluct_field_evol_simpl}) to get $\mathbi{b}$, take the cross product with $\mathbi{v}$ and average. If the mean magnetic field varies only weakly in space and time, $\cal E$ can be represented as a function of $\bar\mathbi{B}$ and its derivatives:
\begin{equation}
  {\cal E}_{\mathrm i} = \hat{\alpha}\bar\mathbi{B} - \hat{\beta} \ \nabla \times \bar\mathbi{B} = \alpha_{\mathrm{ij}} \bar{B}_{\mathrm{j}} + \beta_{\mathrm{ijk}} \frac{\partial \bar{B}_{\mathrm{j}}}{\partial x_{\mathrm{k}}} \label{eqn:emf_linear}
\end{equation}
The quantities $\cal E$ and $\bar\mathbi{B}$ with its derivatives can be calculated directly from the simulation data. The equation (\ref{eqn:emf_linear}) involves two tensor quantities, $\hat{\alpha} = \{\alpha_\mathrm{ij}\}$ and $\hat{\beta} = \{ \beta_\mathrm{ijk} \}$, describing the efficiencies of the amplification and diffusion of the mean magnetic field, respectively. Both coefficients are necessary for the mean field dynamo. Dynamo coefficients $\alpha_{\mathrm{ij}}$ and $\beta_{\mathrm{ijk}}$ can be fitted using statistical methods \citep{kowal05}. We use a very popular and flexible method of the multidimensional linear regression \cite[see, e.g.,][\S15.4]{press97}.

\begin{figure*}[t]  
 \epsscale{1.1}
 \plotone{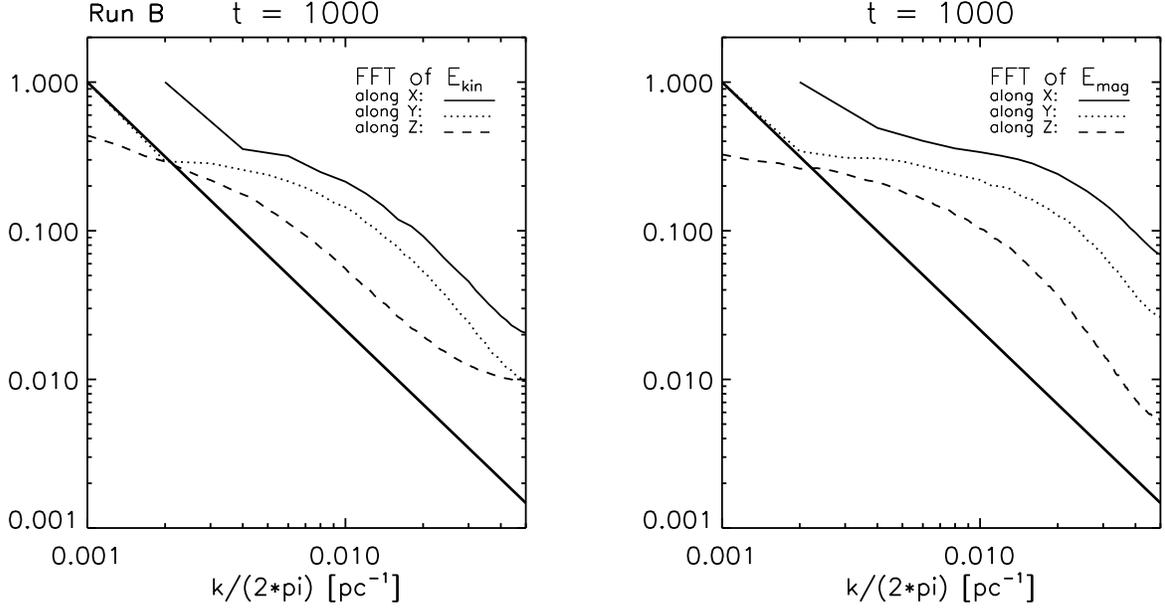}
 \caption{Power spectra of the velocity (left plot) and magnetic (right plot) energy along each direction (thin solid, dotted and dashed lines for X-, Y- and Z-direction respectively) for run B at time 1000~Myr. The thick solid line corresponds to the Kolmogorov power law $E \sim k^{-5/3}$. \label{fig:spectra_b}}
\end{figure*}

%
\subsection{Nonlinear approaches to the dynamo}
\label{sec:nonlinear}

%
\subsubsection{Blackman-Field approach}
\label{sec:blackman}

\cite{blackman02} instead of imposing the form of the ${\cal E}$, solved the dynamo equation using the time evolution of ${\cal E}$
\begin{equation}
\partial_t {\cal E} = \overline{\partial_t \mathbi{v} \times \mathbi{b}} + \overline{\mathbi{v} \times \partial_t \mathbi{b}} .
\end{equation}

Using equations for the evolution of the small-scale velocity and magnetic fields \cite[see Eqs.~4 and 5 in][]{blackman02} and assuming the incompressibility of the flow and isotropy of the resulting velocity and magnetic field correlations for terms linear with $\bar\mathbi{B}$, they obtained an equation for the time evolution of the electromotive force ${\cal E}$
\begin{equation}
\frac{\partial {\cal E}}{\partial t} = \tilde{\alpha} \bar\mathbi{B} - \tilde{\beta} \nabla \times \bar\mathbi{B} + \nu \overline{\nabla^2 \mathbi{v} \times \mathbi{b}} + \lambda \overline{\mathbi{v} \times \nabla^2 \mathbi{b}} + {\bf T}^V + {\bf T}^M ,
\end{equation}
where $\tilde{\alpha} = (1/3) (\overline{\mathbi{b} \cdot \nabla \times \mathbi{b}} - \overline{\mathbi{v} \cdot \nabla \times \mathbi{v}})$ and $\tilde{\beta} = (1/3) \overline{\mathbi{v}^2}$ are dynamo coefficients, $\nu$ and $\lambda$ are the viscosity and magnetic diffusion coefficients respectively, and ${\bf T}^V$ and ${\bf T}^M$ are the triple correlations \cite[see][for description]{blackman02}.

The above and following equations are valid if the divergence of magnetic helicity flux vanishes. Nevertheless, we omit the considerations of magnetic helicity in this paper. We intend to devote a separate paper for the cosmic-ray driven dynamo examined from the point of view of magnetic helicity conservation.

Dynamo coefficients $\tilde{\alpha}$ and $\tilde{\beta}$ do not change if we limit our consideration to the component of ${\cal E}$ parallel to $\bar\mathbi{B}$. In this case we obtain the following simpler expression
\begin{equation}
 \frac{\partial {\cal E}_\parallel}{\partial t} = \tilde{\alpha} \frac{\bar\mathbi{B}^2}{|\bar\mathbi{B}|} - \tilde{\beta} \frac{\bar\mathbi{B} \cdot \nabla \times \bar\mathbi{B}}{|\bar\mathbi{B}|} - \tilde{\zeta} {\cal E}_\parallel .
 \label{eqn:blackman}
\end{equation}
Here $\tilde{\zeta}$ corresponds to the macrophysical dissipation terms \cite[see][]{blackman02}. The coefficients $\tilde{\alpha}$ and $\tilde{\beta}$ are slightly different from the usual dynamo coefficients, because they do not include any characteristic timescale like correlation or relaxation time, and thus have different units.

In the approach described above the authors considered all quadratic terms in the mean field in the applied electromotive force approximation.

\begin{figure*}  
 \epsscale{1.0}
 \plotone{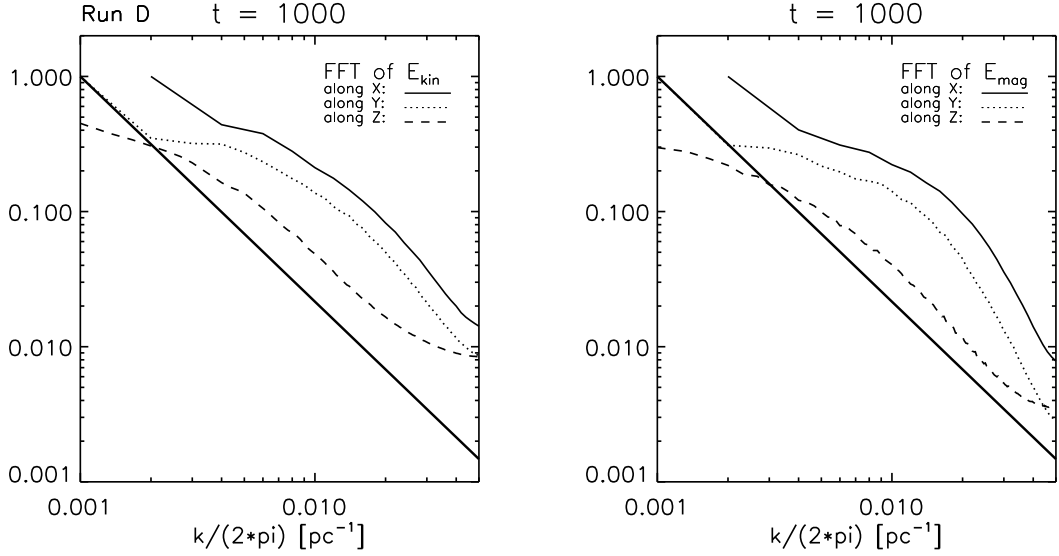}
 \caption{Power spectra of the velocity (left plot) and magnetic (right plot) energy along each direction (thin solid, dotted and dashed lines for X-, Y- and Z-direction respectively) for run D at time 1000~Myr. The thick solid line corresponds to the Kolmogorov power law $E \sim k^{-5/3}$. \label{fig:spectra_d}}
\end{figure*}

\begin{figure*}  
 \includegraphics[angle=90,width=7in]{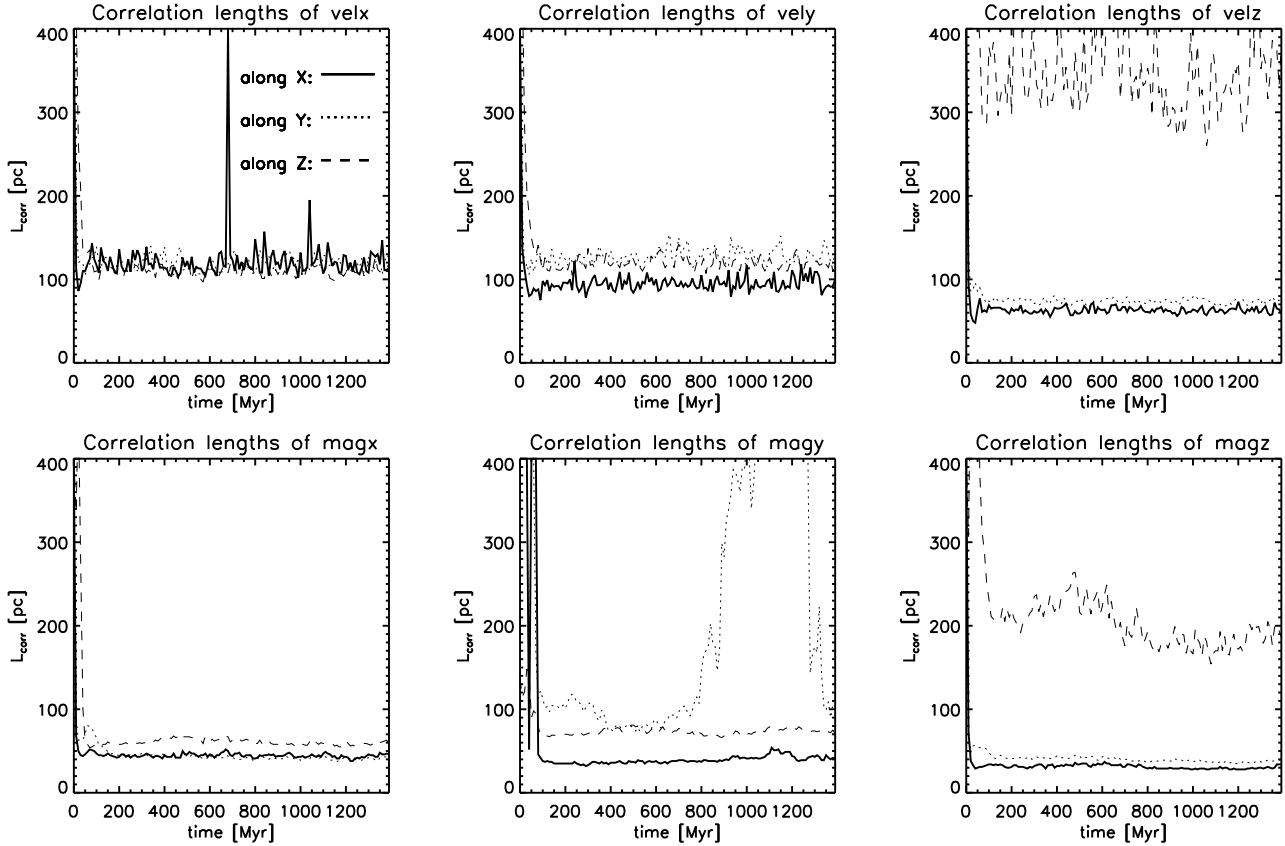}
 \caption{Time variance of the characteristic length-scales for the velocity components (top row) and for the magnetic field components (bottom row) along all three directions (solid, dotted and dashed lines for X-, Y- and Z-directions respectively) in experiment B. On the left, middle and right columns we show X-, Y- and Z-components respectively. \label{fig:corr_b}}
\end{figure*}

\begin{figure}  
 \epsscale{1.0}
 \plotone{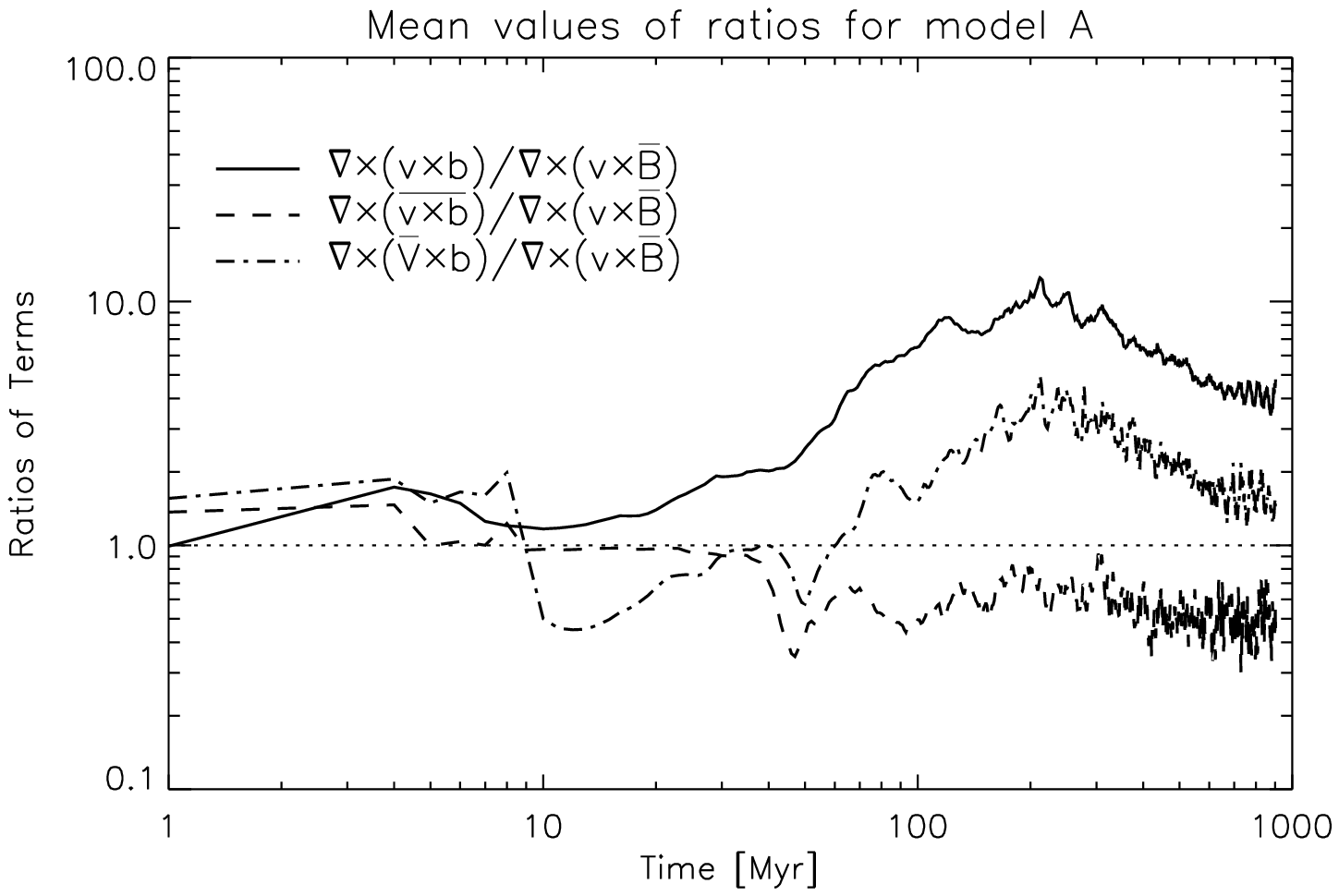}
 \plotone{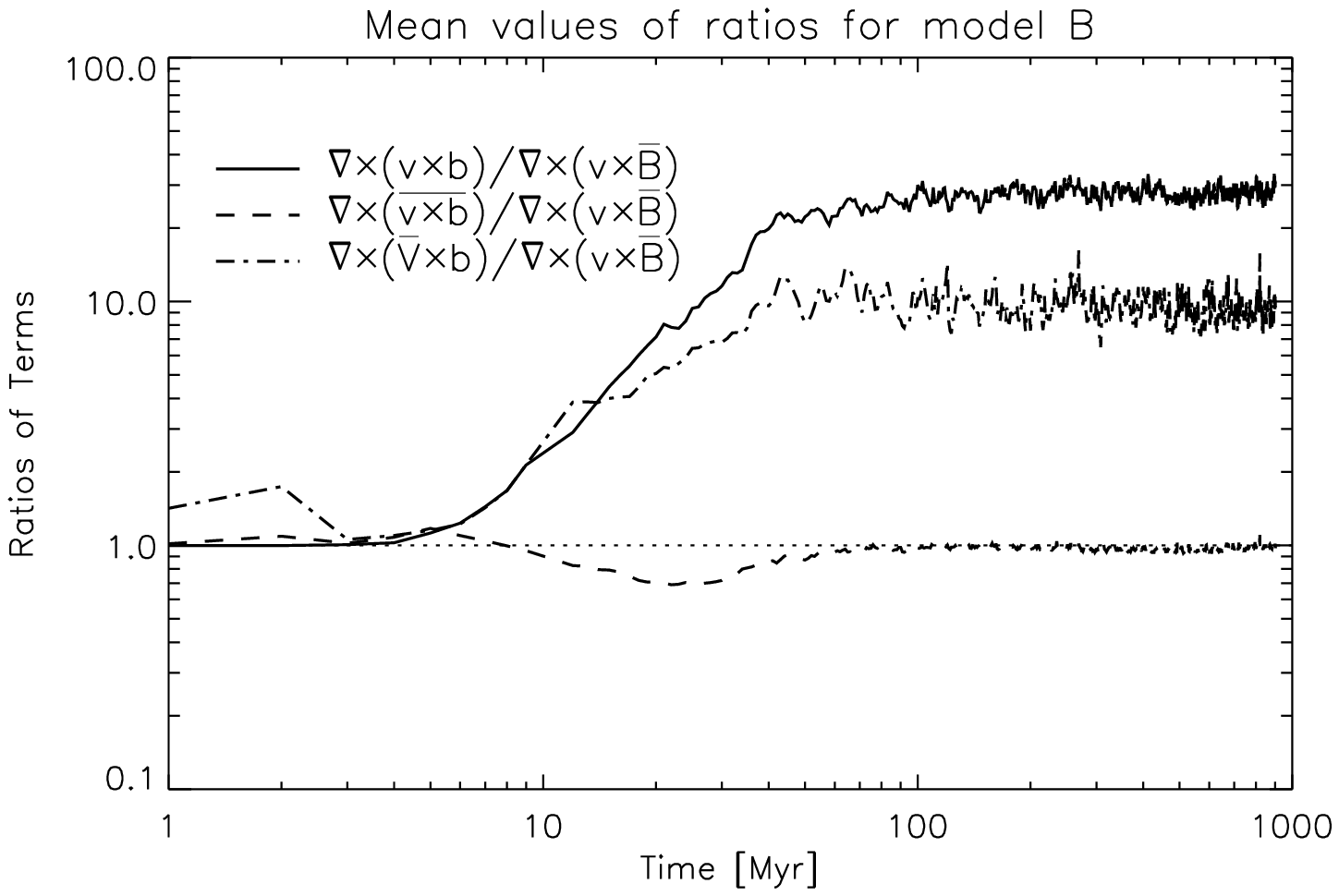}
 \caption{Time variances of the mean values of ratios $\nabla \times (\mathbi{v} \times \mathbi{b}) / \nabla \times (\mathbi{v} \times \bar\mathbi{B})$ (solid line), $\nabla \times (\overline{\mathbi{v} \times \mathbi{b}})/\nabla \times (\mathbi{v} \times \bar\mathbi{B})$ (dashed line) and $\nabla \times (\bar\mathbi{V} \times \mathbi{b})/\nabla \times (\mathbi{v} \times \bar\mathbi{B})$ (dash-dot line) for experiments A (upper plot) and B (lower plot). A condition for a linear approximation of ${\cal E}$ requires these ratios to be much smaller than 1, however this condition is not fulfilled for the presented experiments. \label{fig:terms}}
\end{figure}

\begin{figure}  
 \epsscale{1.0}
 \plotone{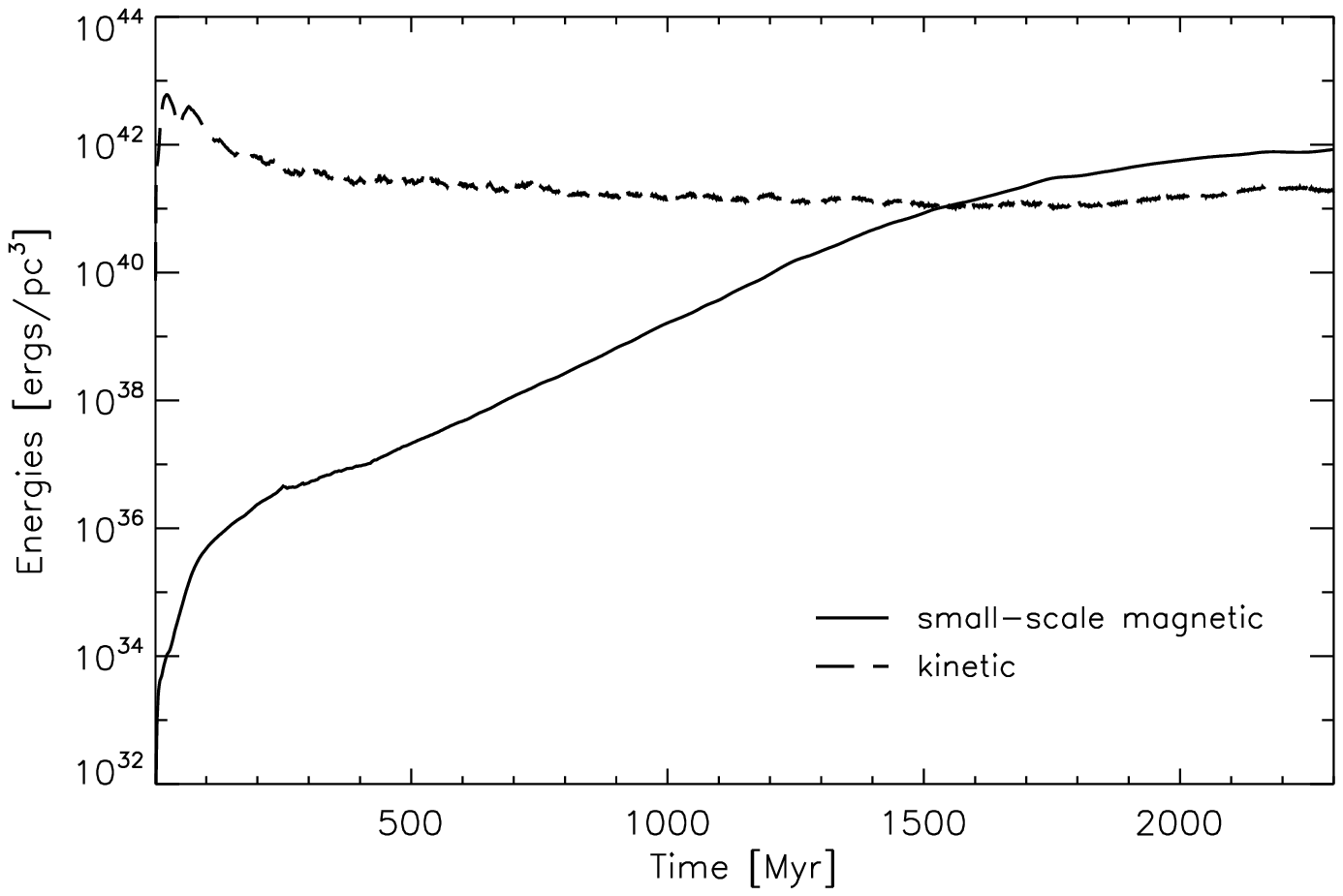}
 \plotone{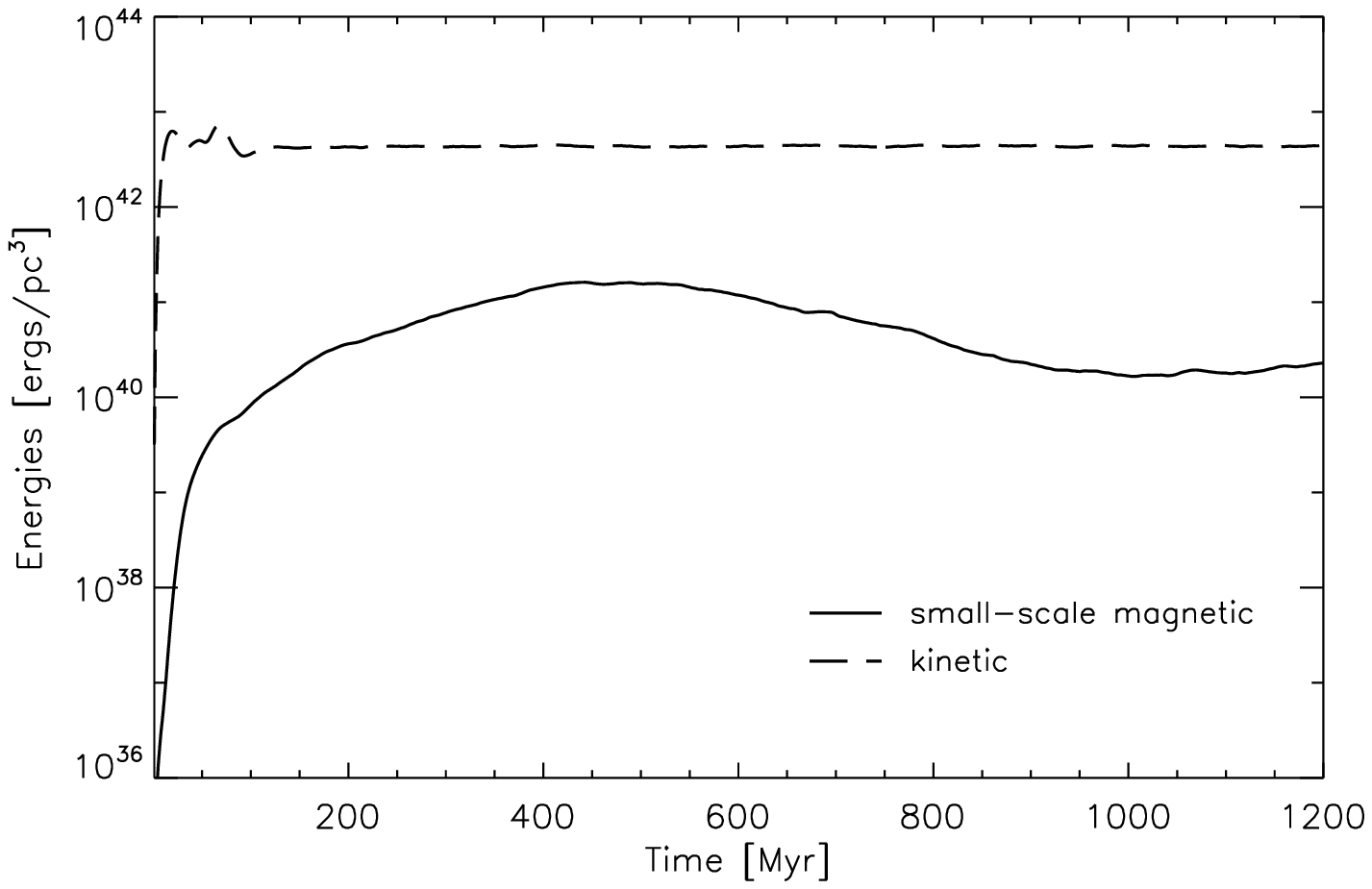}
 \plotone{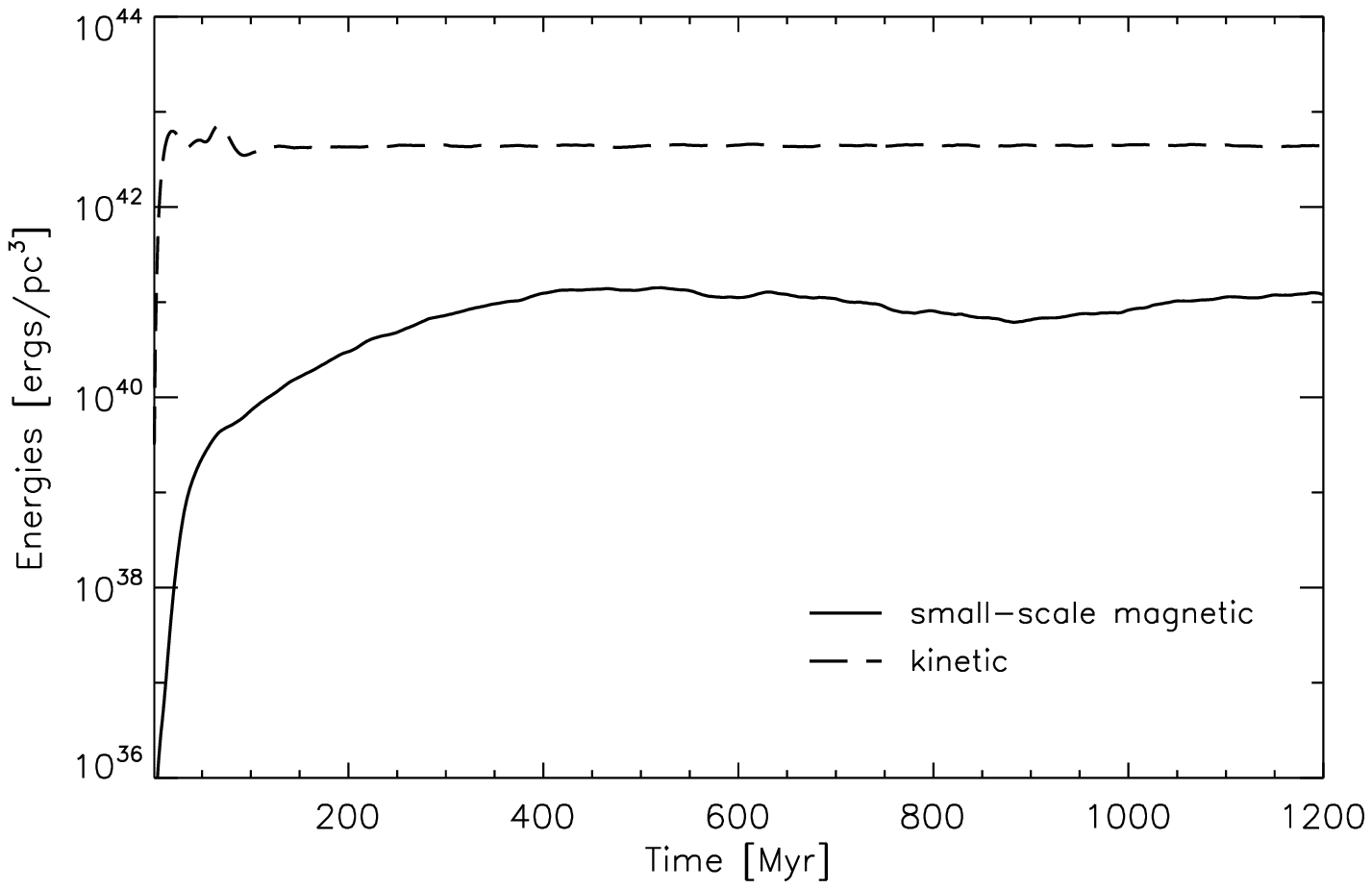}
 \caption{Time evolution of the kinetic turbulent (dashed line) and the small-scale magnetic (solid line) energies for three models: A, B, and C (top, middle and bottom plots respectively). \label{fig:energies}}
 \end{figure}

\subsubsection{Kleeorin-Rogachevski approach}
\label{sec:kleeorin}

The standard local thin disc approximation for dynamo equations for an axisymmetric magnetic field leads to the following equations for the mean radial field $B_r = R_\alpha b_r$ and toroidal field $B_\phi$ for the $\alpha\Omega$-dynamo problem \cite[see][e.g.]{ruzmaikin88,kleeorin03}
\begin{equation}
\frac{\partial b_r}{\partial t} = - \frac{\partial (\alpha (\mathbi{B})
B_\phi)}{\partial z} + \frac{\partial}{\partial z}\left(\eta_A (\mathbi{B})
\frac{\partial b_r}{\partial z}\right) - \frac{\partial \left( \mathbi{V}_A(\mathbi{B}) b_r \right)}{\partial z},
\end{equation}
\begin{equation}
\frac{\partial B_\phi}{\partial t} = D b_r + \frac{\partial}{\partial
z}\left(\eta_B(\mathbi{B})\frac{\partial B_\phi}{\partial z}\right),
\end{equation}
where $D=R_\omega R_\alpha=[r (d\Omega/dr) h^2 / \eta_T][h\alpha_\star/\eta_T]$ is the dynamo number ($h$ is the disc thickness, $\alpha_\star$ is the maximum value of the hydrodynamic part of the $\alpha$-effect), $\eta_A(\mathbi{B})$ and $\eta_B(\mathbi{B})$ are the nonlinear turbulent magnetic diffusion coefficients of poloidal and toroidal magnetic fields, the nonlinear function ${\bf V}_A(\mathbi{B}) \equiv \left[ \eta_A(\mathbi{B}) - \eta_B(\mathbi{B}) \right]\partial_z(\ln B)$, and $\alpha(\mathbi{B})$ is the total nonlinear $\alpha$-effect, which includes its kinetic and magnetic parts
\begin{equation}
\alpha(\mathbi{B}) = \alpha^v + \alpha^m.
\end{equation}
These quantities are determined by the corresponding helicities ($\chi^v$ and $\chi^m$) and quenching functions ($\phi_v$ and $\phi_m$)
\begin{equation}
\alpha(\mathbi{B}) = \chi^v \phi_v(B) + \chi^c(\mathbi{B}) \phi_m(B), \quad
\mathrm{where} \quad B = |\mathbi{B}|.
\label{eqn:alpha_kleeorin}
\end{equation}
\cite{rogachevskii00} derived the quenching functions for an anisotropic turbulence, which are given by Eq.~(10) and (11) in \cite{kleeorin00}. The kinetic helicity is given by $\chi^v = -(\tau/3) \langle \mathbi{v}\cdot(\nabla \times \mathbi{v})\rangle$. The magnetic part of the $\alpha$-effect is introduced by the function $\chi^c(\mathbi{B})$ determined by the evolutionary equation \cite[see Eq.~12 in][]{kleeorin00}.

Eq.~(\ref{eqn:alpha_kleeorin}) with helicities and quenching functions derived by \cite{kleeorin02} contain the main nonlinearities of $\alpha$-effect. They include the conventional and algebraic quenching and dynamic nonlinearity described by the evolution of the magnetic helicity \cite[Eq.~12 in][]{kleeorin02}.

%
\section{Results}

%
\subsection{Tests of the fulfillment of the basic assumptions of dynamo theory}
\label{sec:tests}

\subsubsection{Averaging method and verification of Reynolds rules}

The choice of method for the field decomposition into its large and small-scale parts is a very important step in the practical application of the mean field dynamo theory. The proper averaging procedure should fulfill the Reynolds rules described by Eqs.~(\ref{eqn:reynolds1})-(\ref{eqn:reynolds2}). Otherwise, the assumptions under which we separate the induction equation into two equations describing the evolution of the mean and fluctuating parts of $\mathbi{B}$ could be violated.

Due to the different boundary conditions imposed at different boundaries of our simulation domain (periodic at the azimuthal boundaries, periodic with shear at the radial boundaries and open at the vertical boundaries), we use the simplest averaging method: the averaging over the horizontal planes, as it has been done by Hanasz et al. 2004). The other more advanced methods, e.g. Gaussian smoothing, would require dividing the procedure of averaging into two steps: the backword shearing of the domain to get back a full periodicity at the azimuthal boundaries and the actual averaging. In the case of averaging over the horizontal planes, the former step can be omitted.

\subsubsection{Verification of scale separation and anisotropy in turbulent spectra}
\label{sec:separation}

One of the basic assumptions of the linear and nonlinear dynamo theories \cite[][see also \S\ref{sec:dynamo_theory}]{parker55,moffatt78,raedler80} is the scale separation, which means that all quantities like the velocity and magnetic fields have two characteristic time and length scales: the large and the small ones. Such scales should be apparent as separate peaks in the velocity and magnetic spectra, provided that the computational domain is large enough, showing the characteristic scale lengths of waves in a turbulent region \cite[see Fig.~8.1 in][]{brandenburg05b}.

In Figure~\ref{fig:spectra_b} we show power spectra of kinetic and magnetic energy fluctuations taken at $t=$1000~Myr for model B. We show only one time snapshot because in the case of vanishing explicit resistivity ($\eta=0$) the power spectra of magnetic energy are very similar all over the simulation time. The kinetic energy spectra are obtained on the basis of density and velocity component distributions after subtracting the large scale sheared azimuthal fluid flow. In order to compare the spectra in our models to the Kolmogorov spectrum we present also a line representing the $\rm k^{-5/3}$ slope in the plots. The presented spectra of kinetic energy exhibit in the large wavenumber limit a similarity to the Kolmogorov $\rm k^{-5/3 }$ spectrum. The magnetic spectrum appears to be much flatter in the same wavenumber range. It is apparent that both the kinetic and magnetic spectra are monotonic, thus we can state that there is no separation in the turbulent scales of velocity and magnetic field components.

The fact that the spectra show the dominant power on the box scale has plausibility for the following reasons: (i) The models rely on the buoyancy of cosmic rays and magnetic fields. Cosmic rays, even if injected locally, diffuse effectively along the mean magnetic field to fill the box scales quickly. (ii) Magnetic loops reconnect due to an explicit or numerical resistivity. As it has been originally suggested by Parker (1992), the discussed cosmic-ray driven dynamo relies on the inverse turbulent cascade. The box sizes, signifying the largest available spatial scales, are  limited by the simulation times, which are equal to several months of CPU time in single processor runs for all presented simulations.

In Figure~\ref{fig:spectra_d} we present analogous spectra for model D with large resistivity $\eta=$10~pc$^2$/Myr, at  $t= 1000$~Myr. We can see that the experiment with larger resistivity gives similar kinetic energy spectrum and a steeper magnetic spectrum, which are both close to $\rm k^{-5/3 }$ spectrum, than  model B. This means that the magnetic energy resides more at larger scales in models with larger resistivity, according to expectations.

The differences in spectra of magnetic field and velocity components can be discussed in terms of correlation length of these components along $x$, $y$ and $z$ axes. The temporal evolution of correlation length of all the mentioned components in each direction, for model B, is displayed in Figure~\ref{fig:corr_b}. It is apparent in Figure~\ref{fig:corr_b} that the correlation lengths of $v_x$ and $v_y$ are in all three directions close to 100 pc, while the correlation length of $v_z$ is about 70 pc in $x$ and $y$ direction and oscillates around 350 pc in $z$ direction. The large correlation length of $v_z$ along $z$ direction is a manifestation of a vertical wind powered by the vertical cosmic ray pressure gradient. In the case of magnetic field the correlation length of $B_x$ in all directions is comparable to 50 pc, the correlation length of $B_y$ varies around 40 pc in the $x$ direction,  around 70 pc in the $z$ direction, and varies from 100 pc to more than 400 pc, and again down to 100 pc during the system evolution. Similarly to $v_z$ the correlation length of $B_z$ along the $z$ direction is much larger than the correlation length of $B_z$ in the $x$ and $y$ directions. This is again a signature of vertical, nonuniform wind, which stretches the magnetic field in the vertical direction.

%
\subsubsection{Tests of validity of the linear approximation}

For the next test we check the relative magnitudes of all terms in the ${\cal E}$ calculated from our experiments A and B (see \S\ref{sec:dynamo_theory}). The ratios of electromotive force terms calculated according to Eq.~(\ref{eqn:fluct_field_evol}) are presented in Figure~\ref{fig:terms} for models A (top) and B (bottom). As it was shown in \S\ref{sec:conditions} the linear dynamo theory neglects all quadratic terms of the mean field in the EMF equation so they remain only with the term $\nabla \times (\mathbi{v} \times \bar\mathbi{B})$ (see Eq.~\ref{eqn:fluct_field_evol_simpl}). To check how large are the three other terms ($\nabla \times (\mathbi{v} \times \mathbi{b})$, $\nabla \times (\overline{\mathbi{v} \times \mathbi{b}}$, $\nabla \times ({\bar\mathbi{V}} \times \mathbi{b})$) in comparison with this term we divide them  by the first one.  It is apparent that  after 10~Myr for the model A the highest ratio is obtained for the term $\nabla \times (\mathbi{v} \times \mathbi{b})$, which grows even ten times higher than the term  $\nabla \times (\mathbi{v} \times \bar\mathbi{\bf B})$. The second term $\nabla \times (\bar\mathbi{V} \times \mathbi{b})$ grows few times higher than the first term. The third term $\nabla \times(\overline{\mathbi{v} \times \mathbi{b}})$ shows ratios changing from 0.1 to a few times larger. The evolution of term ratios in model B (Fig.~\ref{fig:terms}, bottom plot) is very  similar to the evolution of two other models C and D, so we present only this case in the Figure. In Figure~\ref{fig:terms} (bottom) we can see that in the beginning all terms studied are close to the term  $\nabla \times (\mathbi{v} \times \bar\mathbi{B})$. Later on, the  term $\nabla \times (\overline{\mathbi{v} \times \mathbi{b}})$ has similar value to  $\nabla \times (\mathbi{v} \times \bar\mathbi{B})$. The ratio of the  terms $\nabla \times (\mathbi{v} \times \mathbi{b})$ and  $\nabla \times(\overline{\mathbi{V}} \times \mathbi{b})$ changes their magnitude from about 0.8 in the beginning to above 10 in the end of the evolution. Such high values of the normally neglected EMF terms  mean that in our numerical  experiments of the Parker instability, driven by the cosmic rays and shear, the resulting electromotive force is not in a linear regime. For these reasons one should consider inclusion of all the terms in the calculations of the dynamo coefficients, as it was proposed by \cite{blackman02} in their calculations  of the quenched evolution of EMF. Although as it will appear later, the quality of the approximation of the mean electromotive force will not become satisfactory.

The next issue is related to the assumption that the kinetic turbulent energy is much larger than the energy of the small-scale magnetic field \cite[see][]{kleeorin03}. Figure~\ref{fig:energies} presents that in model A, since 1000~Myr, both energies are comparable. In the case of experiments B and C the situation is different, showing that the magnetic turbulent energy is smaller than the kinetic one during the whole period of calculations.  For this reason in calculations of the dynamo coefficients according to \cite{kleeorin03} we apply their approximations for models B--D.

\begin{figure}  
 \epsscale{1.0}
 \plotone{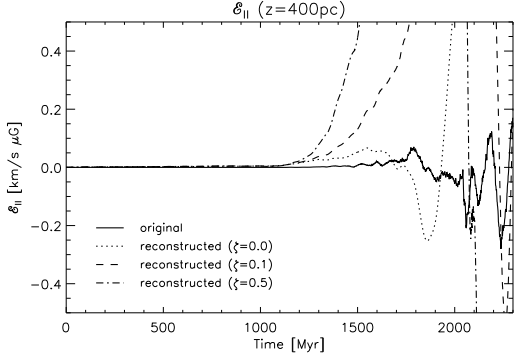}
 \caption{Comparison of the time evolution of  the component of ${\cal E}$ parallel to the local mean magnetic field reconstructed by the Blackman approach for three values of $\tilde\zeta$ (0.0 -- dotted line, 0.01 -- dashed line, 0.05 -- dash-dotted line) to $ {\cal E}_\parallel$ taken directly from the experiments (solid line) for model A for Z=400pc. \label{fig:blackman_a}}
\end{figure}

%
\subsection{The fitting of the different dynamo coefficients to the calculated EMF}
\label{sec:fitting}

%
\subsubsection{Implementation of the Blackman-Field approach}

\begin{figure}  
 \epsscale{1.0}
 \plotone{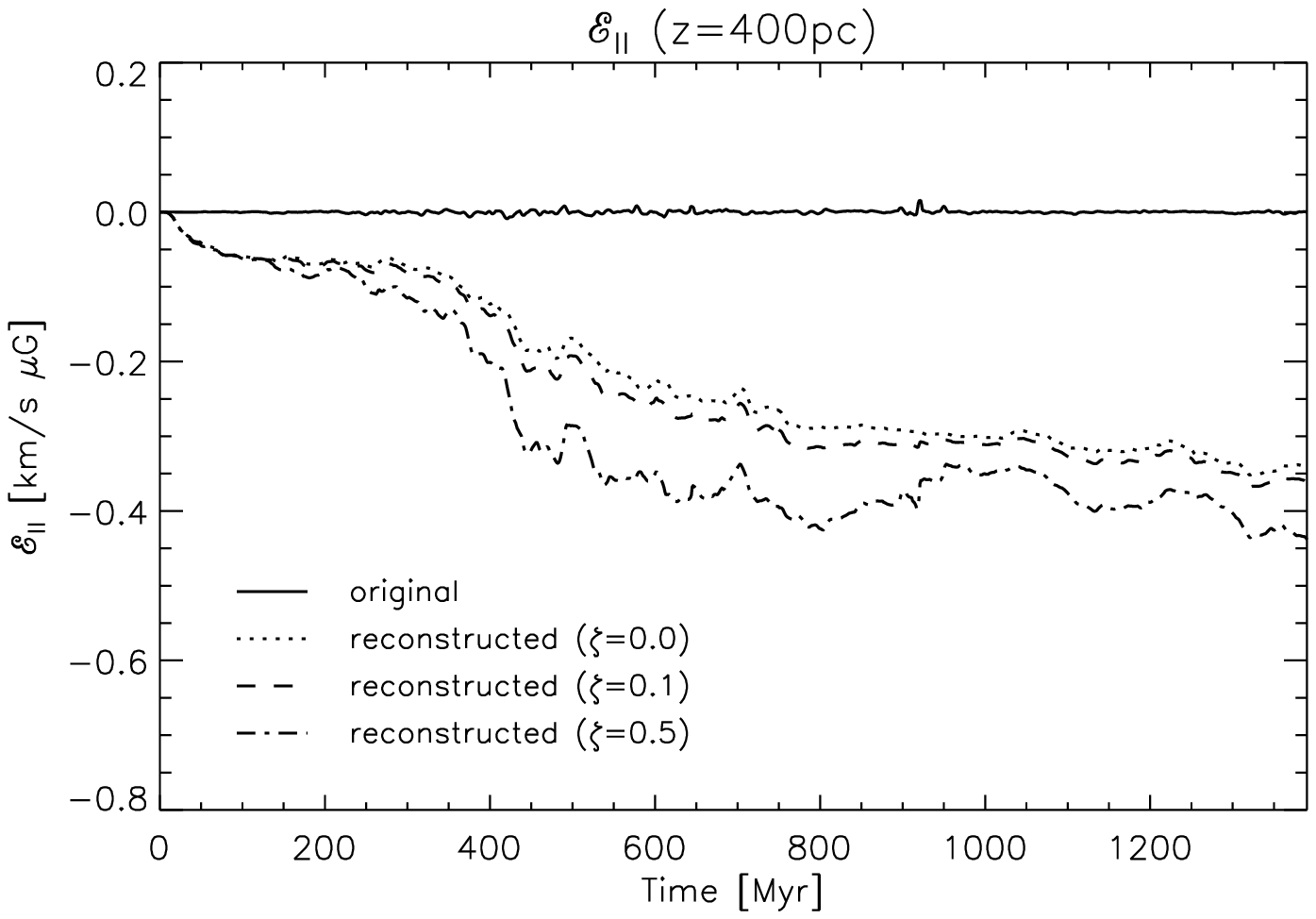}
 \plotone{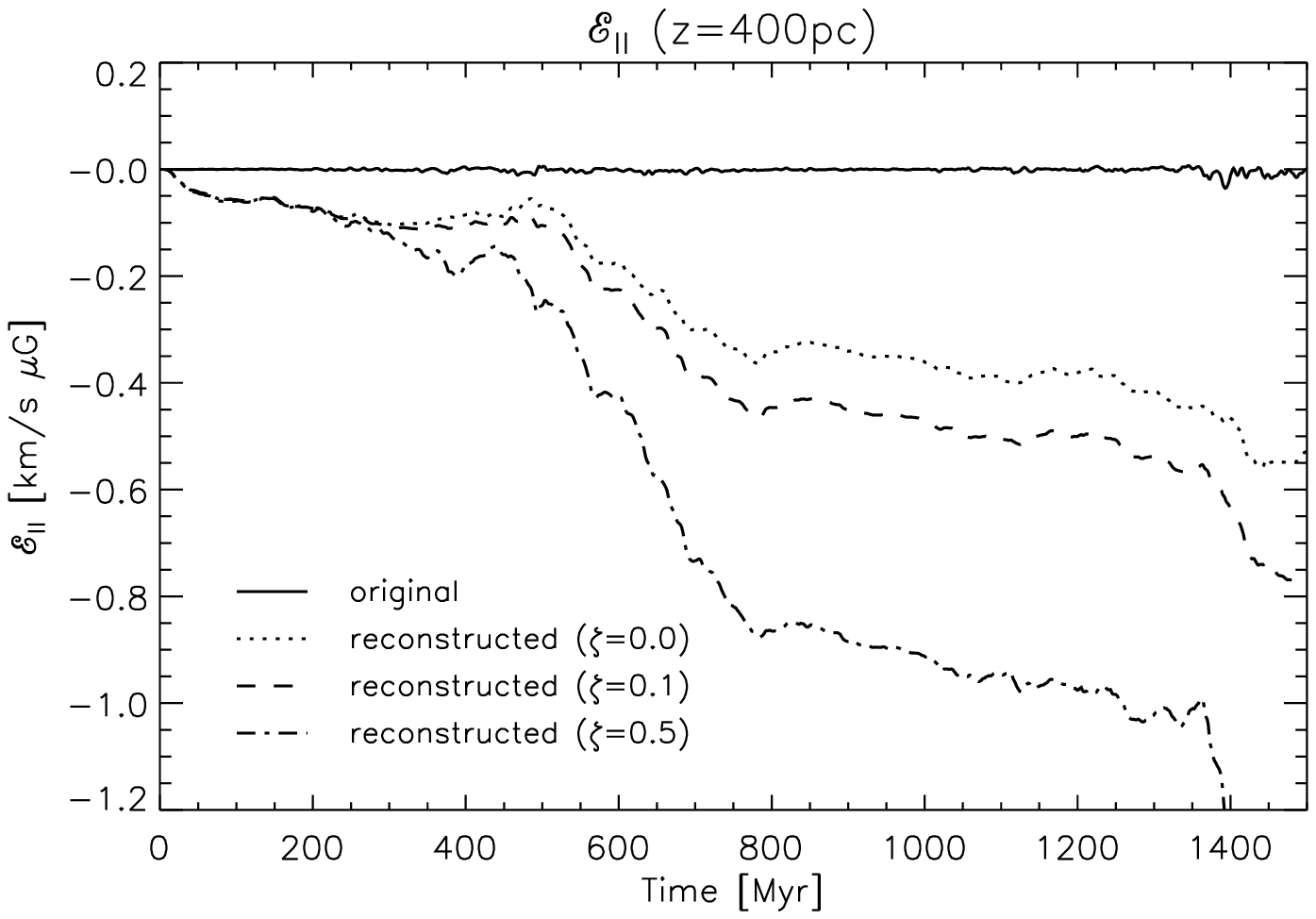}
 \caption{Comparison of the time evolution of  the component of ${\cal E}$ parallel to the local mean magnetic field reconstructed by the Blackman approach for three values of $\tilde\zeta$ (0.0 -- dotted line, 0.01 -- dashed line, 0.05 -- dash-dotted line) to $ {\cal E}_\parallel $ taken directly from the experiments (solid line) for models B and C (upper and lower plot, respectively) for Z=400pc. \label{fig:blackman_bc}}
\end{figure}

The next goal is to fit the dynamo coefficients calculated with the help of the \cite{blackman02} prescription to the electromotive force resulting from our numerical models A, B and C  \cite[see also][]{kowal05}. The authors (BF02) applied the nonlinear dynamo theory including all quadratic terms in the mean field in the EMF approximation (see \S\ref{sec:blackman}). We shall check if their approximation gives similar evolution of the EMF to the original one. We apply Eq.~(\ref{eqn:blackman}) to obtain the time derivative of the component of the EMF parallel to the local magnetic field, the values of coefficients $\tilde{\alpha}$ and $\tilde{\beta}$ and the remaining terms in that equation. Figs.~\ref{fig:blackman_a} and \ref{fig:blackman_bc} present the time evolutions of the original and the reconstructed EMF for models A, B and C in the chosen time periods.

The large discrepancies  of the original and reconstructed EMF signify a weak application of the Blackman model to the results obtained in the simulations. We shall further discuss in \S\ref{sec:integration_blackman}. the results of reconstruction of mean magnetic field components, based on Blackman \& Field approach.

\begin{figure}  
 \epsscale{1.0}
 \plotone{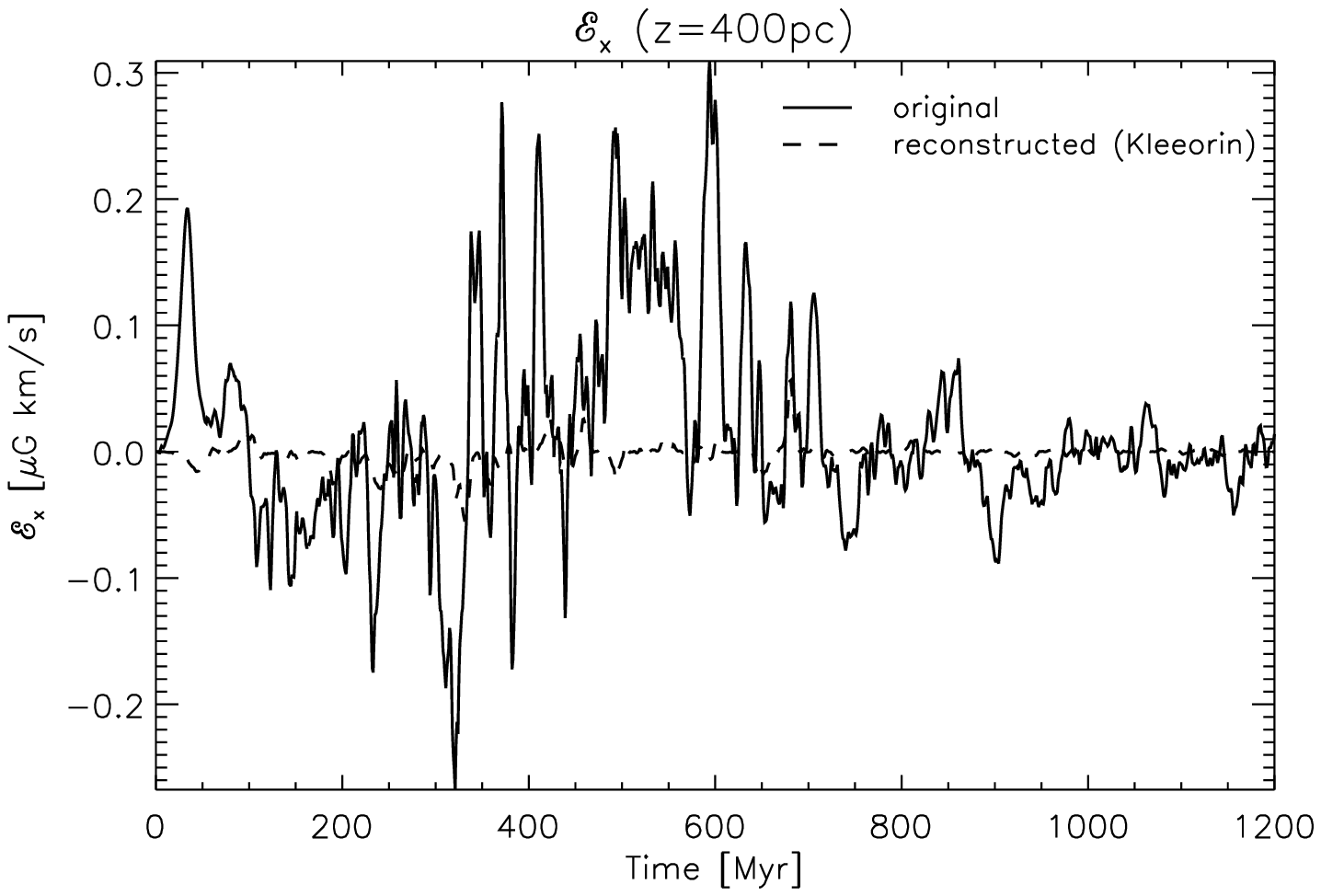}
 \plotone{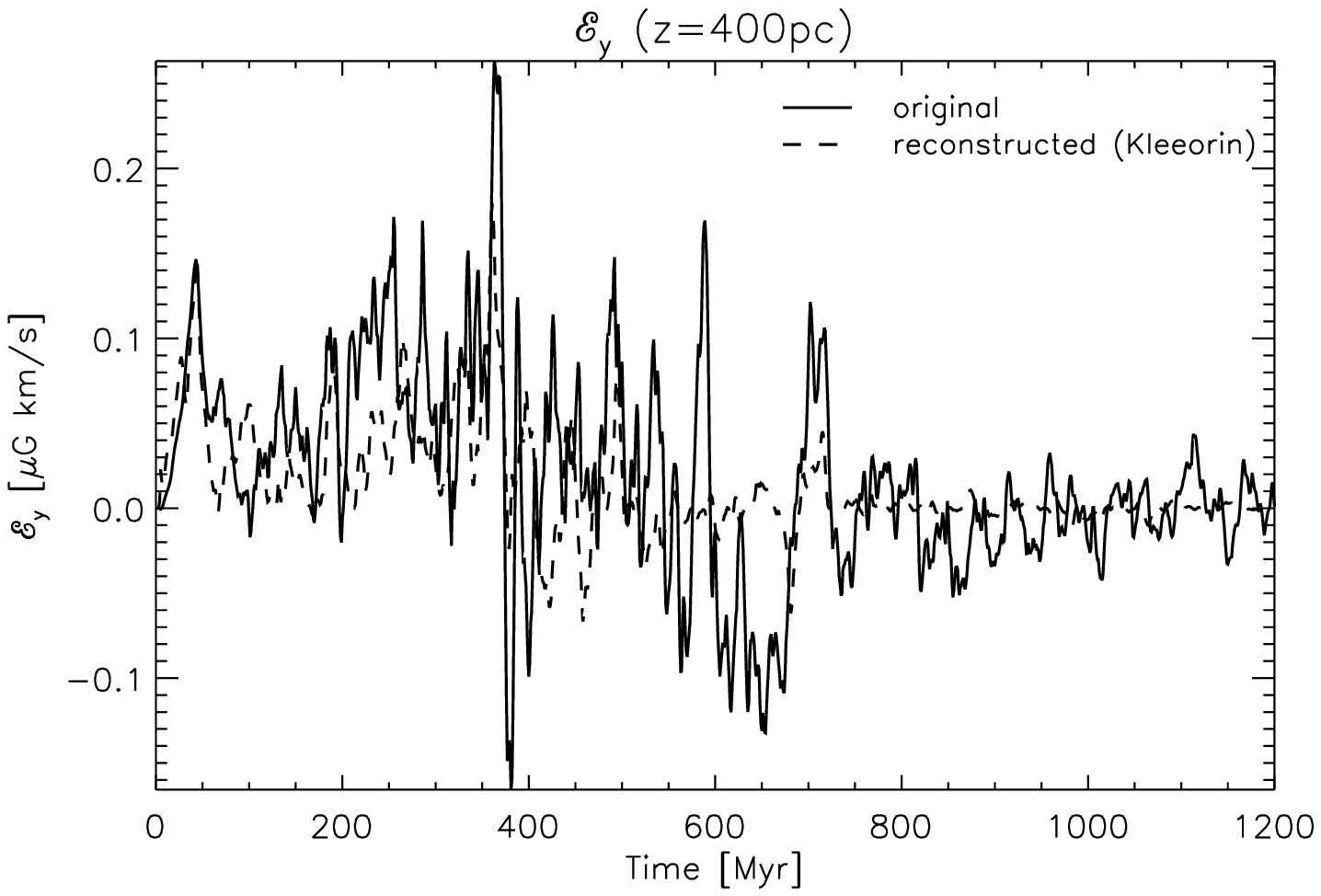}
 \plotone{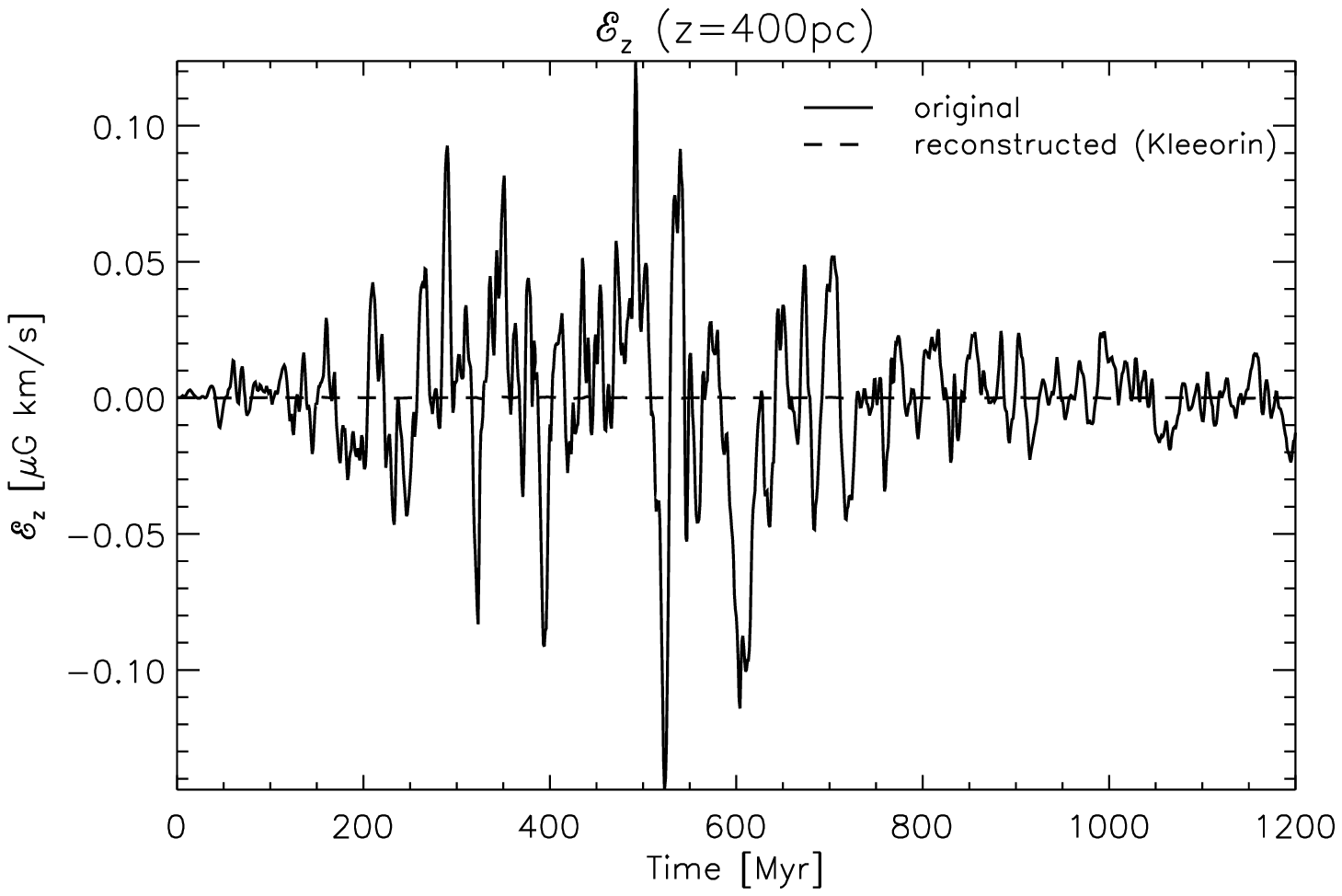}
 \caption{Comparison of the time evolution of X-, Y-, and Z-component of ${\cal E}$ (top, middle and bottom plots respectively) reconstructed by the Kleeorin-Rogachevski approach (dashed lines) to those taken directly from the experiments (solid lines) for example model B for Z=400pc. \label{fig:kleeo3}}
\end{figure}

\subsubsection{Implementation of the Kleeorin-Rogachevski approach}

The second analysis concerns the prescription of the dynamo coefficients given in the paper \cite{kleeorin03} (see \S\ref{sec:kleeorin}). The authors apply in the electromotive force prescriptions only the term $\overline{\mathbi{v} \times \mathbi{b}}$ and neglect the terms $\mathbi{V} \times \mathbi{b}$, $\mathbi{v} \times \bar\mathbi{B}$, and $\mathbi{ v} \times \mathbi{b} $, but they take into account the anisotropy of turbulence \cite[see also][]{rogachevskii01}. \cite{kleeorin03} applied their quenched form of the dynamo coefficients to the large-scale dynamo simulations. In \S\ref{sec:kleeorin} we present the set of equations from their paper, which we include in our calculations. Again we use our model B in order to calculate both magnetic and kinetic $\alpha$, as well as the different forms of the diffusion coefficient. We add all resulting terms to get the components of the electromotive force according to \cite{kleeorin03}.

The time evolution of the actual (the solid line) and reconstructed (the dashed line) electromotive force components averaged over the planes are presented in Figure~\ref{fig:kleeo3}. One can see that differences between the actual and the reconstructed electromotive forces are remarkable, however, their time averaged values seem to match the reconstructed electromotive forces, at least in the y direction. The discussion of reconstructed mean magnetic field basing on the Kleeorin \& Rogachevski approach will be presented in \S\ref{sec:integration_kleeorin}.

%
\subsection{Integration of the mean magnetic field from EMF}
\label{sec:integration}

In the previous section we show that, the Blackman-Field approach does not reconstruct the electromotive force properly, while the Kleeorin-Rogachevski approach provides in some cases electromotive forces that follow approximately time-averaged electromotive forces taken from experiments. Nevertheless, these approaches permit for obtaining some useful informations about the electromotive force and its properties. For instance, integrating Eq.~(\ref{eqn:mean_field_evol}) we can obtain a qualitative information about the amplification of the mean magnetic field. We also can analyze, how much of the mean magnetic field can be restored by the integration of the electromotive forces reconstructed by each approximation.

We perform the time integration of Eq.~(\ref{eqn:mean_field_evol}) substituting the term $\overline{\mathbi{v} \times \mathbi{b}}$ by either original or reconstructed electromotive force ${\bf \cal E}$:
\begin{equation}
 \frac{\partial \bar\mathbi{B}}{\partial t} = \nabla \times \left[ {\bf \cal E} + \bar\mathbi{V} \times \bar\mathbi{B} - \eta \nabla \times \bar\mathbi{B} \right] \, .
 \label{eqn:mean_field_integ}
\end{equation}
In addition, we could neglect the second term incorporating the large-scale fields $\bar\mathbi{V}$ and $\bar\mathbi{B}$, because in general, these field should be weakly varying in space if the procedure of separation undergo the Reynolds rules. However, in our analysis, we incorporated the plane averaging procedure, which could give substantially large vertical derivatives, thus in the above equation we take the large-scale component of velocity $\bar\mathbi{V}$ directly from models. In this way, the only unknown variable is the large-scale component of magnetic field $\bar\mathbi{B}$. If assumptions of the mean field theory are fullfiled, the Eq.~(\ref{eqn:mean_field_integ}) should reconstruct the mean magnetic field correctly. The numerical integration is performed with the use of the fourth order Runge-Kutta method \cite[see e.g.][]{press97}.

\begin{figure*}  
 \epsscale{0.35}
 \plotone{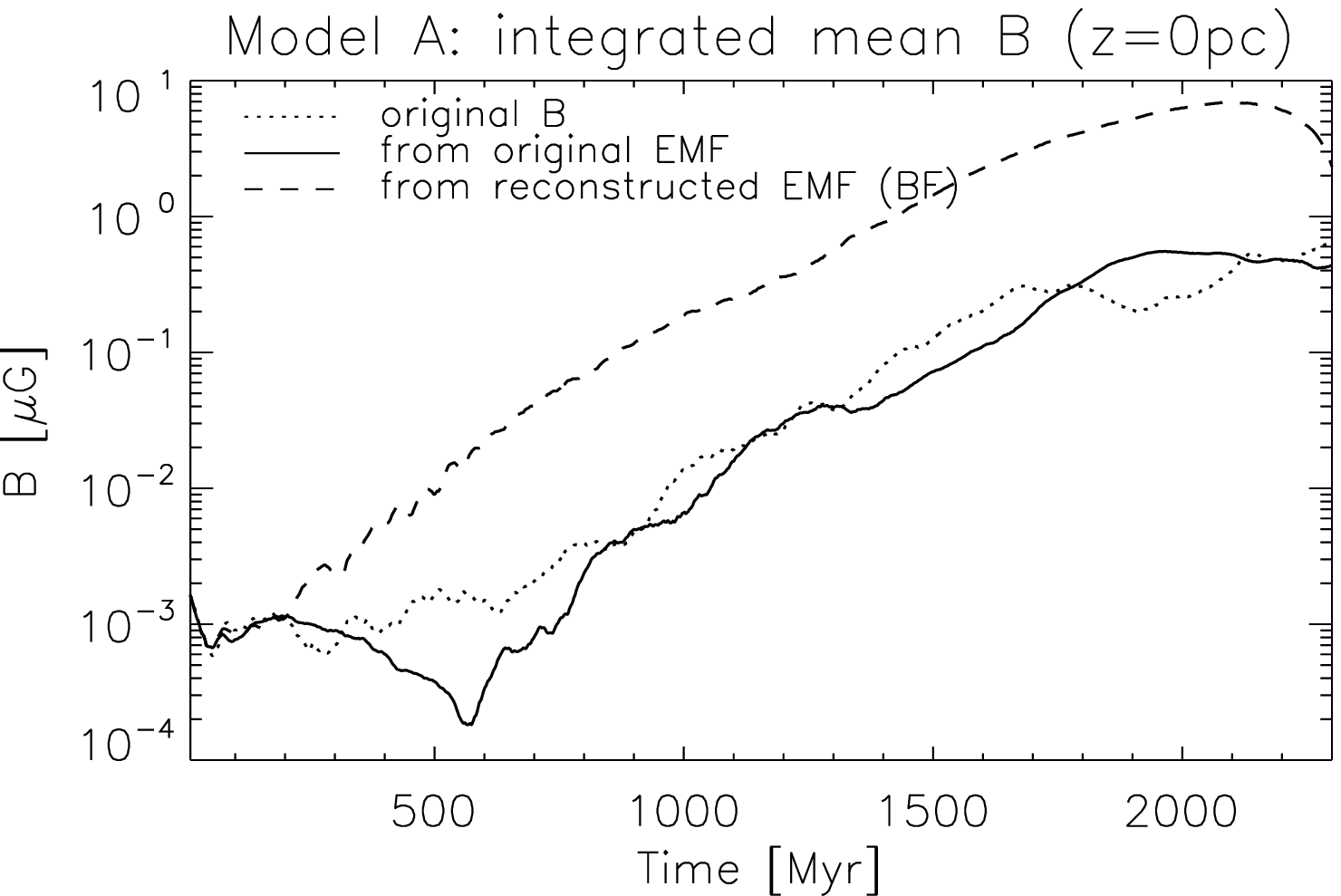}
 \plotone{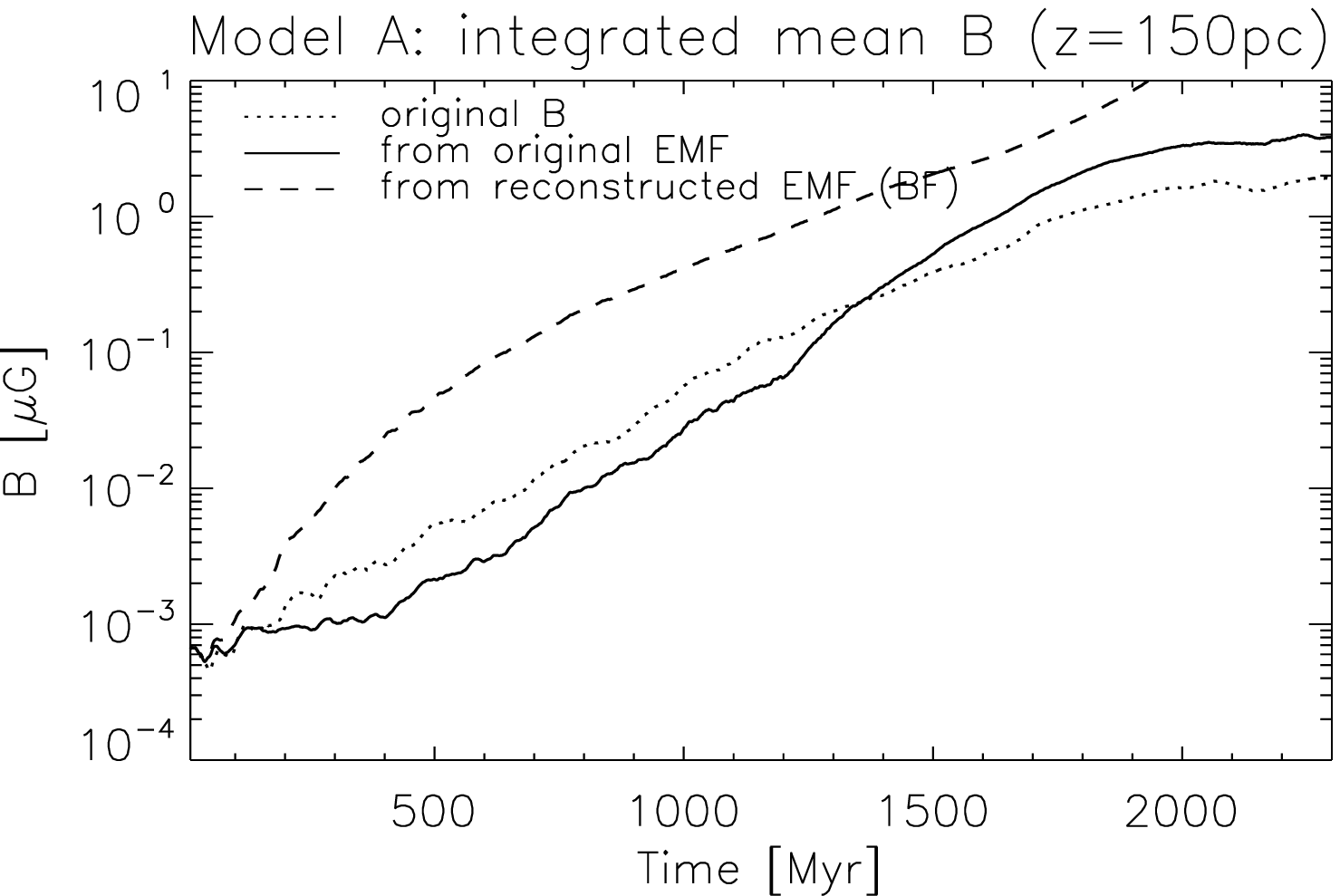}
 \plotone{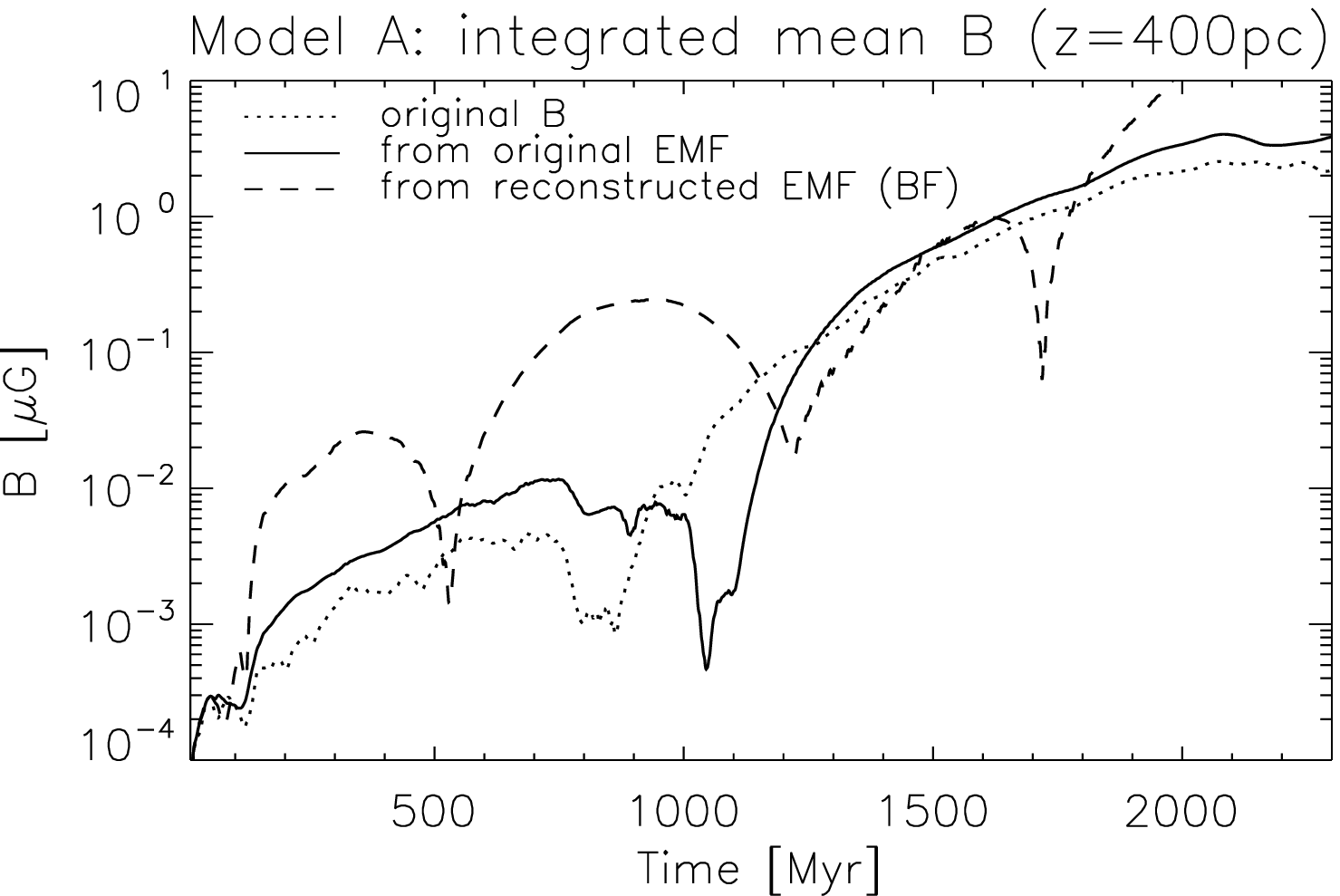}
 \plotone{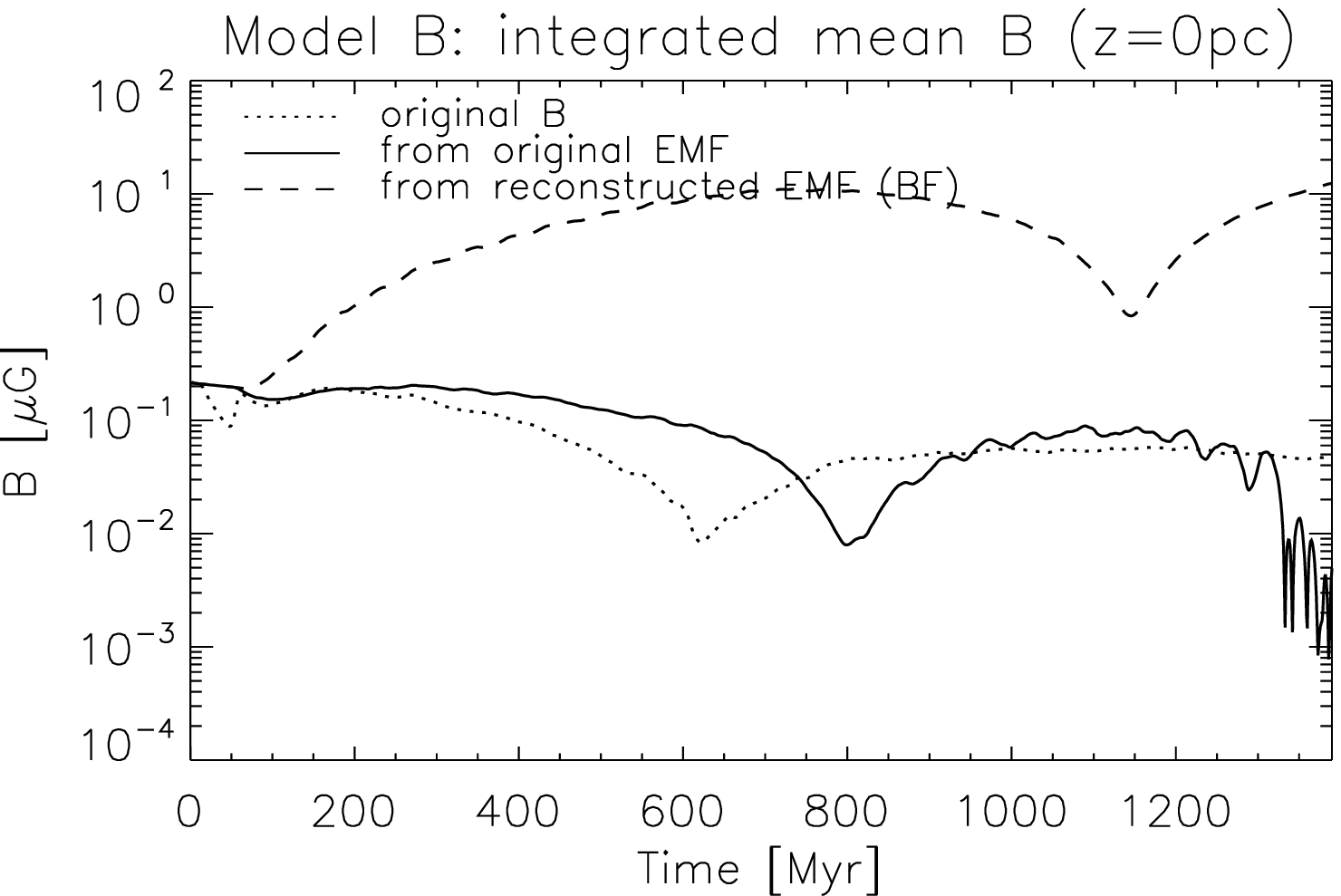}
 \plotone{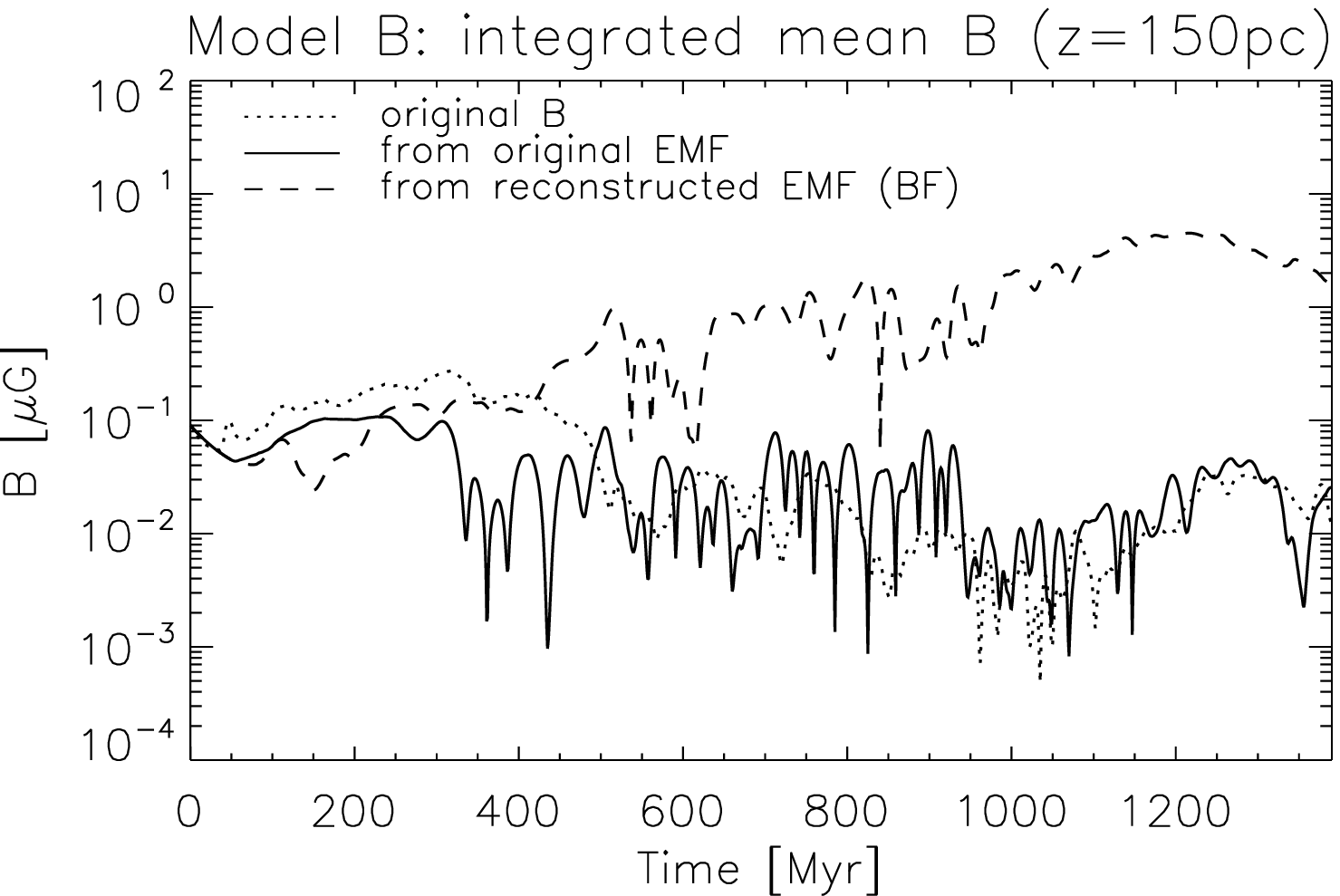}
 \plotone{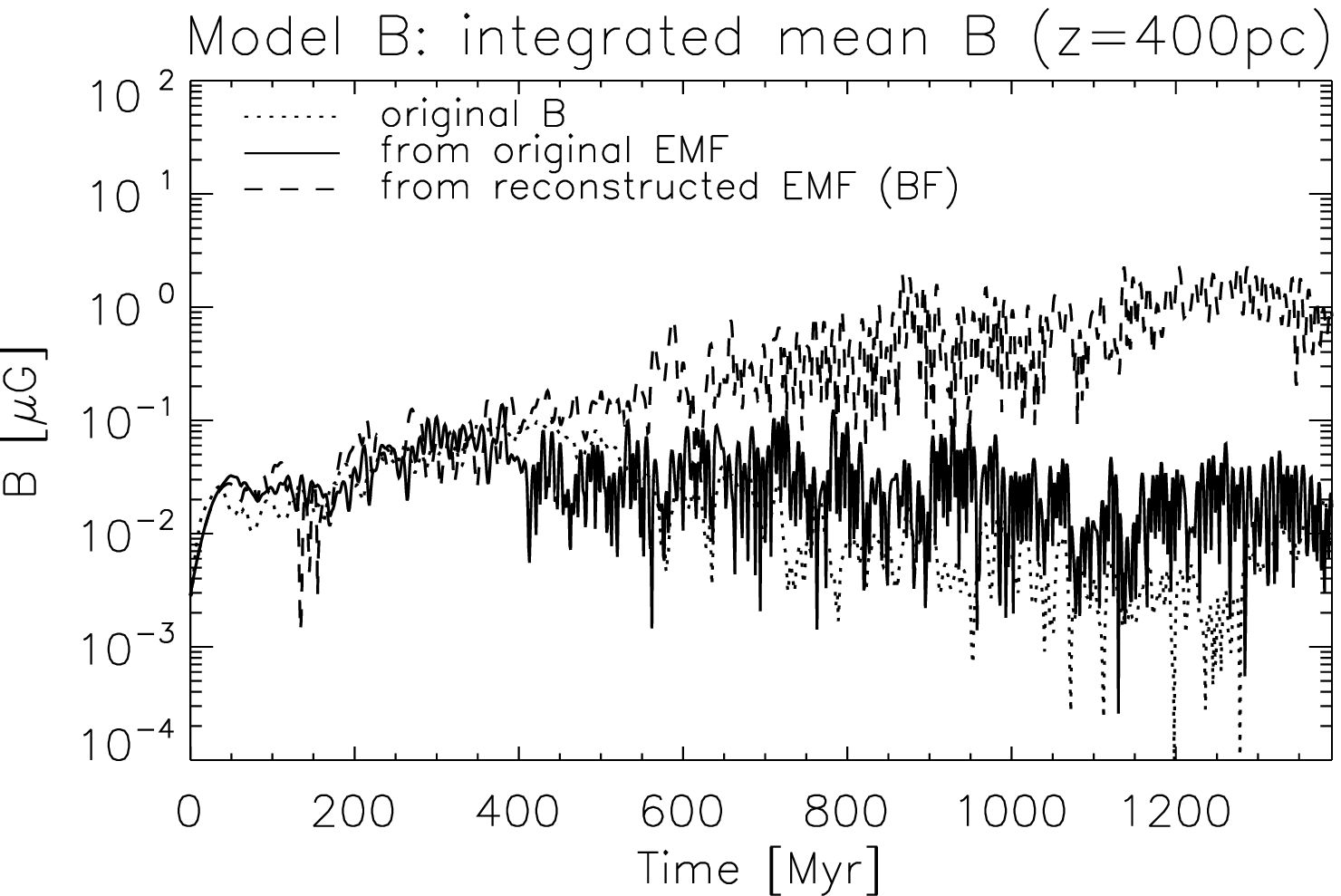}
 \plotone{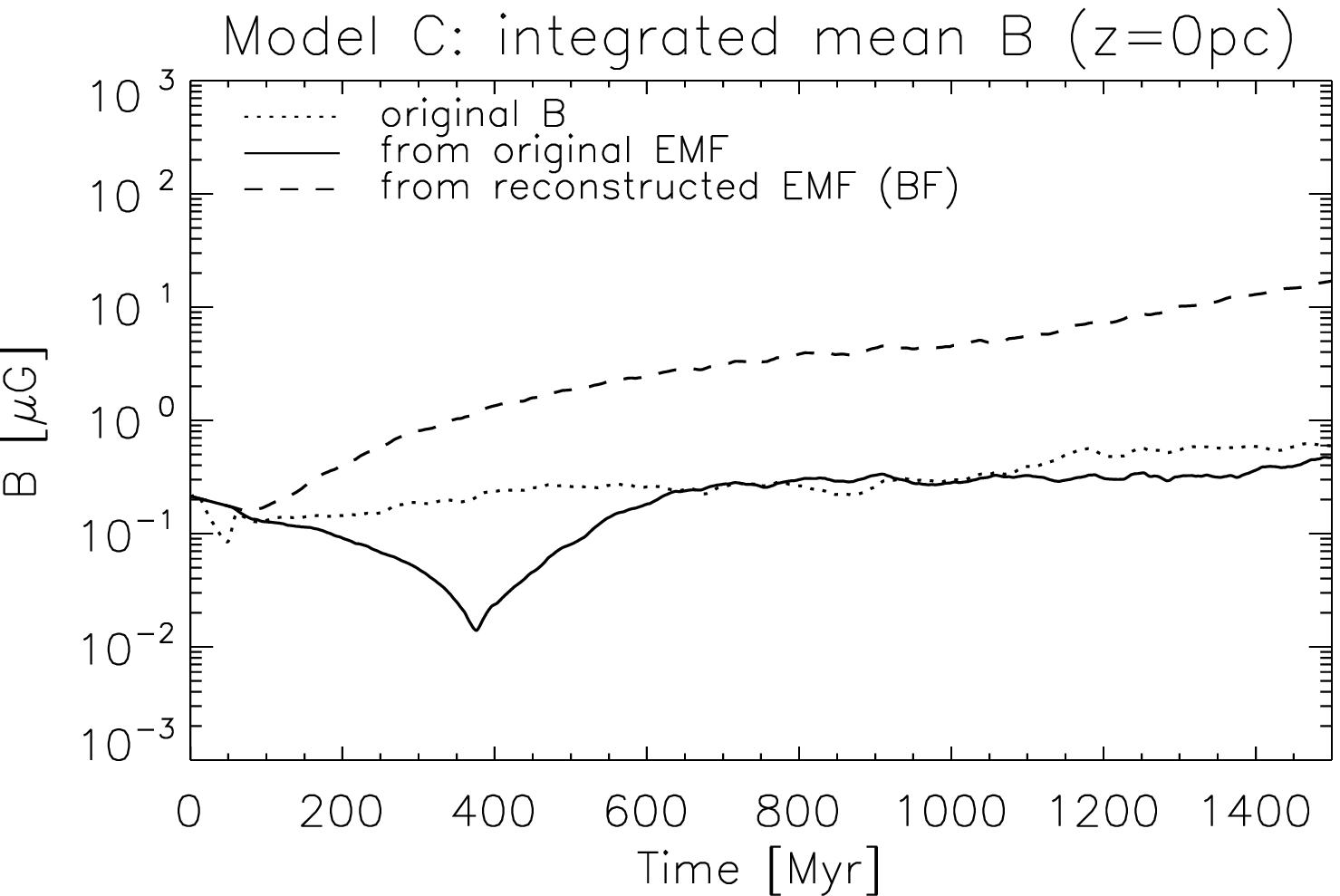}
 \plotone{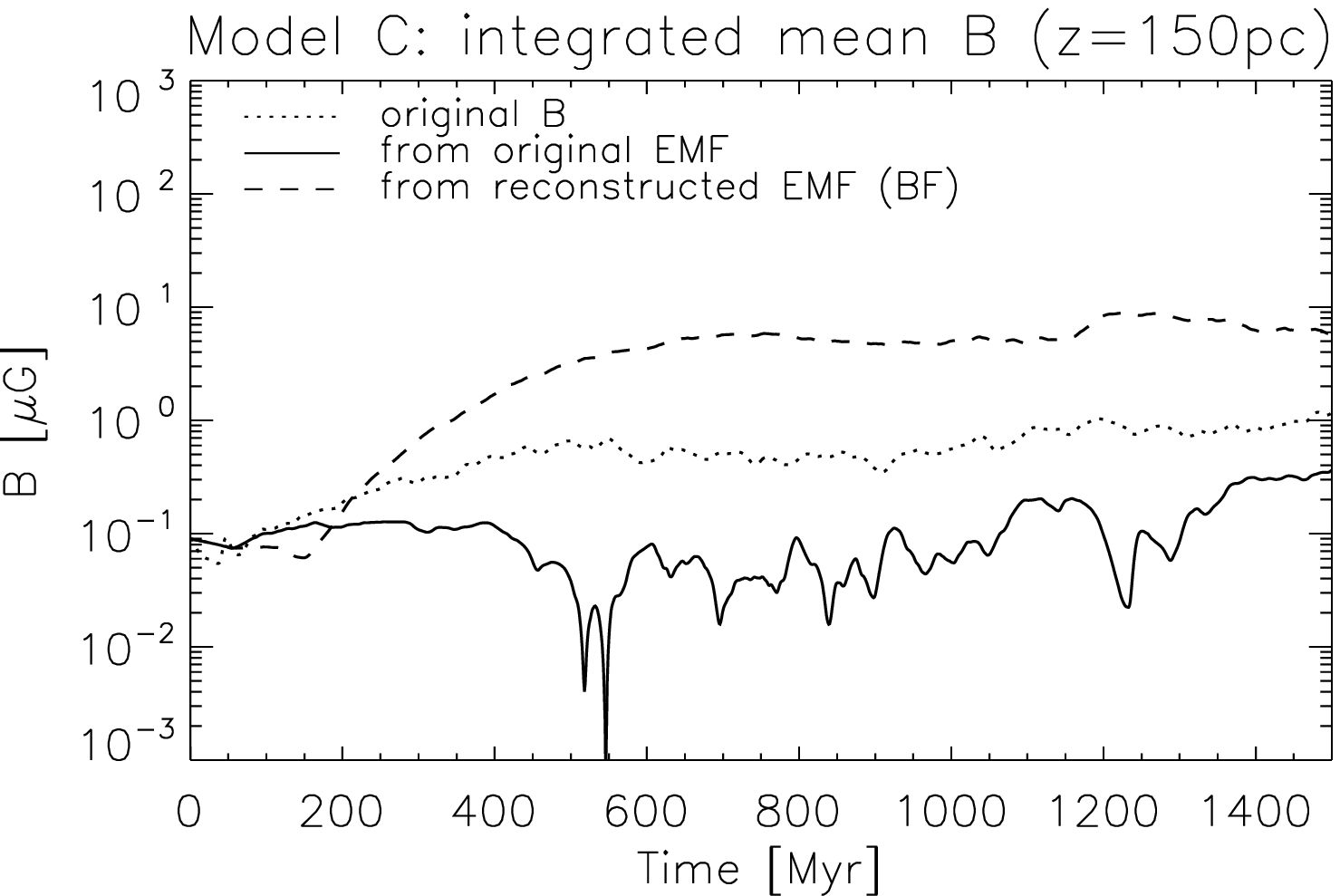}
 \plotone{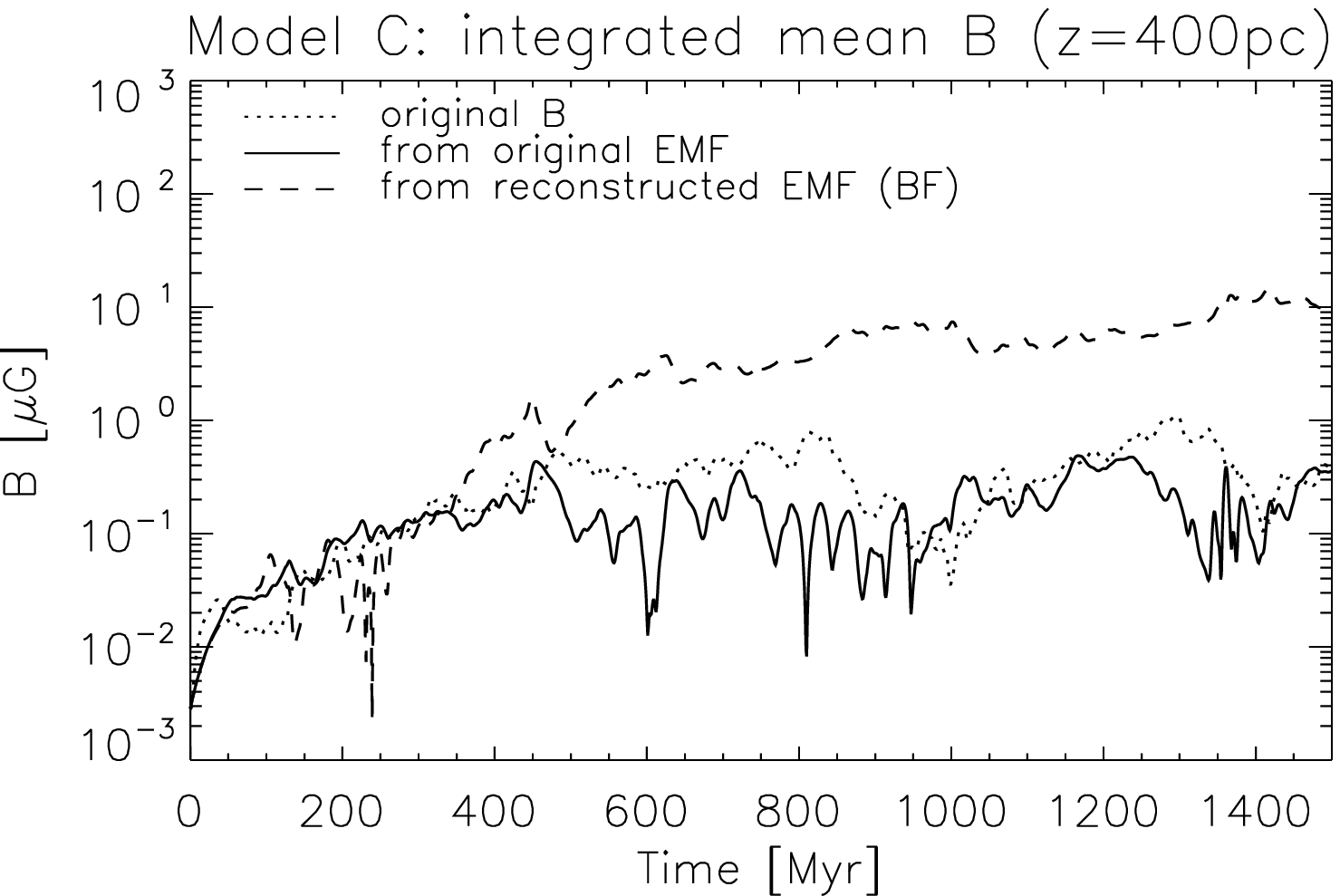}
 \plotone{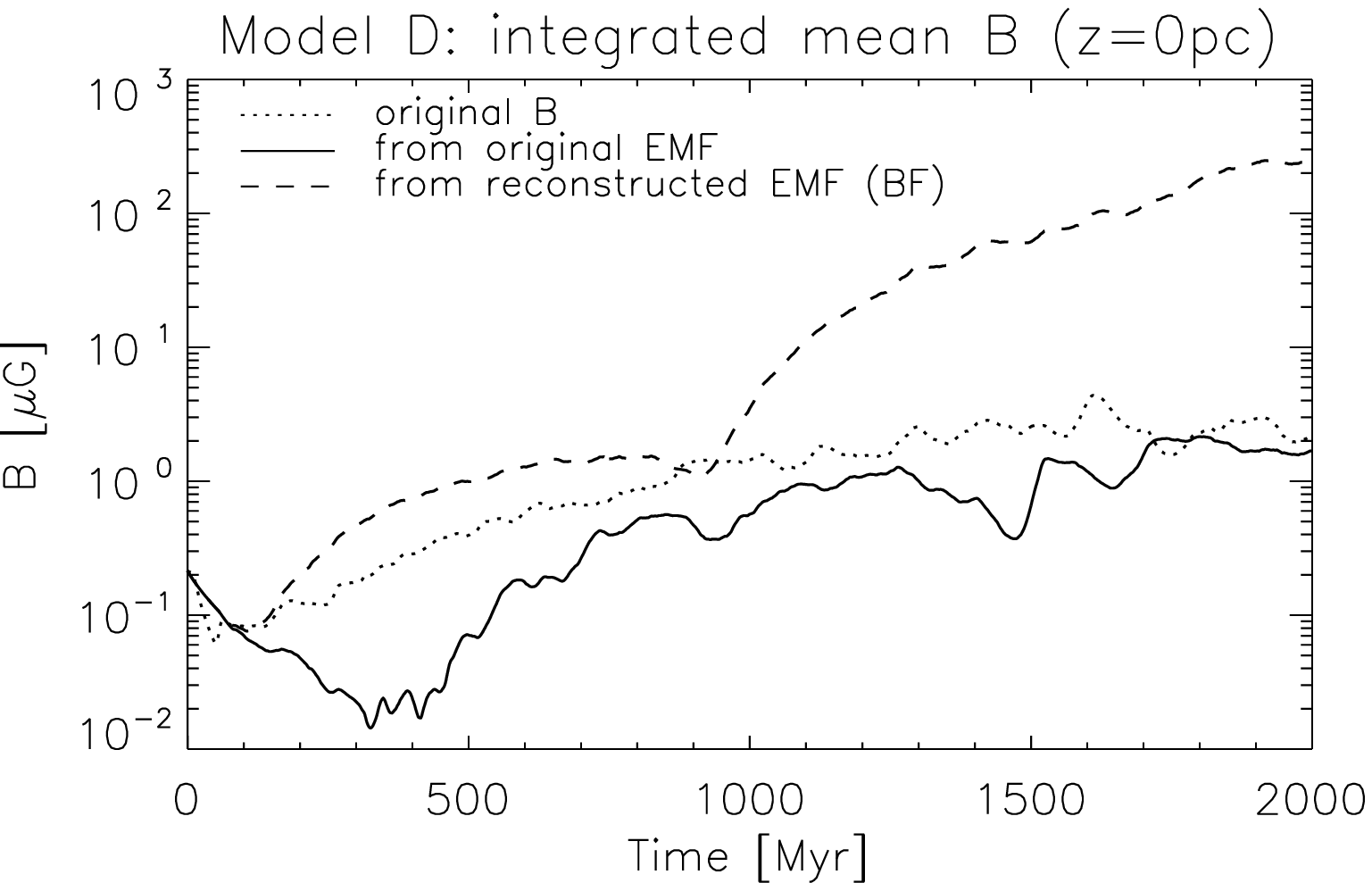}
 \plotone{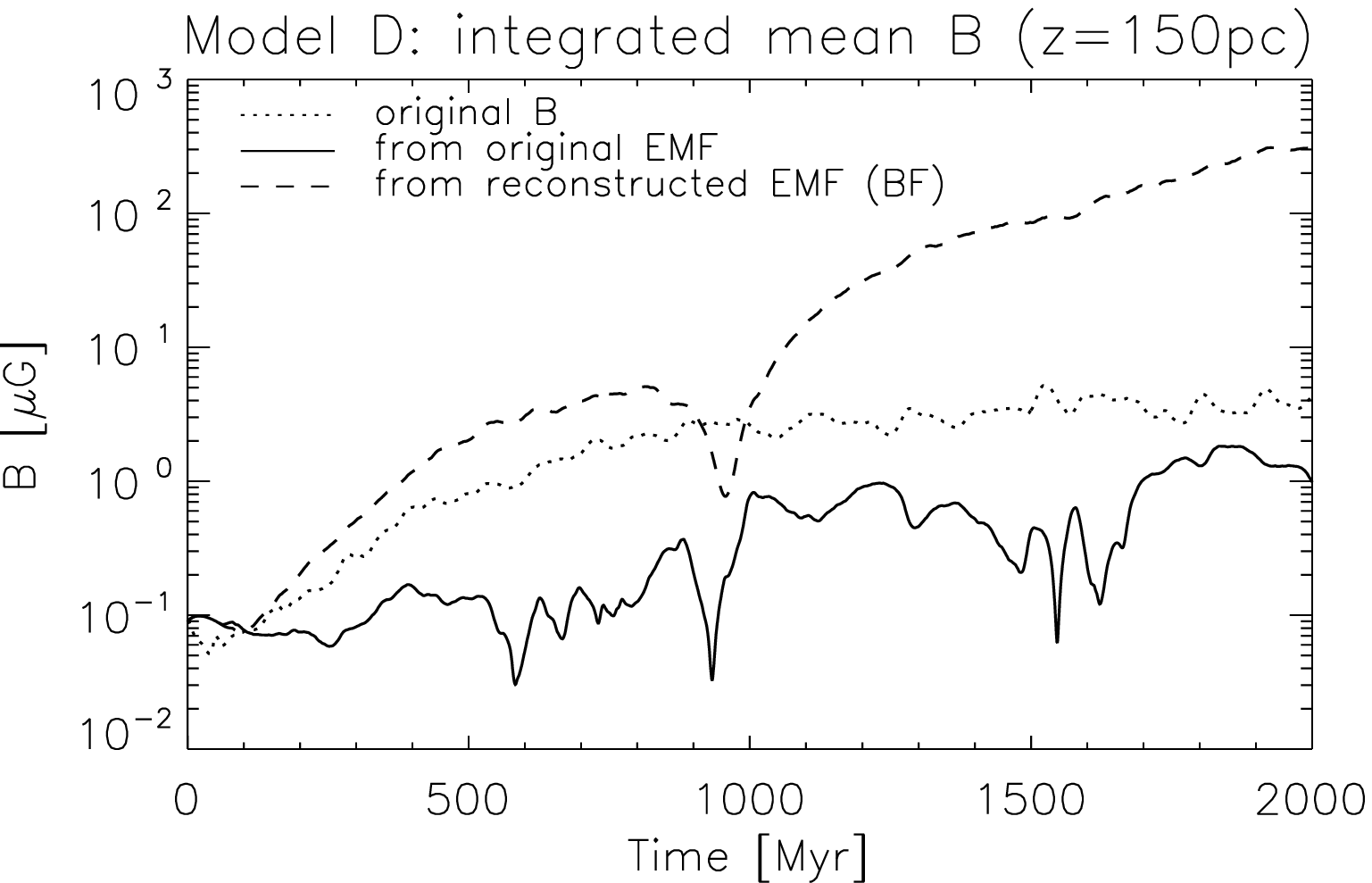}
 \plotone{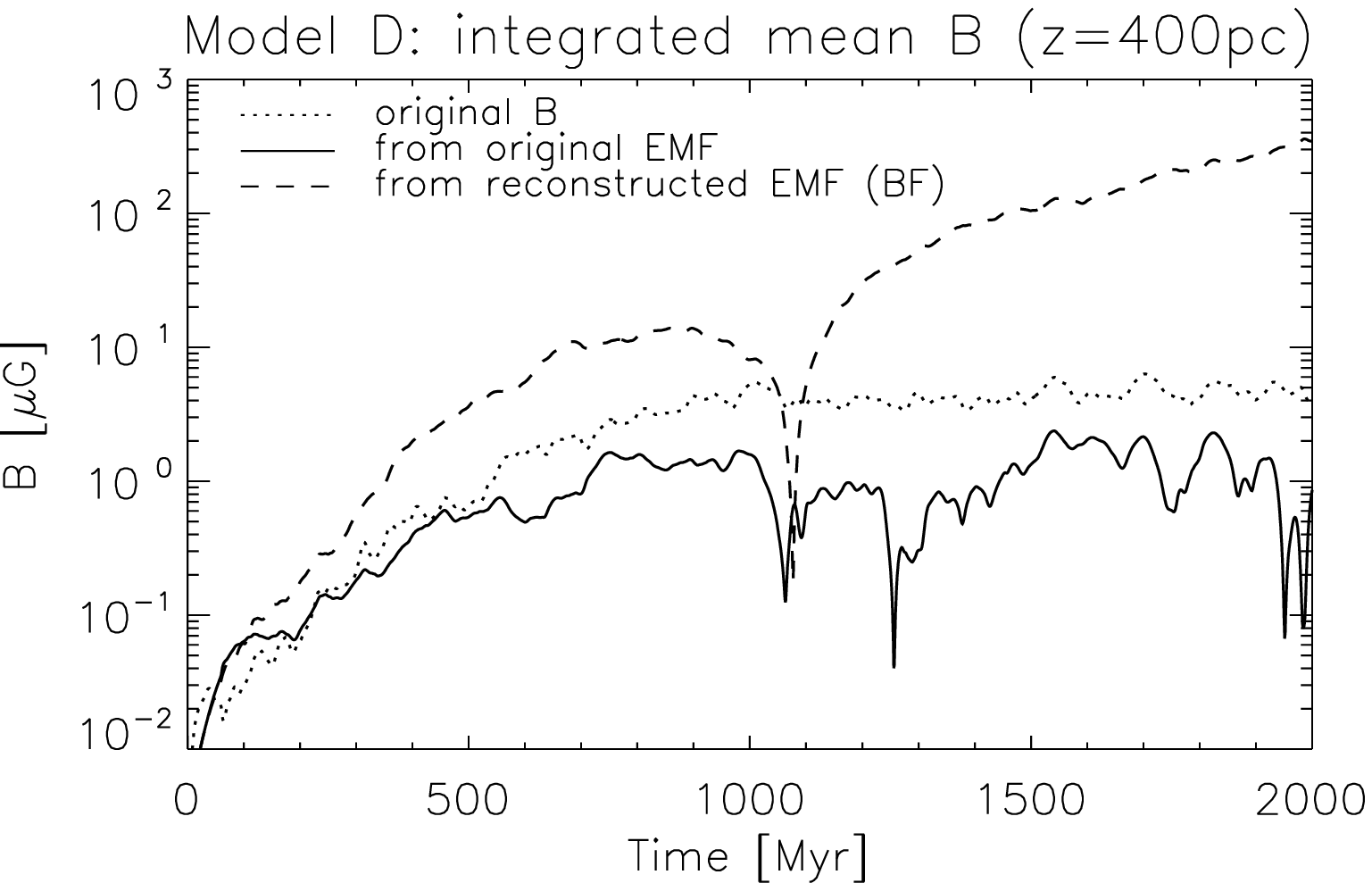}
 \caption{Time evolution of the absolute values of the mean magnetic field integrated from original (solid lines) and reconstructed (dashed lines) ${\cal E}$ for three vertical positions Z=0, 150, 400pc (left, middle and right columns, respectively) for models A, B, C and D (from top to bottom row). The reconstructed  ${\cal E}$ was calculated using the Blackmann-Field approach. We also show the original mean magnetic field taken directly from simulations (dotted lines). \label{fig:int_mag_blackman}}
\end{figure*}

\begin{figure*}  
 \epsscale{0.35}
 \plotone{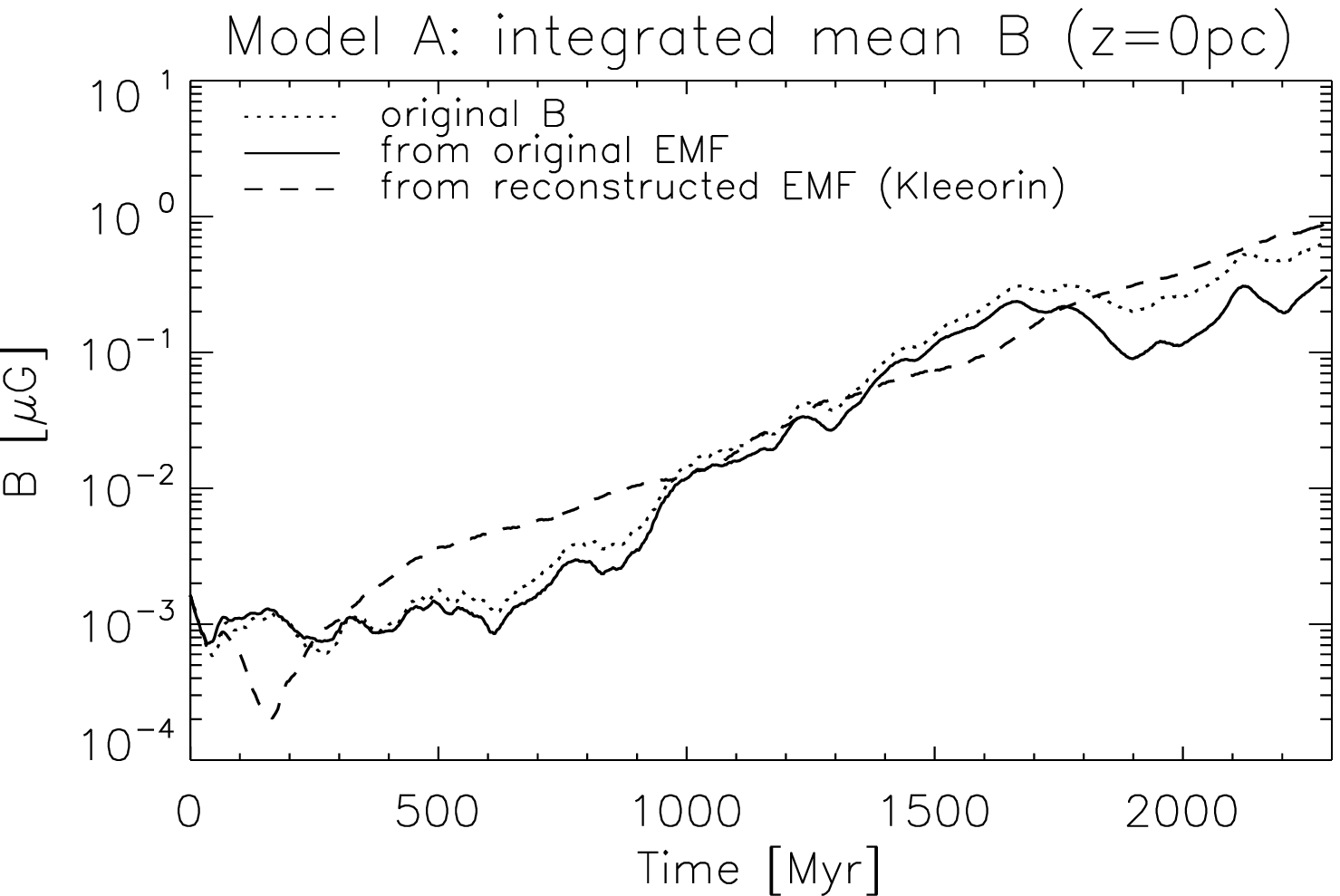}
 \plotone{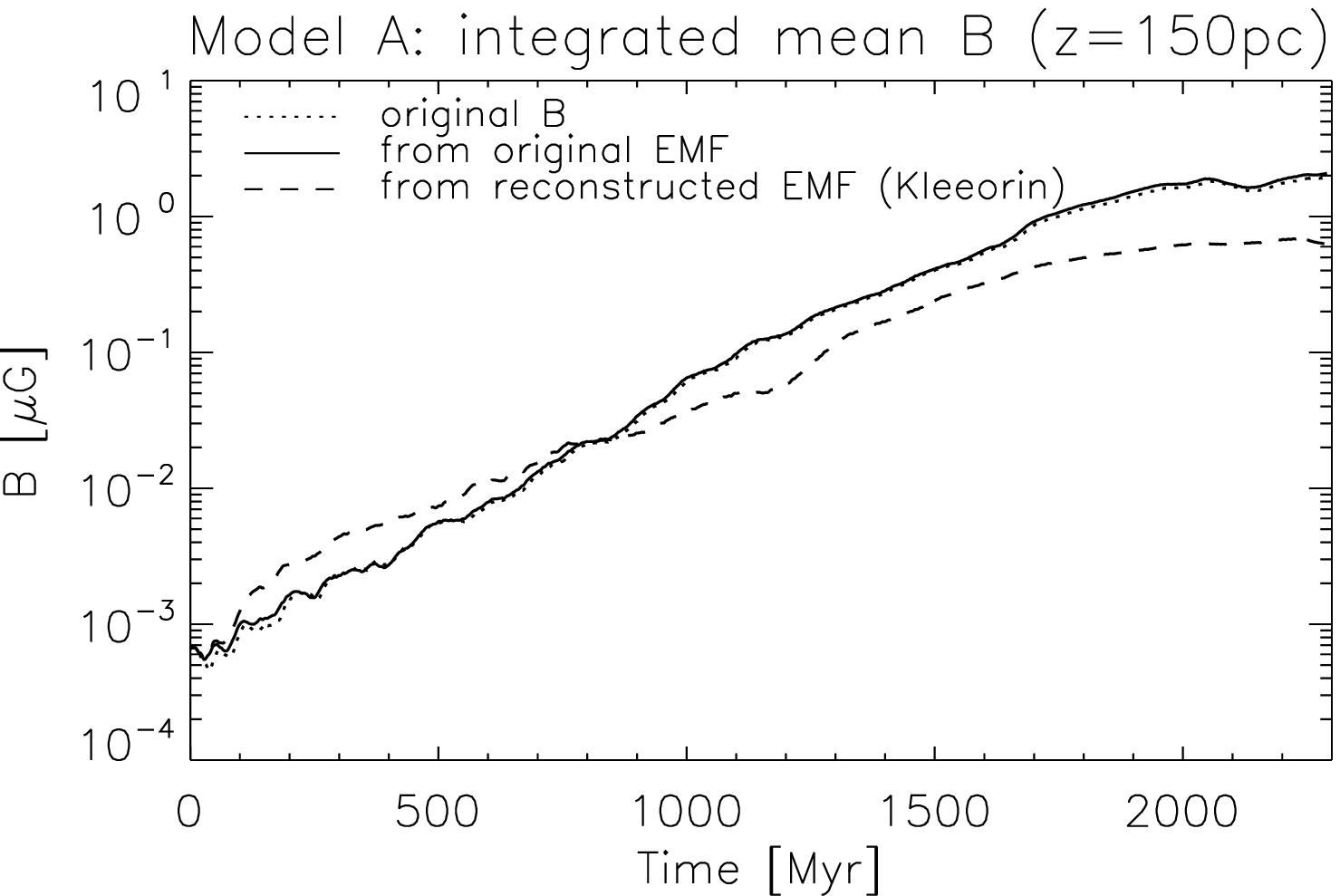}
 \plotone{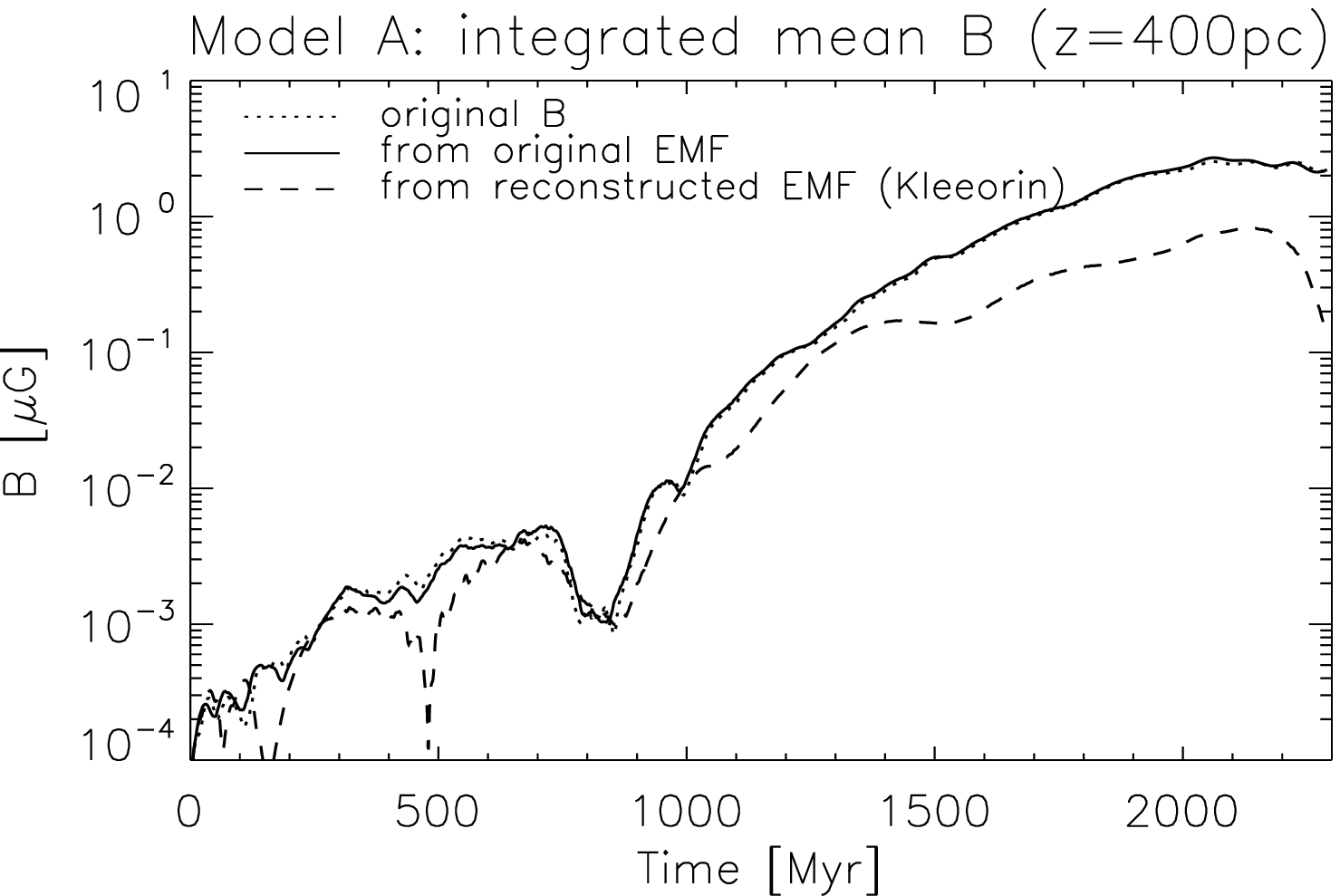}
 \plotone{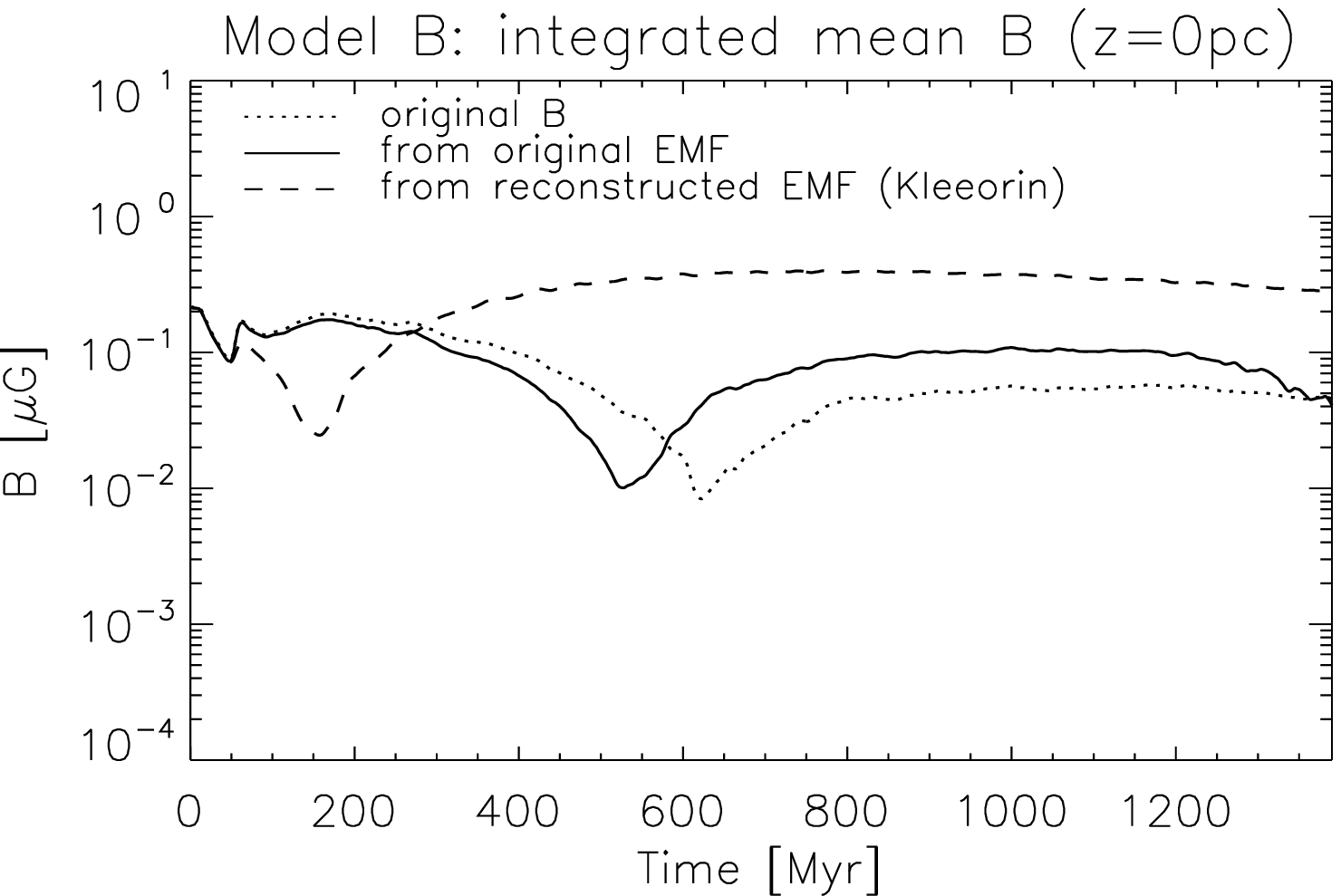}
 \plotone{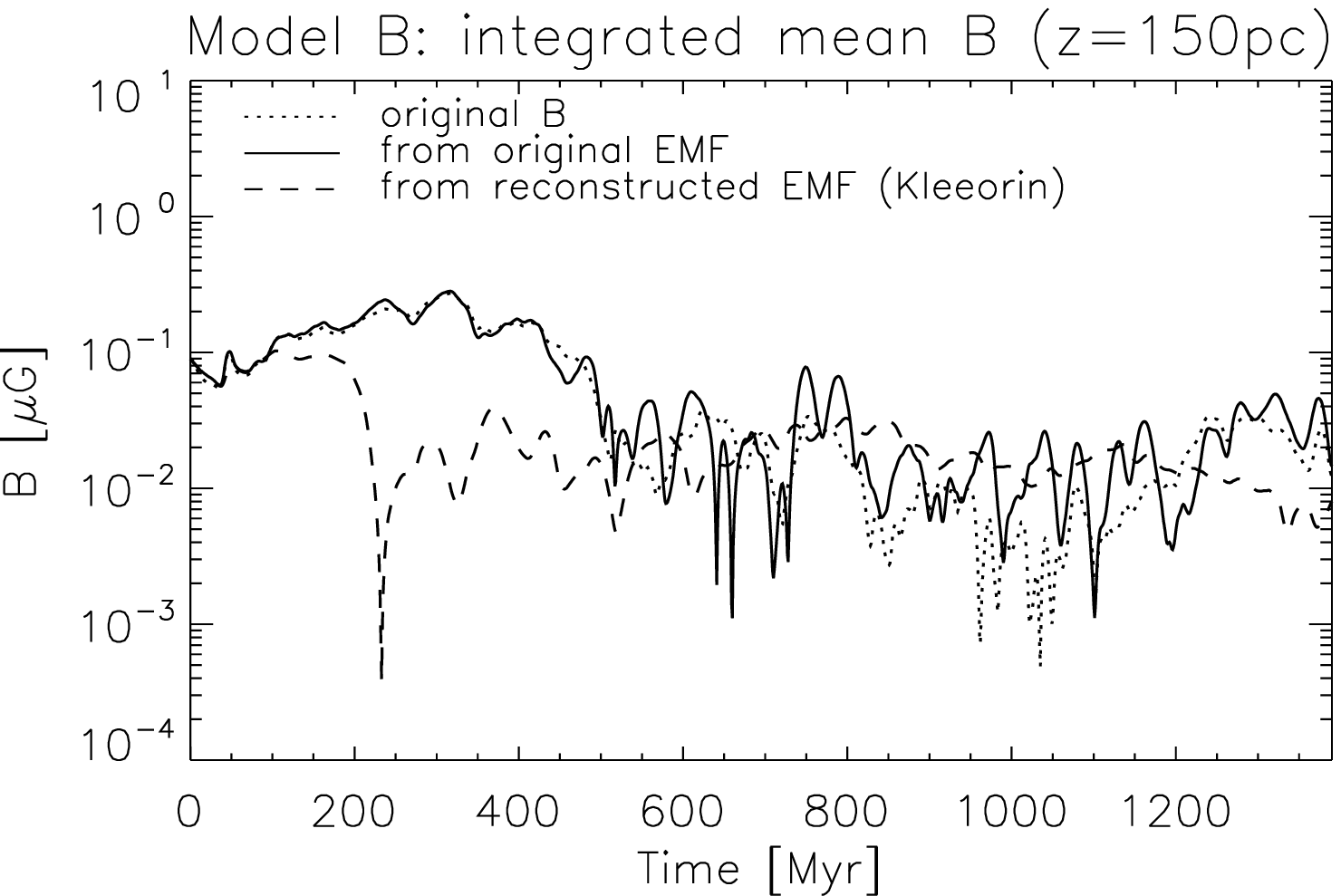}
 \plotone{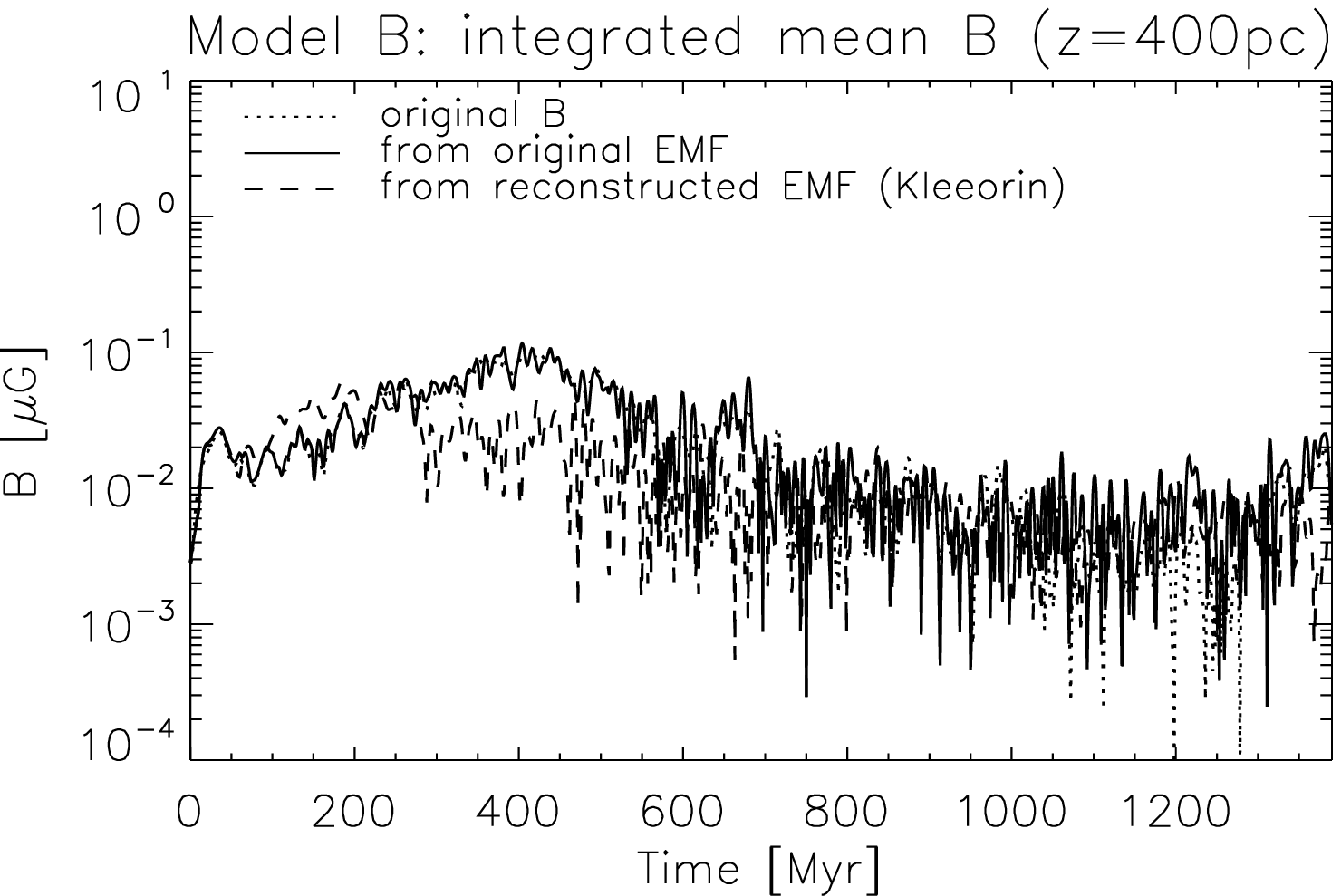}
 \plotone{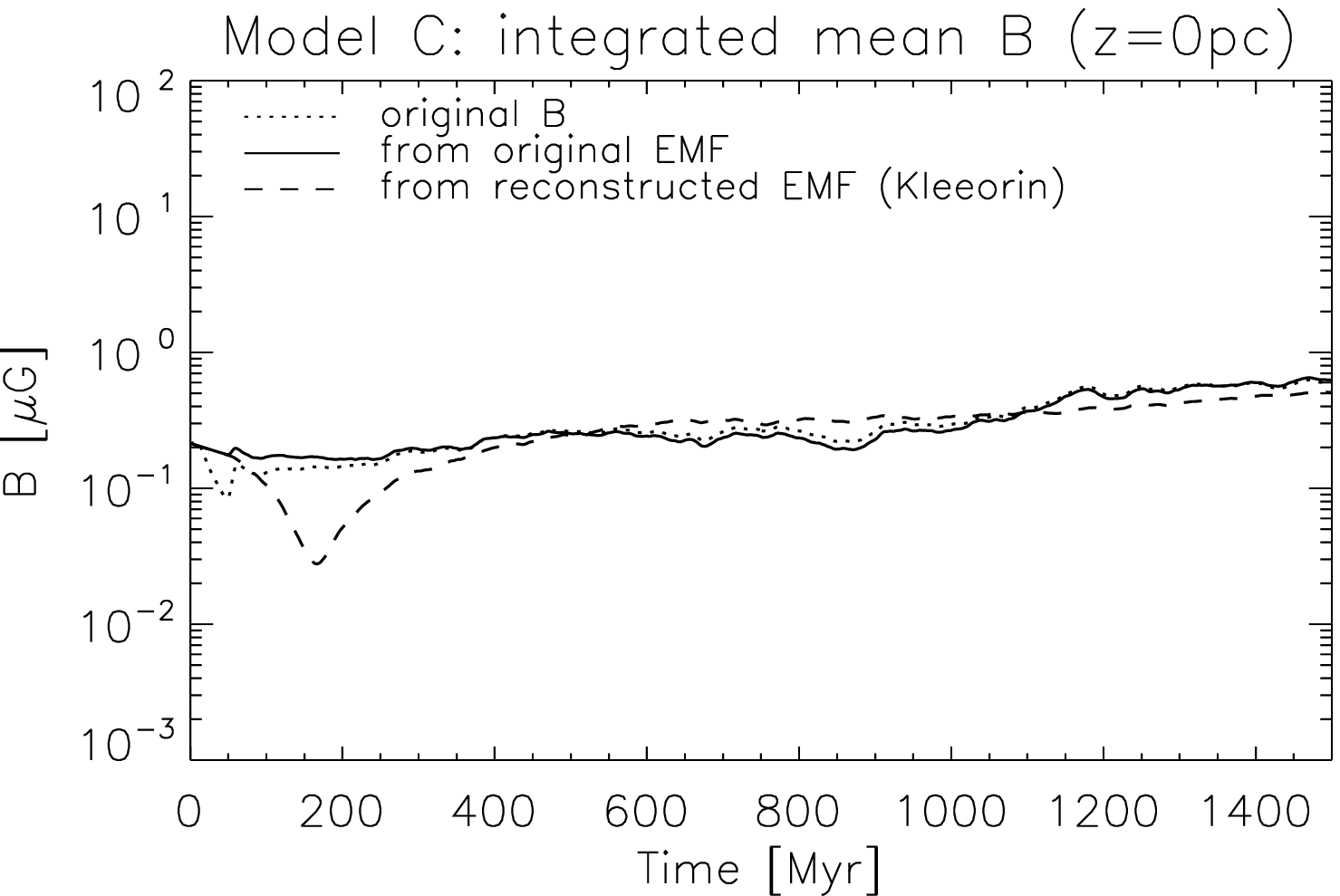}
 \plotone{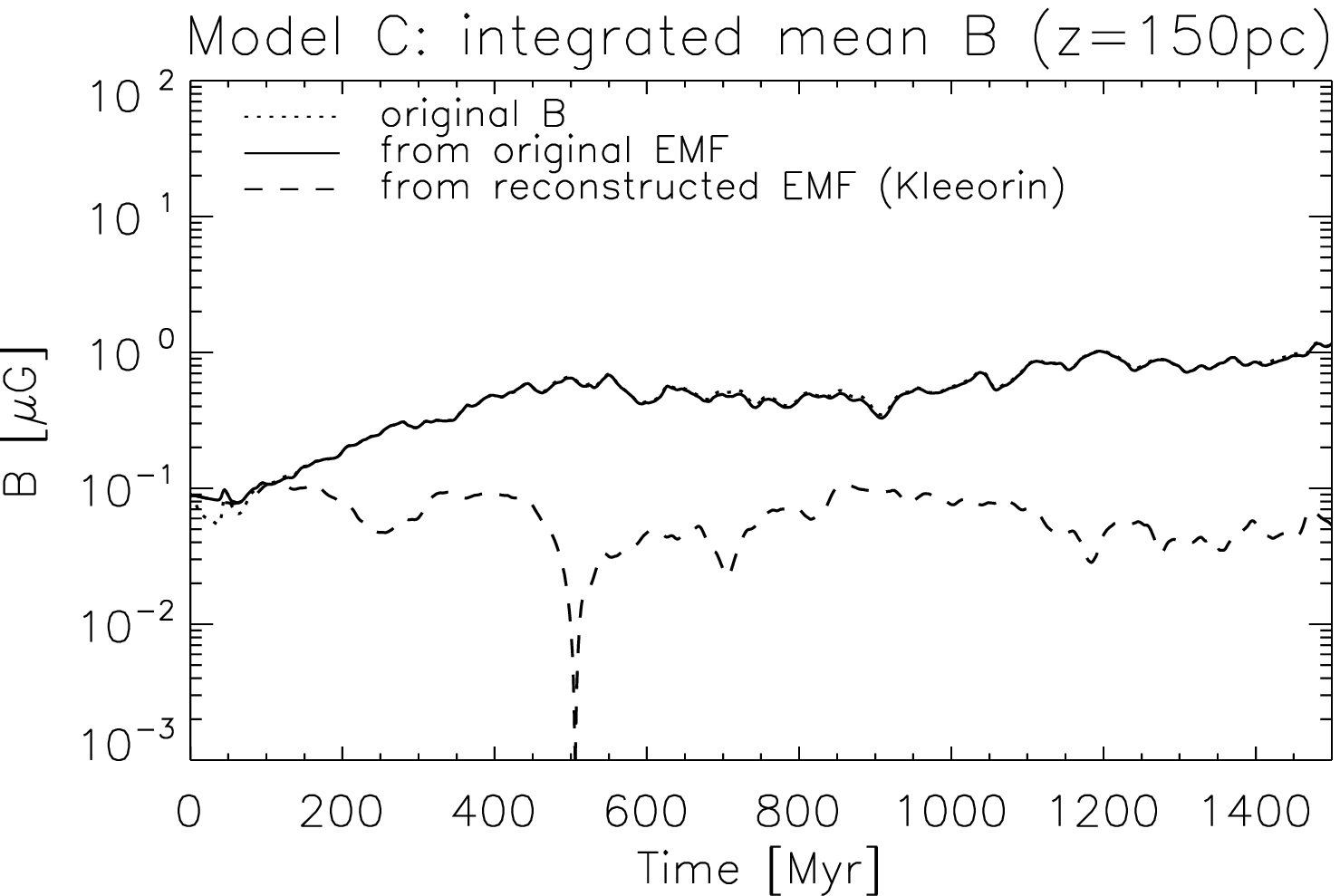}
 \plotone{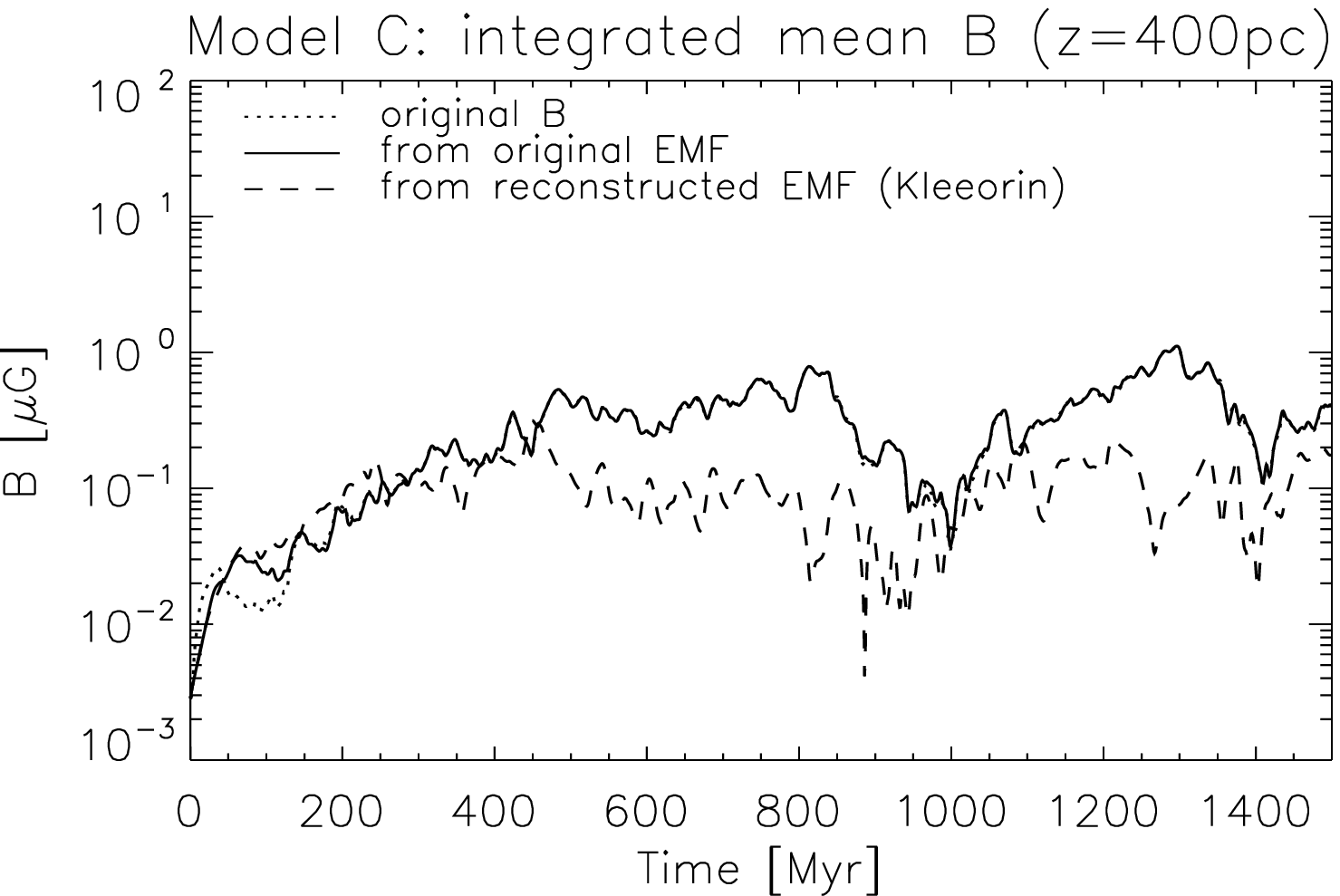}
 \plotone{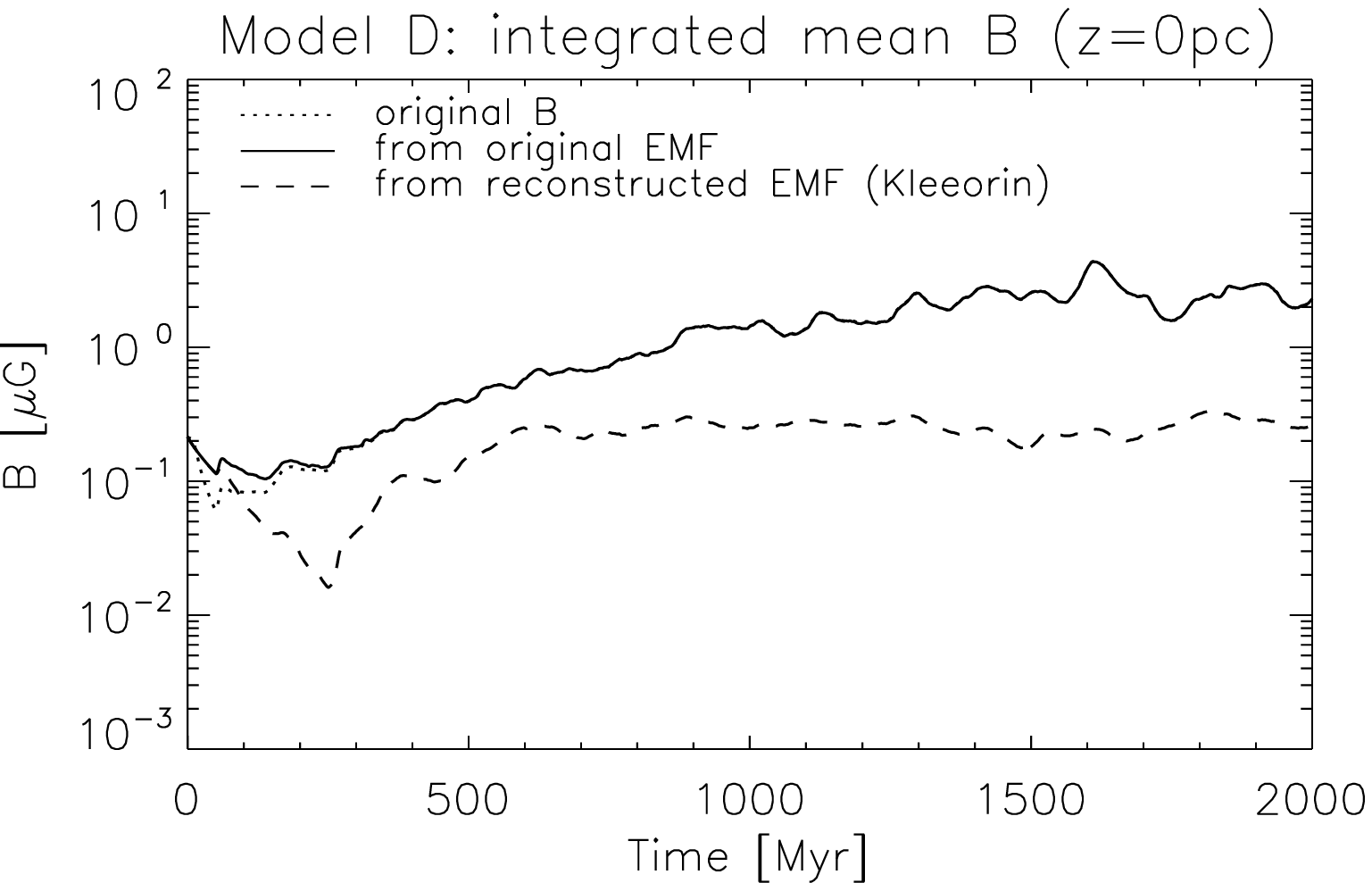}
 \plotone{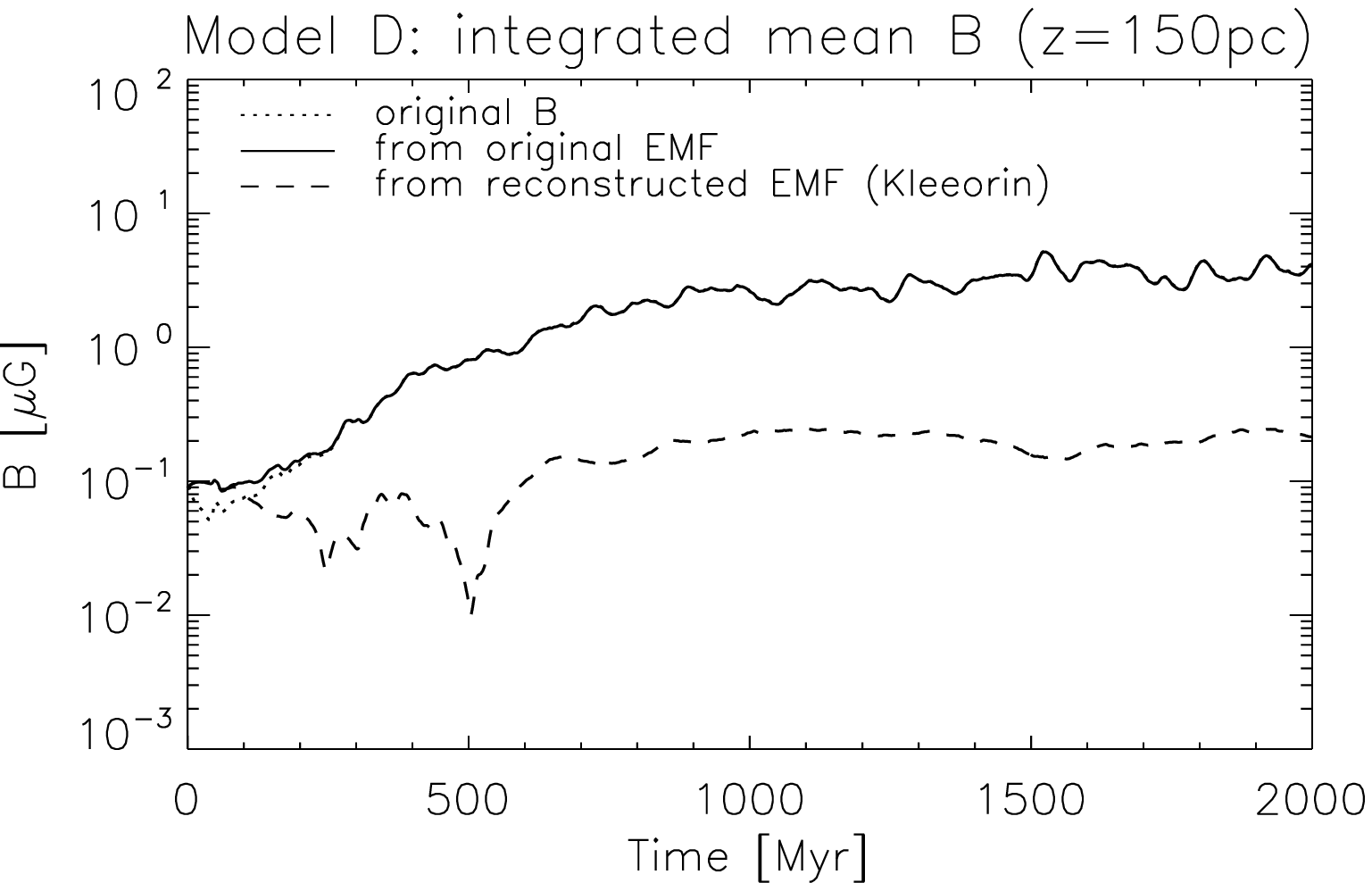}
 \plotone{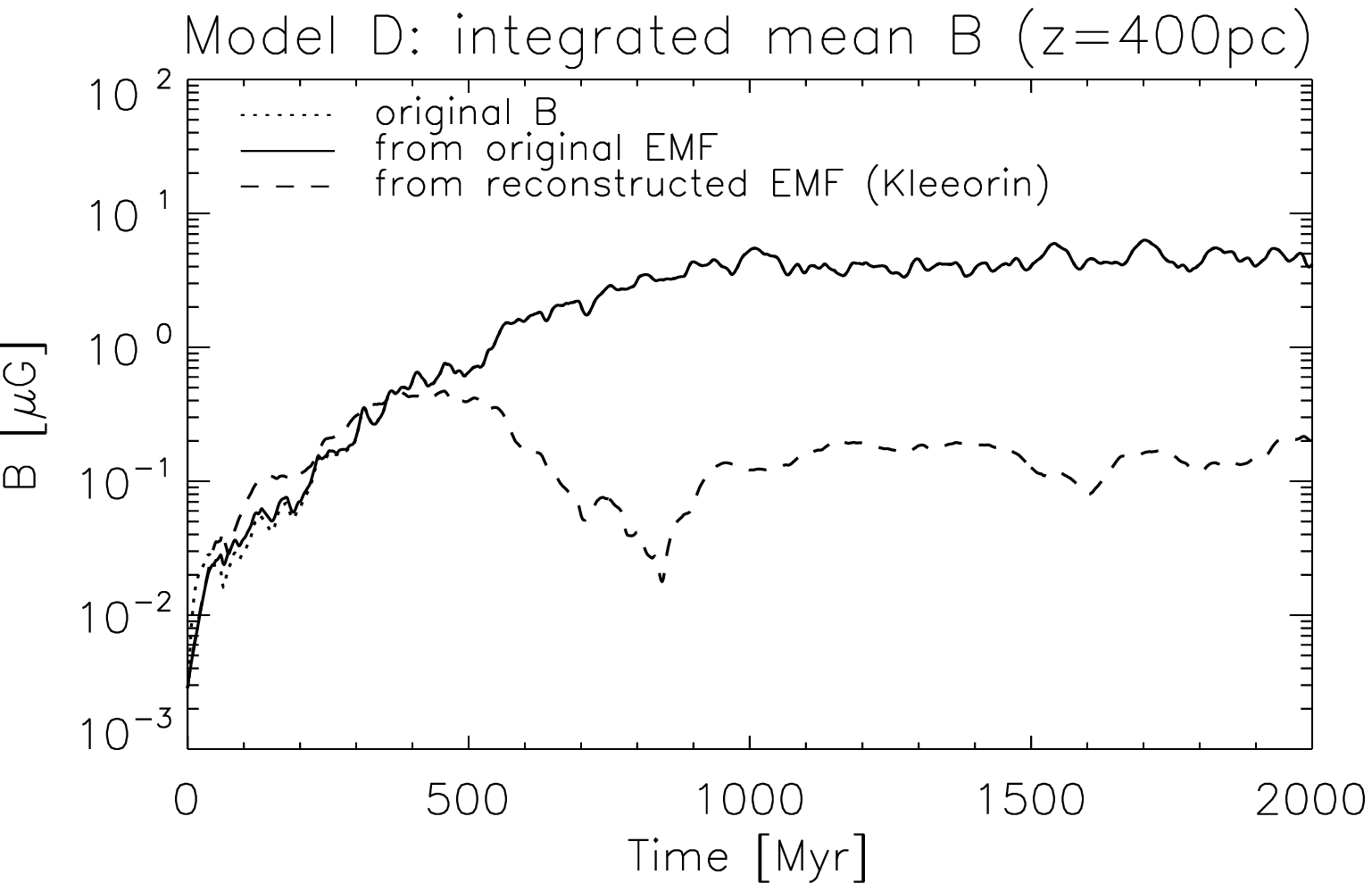}
 \caption{Time evolution of the absolute values of the mean magnetic field integrated from original (solid lines) and reconstructed (dashed lines) ${\cal E}$ for three vertical positions Z=0, 150, 400pc (left, middle and right columns, respectively) for models A, B, C and D (from top to bottom row). The reconstructed  ${\cal E}$ was calculated using the Kleeorin-Rogachevski approach. We also show the original mean magnetic field taken directly from simulations (dotted lines). \label{fig:int_mag_kleeorin}}
\end{figure*}

%
\subsubsection{Reconstruction from the Blackman-Field approach}
\label{sec:integration_blackman}

In the Blackman-Field approach the information about the electromotive force is incomplete, because this approach reconstructs only the component of the ${\cal E}$ parallel to the local mean magnetic field $\bar\mathbi{B}$. The perpendicular component of ${\cal E}$ is lost and cannot be recovered in this approach. Moreover, the ${\cal E}_\parallel$ is now a scalar field. Due to this limitation we need to add an additional step in which all three X, Y and Z-components of ${\bf \cal E}_\parallel$ must be reconstructed. The simplest way to perform a projection of the ${\cal E}_\parallel$ on the direction of $\bar\mathbi{B}$ taken directly from the simulations. We should note, that by taking only ${\cal E}_\parallel$ into integration, even that calculated directly from the simulations, does not guarantee a perfect reconstruction of the $\bar\mathbi{B}$.

The resulting mean magnetic fields integrated from EMFs, which were reconstructed by the Blackman-Field approach are shown in Figure~\ref{fig:int_mag_blackman}. This figure presents the integrated mean magnetic field for all four models: A, B, C, and D (from top to bottom) for three different coordinates Z=0, 150 and 400pc (left, middle, and right columns, respectively). We see that all integrations give different evolutions, which are not always in the agreement with the original $\bar\mathbi{B}$. For instance, in Figure~\ref{fig:int_mag_blackman} we see, that the integration of the EMF taken directly from models A--D delivers a good agreement with the original $\bar\mathbi{B}$ only during the short initial period of the evolution. Later on, the reconstruction of ${\cal E}_\parallel$ from the Blackman-Field approach provides significantly larger growth of the mean fields, especially for z=0 and 150pc. When the altitude is higher, the dashed lines corresponding to the intergration of the Blackman-Field's ${\cal E}_\parallel$ show much larger discrepancies than the solid lines for integration of the original ${\cal E}_\parallel$, but they still unveil the growth of the mean magnetic field $\bar\mathbi{B}$. In models C and D (two lower rows in Figure~\ref{fig:int_mag_blackman}), the better compatibility with the original $\bar\mathbi{B}$ comes from the integration of ${\cal E}_\parallel$ taken from the simulation. In these two models, the Blackman-Field approach simply overamplifies the authentic evolution of the mean magnetic field, producing up to two orders of magnitude larger $\bar\mathbi{B}$. For higher altitudes (plots on the right column), the evolutions are again comparable, although we still see a clear tendency to overamplify the magnetic field by the Blackman-Field approach. The model B is exceptional, because the amplification of $\bar\mathbi{B}$ takes place only in the initial stage of the evolution. Later on, the strength of the mean magnetic field declines. Despite that fact, the integration of the actual ${\cal E}$ gives almost constant mean magnetic field at the midplane, while only a small growth of $\bar\mathbi{B}$ at the initial stage at higher heights. From time T=400Myr we do not see any essential amplification of $\bar\mathbi{B}$. The Blackman-Field approach gives again the amplification of magnetic field at the initial stage. Later on, it varies or even declines, what is well seen at higher altitudes.

The above analysis draws a conclusion that the Blackman-Field approach reconstructs the ${\cal E}_\parallel$ preserving its property of amplification of the magnetic field. However, this approach often overamplify the  mean magnetic field, even in the situations when the strength of the actual magnetic field decreases. Moreover, we see another conclusion related maybe not directly to the Blackman-Field approach, but rather to the properties of the electromotive force. Model A shows that the parallel component of the electromotive force contributes to the amplification of $\bar\mathbi{B}$ later than the actual amplification is observed. This is justified by the fact, that the electromotive force reconstructed from its component parallel to $\bar\mathbi{B}$ is incomplete. This could indicate, that the amplification  of $\bar\mathbi{B}$ during the initial stage comes mainly from the remaining perpendicular component of ${\cal E}$.

%
\subsubsection{Reconstruction from the Kleeorin-Rogachevski approach}
\label{sec:integration_kleeorin}

The mean magnetic fields integrated from the electromotive force reconstructed by the Kleeorin-Rogachevski approach are shown in Figure~\ref{fig:int_mag_kleeorin} for the same models and altitudes as previously. Here,  the result of integration is more consistent with the actual evolution of $\bar\mathbi{B}$, because we integrate full ${\cal E}$, and not only its component parallel to $\bar\mathbi{B}$. In the top row of Figure~\ref{fig:int_mag_kleeorin} we see the mean magnetic fields integrated for model A, which we also take into account in order to be consistent with the BF02 presentation. The Kleeorin-Rogachevski approach reproduces the mean magnetic field relatively well at z=0pc, however, only for the initial several hundred Myrs of evolution. Later on, the discrepancies  are larger. The growth rate of $\bar\mathbi{B}$ {resulting from} this approach starts to decrease, while the growth  rate of the actual field is relatively constant. At the altitudes 150 and 400pc the growth of $\bar\mathbi{B}$ is very consistent with that integrated from ${\cal E}$ up to time about 1500pc. Later, the approach gives again weaker magnetic field, by about one orders of magnitude.

In model B (second row in Fig.~\ref{fig:int_mag_kleeorin}), the mean magnetic field integrated from the original ${\cal E}$ and the mean magnetic field taken directly from simulations are more or less in agreement in the whole  time period, while the reconstructed  curves show much larger discrepancies, especially at z=0pc. This model is less interesting, because we do not observe an amplification of $\bar\mathbi{B}$ here.   In  model  C  the reconstruction of the mean magnetic field  given by the original ${\cal E}$  is really very good, while the curves reconstructed according Kleeorin-Rogachevski approach are good only for the height z=0pc. At two higher z  (150 and 400pc) show that this approach gives much lower values of the mean magnetic field. This approach fails completely also in model D, where the large resistivity responsible for the strong diffusion leads to a fast growth of $\bar\mathbi{B}$.  Here, the Kleeorin-Rogachevski reconstruction gives only marginal amplification of the mean magnetic field.

%
\section{Discussion}
\label{sec:discussion}

In the present project the model of the CR-driven galactic dynamo is analyzed in order to determine its linear or non-linear character.  We analyze the conditions for validity of linear approximation of the dynamo equation in our model. First, we check the main assumptions of the dynamo theory: the Reynolds rules, the separation of the large and small scales -- both magnetic and kinetic. Then we compare the ratios of different terms appearing in the equation for time evolution of the fluctuating part of the magnetic field. We applied averaging over the horizontal planes. Further investigations revealed that the condition of scale separation is violated in our numerical model. In this sense our models   do not fulfill the assumptions  of dynamo theory (see \S\ref{sec:conditions}). As far as the scale separation is concerned, our magnetic and kinetic spectra computed for all chosen numerical models (A, B, C and D) are flat and are characterized by the same slope at all scales. This means that in our experiment no scale separation occurs. The next essential problem concerns the comparison of the different EMF terms. The plots shown in \S\ref{sec:tests} indicate the similarity among all experiments. The terms, which are normally neglected in the linear dynamo theory should  be  taken into account (see BF02 for comparison). For this reason the discrepancy between actual EMF  and the reconstructed EMF is present in majority of experiments.

We have  check subsequently, how the dynamo nonlinear approximations presented in the literature \cite[BF02 and][]{kleeorin03} fit the electromotive force obtained from our models (A, B, C and D). We compute BF02 electromotive force and present that it does not follow the time-evolution of the EMF obtained directly from our models. In particular,  their values diverge from the original ones, and ensure too fast growth of energy of the reconstructed mean magnetic field. The discrepancies may result from the truncation of the expansion series and a lack of the scale separation, the anisotropy of the turbulence (see \S\ref{sec:tests}), or the compressibility of the gas.

The approximations of \cite{rogachevskii00,rogachevskii01,rogachevskii03} are used by many authors \cite[e.g.][]{kleeorin03,brandenburg04,brandenburg05a}. That is why we also check if their anisotropic and quenched dynamo coefficients fit our modeled EMF. The authors applied \cite{raedler80} prescriptions, but neglected three of four terms of the EMF. It is easy to notice (see \S\ref{sec:fitting}) that the electromotive force obtained from our numerical model of the cosmic ray driven dynamo is different from the Kleeorin one. The obtained growth of the energy of the mean magnetic field reconstructed from the Kleeorin-Rogachevski approach is satisfactory only in models A and B and is too slow in other models with larger resistivity in comparison with numerical experiments.

 As we mentionned already, we did not discuss the effects of magnetic helicity
transport and its relation to the alpha effects, although the terms related to
the magnetic helicity current have been included in both Blackman-Field and in
Kleeorin-Rogachevskii approaches through the magnetic part of the
$\alpha$-tensor. Considerations of magnetic helicity conservation is planned as
an extension of the present work.
The role of magnetic helicity current in our dynamo model can be
anticipated from the paper by \cite{shukurov06}, where helicity losses through
galactic fountain flow are examined in context of catastrophic quenching of
galactic dynamos. These authors demonstrate that a vertical galactic wind can
advect magnetic helicity and make it possible for the large-scale magnetic field
to reach the strength of 10 \% of the equipartition value.

The simmilarity of our model and the galactic fountain model is quite obvious.
In both models convective motions are present, although their origin is
different. In our case cosmic ray buoyancy, and in fountain model the buoyancy
of hot gas drive convective motions reaching vertical speeds of 200 km/s. We
point out to the apparent advection of magnetic fluctuations at large
altitudes, which indicate the possibility of carrying the magnetic helicity
current. Furthermore, the qualitative behaviour of our models is simmilar to
those by \cite{shukurov06}. The characteristic growth and decay phases are
apparent in our dynamo models with low resistivity and in galactic fountain
models with low vertical winds. The saturation of magnetic field growth at
about  10\% of equipartition value is also a feature of both types of models for
larger resistivity values and moderate verical winds respectively.

%
\section{Conclusions}
\label{sec:conclusions}

Our paper presents the analysis of the nonlinear electromotive forces in the model of the  cosmic-ray driven dynamo in the galactic disk. Our results may be summarized by the following conclusions:
\begin{enumerate}
\item {Neither the velocity nor the magnetic field scale separation occurs in our model.}

\item { The electromotive forces in the  cosmic-ray driven dynamo model are nonlinear, but none of the two examined nonlinear approaches is capable of reproducing electromotive forces in the numerical experiments correctly.}

\item {Various nonlinear prescriptions of the dynamo coefficients have been proposed by other outhors, however, they are not capable of reconstructing the electromotive force resulting from experiments of cosmic-ray driven dynamo. Moreover the reconstructions of the magnetic field  produce too fast or too slow growth of the magnetic energy in comparison with the results of cosmic-ray driven dynamo numerical experiments.}

\end{enumerate}

 Extension of the present work, including considerations of
magnetic helicity conservation will be presented in the forthcoming paper.

\acknowledgements
This work was partly supported by the Polish Committee for Scientific Research from the grants 1 P03D 004 26 and 1 P03D 002 28.

\end{document}